\documentclass[aps,pre,reprint,secnumarabic,nofootinbib,floatfix,a4paper,
longbibliography,notittlepage,superscriptaddress]{revtex4-1}

\usepackage{graphicx,setspace}
\usepackage{amsfonts,amsmath,amssymb,bm}
\usepackage[dvips,usenames]{color}

\begin{document}

\title{Effective squirmer models for self-phoretic chemically active 
spherical colloids
}

\author{M.N. Popescu}
\email[corresponding author: ]{popescu@is.mpg.de}
\affiliation{Max-Planck-Institut f\"{u}r Intelligente Systeme, 
Heisenbergstr. 3, D-70569
Stuttgart, Germany}
\affiliation{IV. Institut f\"ur Theoretische Physik, 
Universit\"{a}t Stuttgart,
Pfaffenwaldring 57, D-70569 Stuttgart, Germany}
\author{W.E. Uspal}
\affiliation{Max-Planck-Institut f\"{u}r Intelligente Systeme, 
Heisenbergstr. 3, D-70569
Stuttgart, Germany}
\affiliation{IV. Institut f\"ur Theoretische Physik, 
Universit\"{a}t Stuttgart,
Pfaffenwaldring 57, D-70569 Stuttgart, Germany}
\author{Z. Eskandari}
\affiliation{Max-Planck-Institut f\"{u}r Intelligente Systeme, 
Heisenbergstr. 3, D-70569
Stuttgart, Germany}
\affiliation{IV. Institut f\"ur Theoretische Physik, 
Universit\"{a}t Stuttgart,
Pfaffenwaldring 57, D-70569 Stuttgart, Germany}
\author{M. Tasinkevych}
\affiliation{Centro de F{\'i}sica Te{\'o}rica e Computacional, 
Departamento de F{\'i}sica, Faculdade de Ci{\^e}ncias, 
Universidade de Lisboa, Campo Grande, P-1749-016 Lisboa, 
Portugal}
\author{S. Dietrich}
\affiliation{Max-Planck-Institut f\"{u}r Intelligente Systeme, 
Heisenbergstr. 3, D-70569
Stuttgart, Germany}
\affiliation{IV. Institut f\"ur Theoretische Physik, 
Universit\"{a}t Stuttgart,
Pfaffenwaldring 57, D-70569 Stuttgart, Germany}

\date{\today}

\begin{abstract}
Various aspects of self-motility of chemically active colloids in Newtonian 
fluids can be captured by simple models for their chemical activity plus 
a  phoretic slip hydrodynamic boundary condition on their surface. For particles 
of simple shapes (e.g., spheres) -- as employed in many experimental studies -- 
which move at very low Reynolds numbers in an unbounded fluid, such models of 
chemically active particles effectively map onto the well studied so-called 
hydrodynamic squirmers [S. Michelin and E. Lauga, J. Fluid Mech. \textbf{747}, 
572 (2014)]. Accordingly, intuitively appealing analogies of ``pusher/puller/neutral'' 
squirmers arise naturally. Within the framework of self-diffusiophoresis we 
illustrate the above mentioned mapping and the corresponding flows in an unbounded 
fluid for a number of choices of the activity function (i.e., the spatial distribution 
and the type of chemical reactions across the surface of the particle). We use the 
central collision of two active particles as a simple, paradigmatic 
case for demonstrating that in the presence of other particles or 
boundaries the behavior of chemically active colloids may be \textit{qualitatively} 
different, even in the far field, from the one exhibited by the corresponding 
``effective squirmer'', obtained from the mapping in an unbounded fluid. This 
emphasizes that understanding the collective behavior and the dynamics under 
geometrical confinement of chemically active particles necessarily requires to 
explicitly account for the dependence of the hydrodynamic interactions on 
the distribution of chemical species resulting from the activity of the particles.
\end{abstract}



\maketitle

\section{\label{intro} Introduction}

During the last decade there has been significant interest in the development of 
chemically active, micron-sized particles (or drops) which are capable of 
moving within a fluid environment by promoting chemical reactions which 
involve the 
surrounding solution. Such particles exhibit motility in the absence of external 
forces or torques acting on them or on the fluid. Various types of 
motile, chemically active particles have been proposed and studied 
experimentally (see, e.g., Refs. 
\cite{Ismagilov2002,Paxton2004,ozin2005,Paxton2006,Paxton2006a,solovev2009,
mirkovic2010,Fournier-Bidoz2005,Howse2007,volpe11,Golestanian2012,Bechinger2013a,
Bechinger2013b,Fisher2014,Golestanian2014,Bechinger2014,Ma2016,
Herminghaus2014,Seemann2016,Kroy2016,Bechinger2016}). The mechanisms of motility, 
in particular for self-phoresis (on which we shall focus here), have been the topic 
of numerous theoretical studies (see, e.g., 
Refs.  \cite{Golestanian2005,Golestanian2007,Kapral2007,Julicher2009,Popescu2011EPL,
Seifert2012a,Seifert2012b,Kapral2013,Koplik2013,Lowen2011,
Michelin2015,Gommper2015,Popescu2016,Stark2016,deGraaf2015,Gleb2017,Lammert2016,
Brown2017}). Thorough and insightful reviews of the developments in the area 
of man-made motile colloids, as well as in the related one of biological 
microswimmers, are provided by Refs. 
\cite{Lauga2009,Ebbens2010,SenRev,Gommper2015_rev,Bechinger2016_rev,Posner2017}.

Similar to the case of classic phoresis -- in which gradients of thermodynamic 
fields (such as chemical potentials or temperature) are imposed externally -- 
self-phoretic motion results from the distinct interactions between the 
particle and the various molecular species, i.e., reactant and reaction product 
molecules, which are inhomogeneously distributed in the solution due to the 
chemical reactions promoted on parts of the surface of the 
particle \cite{Derjaguin1966,Anderson1989}. The same interactions (due to the 
action-reaction principle) lead also to hydrodynamic flow of the solution 
which consists of solvent, reactant, and reaction products. The spatially 
varying number densities 
of the reactant and product molecules and the hydrodynamic flow of the solution will 
be referred to as chemical and hydrodynamic fields associated with the active 
particle, respectively; as noted above, the two fields are coupled.

The typical experimental realizations of self-phoresis involve aqueous solutions, 
molecular solutes, and micrometer-sized particles moving at speeds of the order 
of a particle diameter per second. Therefore, we focus the discussion on the 
case of Newtonian fluids and to the case in which the P{\'e}clet number of the 
solutes and the Reynolds number of the hydrodynamic flow are very small 
\cite{Ebbens2010,SenRev,Bechinger2016_rev,Posner2017}. 
In this case the transport of molecular species by diffusion dominates advection 
and viscous friction dominates over inertial effects as far as hydrodynamics 
is concerned. Moreover, in many cases the spatial range of the interactions 
between the molecular species and the particle is much smaller than the size of the 
particle. This allows one to express the aforementioned coupling in terms of a 
``phoretic slip'' hydrodynamic boundary condition at the surface of the 
colloidal particle: there the flow velocity (relative to the particle) is 
proportional to the gradients of the number densities of the solutes along the 
surface of the particle \cite{Derjaguin1966,Anderson1989}. 

This phoretic-slip formulation significantly reduces the complexity of 
determining the hydrodynamic field associated with the motion.\footnote{Even 
if analytical solutions are not available (as in general it is the case 
due to, e.g., a non-spherical shape of the particle or reduced symmetries of the system),  
the phoretic slip approximation allows one to employ efficient numerical methods, 
such as the Boundary Element Method (BEM) \cite{pozrikidis02}, 
which involve only integrals over the surface of the particle (and of the 
confining boundaries, if present).} For instance, for spherical particles with 
axially symmetric surface properties (on which we focus here) 
immersed in an unbounded fluid, the flow can be inferred directly from the 
available solutions of the Stokes equations \cite{HaBr73}. In the 
context of motility of microorganisms, this leads to the well known ``squirmer'' 
model proposed by Lighthill and Blake \cite{Lighthill1952,Blake1971} (see also the 
recent generalization obtained in Ref. \cite{Lauga2014}). The squirmer model 
successfully captures many of the qualitative features exhibited by swimming 
microorganisms in unbounded fluids or near surfaces 
\cite{Lauga2006,Lauga2008,Goldstein2010,Lauga2014b,Gommper2015}). Concepts such as 
``pullers'' and ``pushers'' have emerged from this model and have turned out to be 
physically insightful concerning, e.g., the understanding of various behaviors 
exhibited by swimming microrganisms near boundaries
\cite{Lauga2006,Lauga2008,Lauga2014b,Gommper2015,Yeomans2016,Ignacio2010,
Ishimoto2013,Holm2016,Brown2016,Spagnolie2012,Spagnolie2015,Shelley2014}, 
the hydrodynamic interactions between microrganisms,  
\cite{Pedley2006,Alexander2017,Ignacio2010b}, or the aggregation and ordering 
behavior in suspensions of squirmers \cite{Ignacio2013,Sano2016,Shelley2008,Lauga2016}. 
In the context of artificial, man-made active particles, the effective mapping 
onto squirmers noted above has been explicitly carried out for a spherical ``hot'' 
colloid \cite{Majee2013} or a spherical particle with spatially varying phoretic 
mobility \cite{Michelin2014,Liverpool2016} (see, c.f., Sec. \ref{squirm}).

Recent studies have shown that when active particles move near walls, 
fluid interfaces, or in the vicinity of other -- active or inert -- particles 
(i.e., situations which typically do occur in experiments, see, e.g., 
Refs. \cite{Paxton2004,SenRev,Golestanian2007,Bechinger2014,Baraban2012,
Bechinger2013b})
they may exhibit complex behaviors, such as surface-bound steady 
states \cite{Pine2013,Uspal2015a,Koplik2016,Brown2016b}, long-ranged effective 
interactions \cite{Leshansky1997,Alvaro2016}, ``guidance'' 
by topographical or chemical features 
\cite{Howse2015,Simmchen2016,Uspal2016,Popescu2017a,Uspal2018a}, or enhanced 
velocity under 
geometrical confinement \cite{Wei2016,Popescu2009}. If in addition they are 
exposed to external flows or force fields, a very rich and interesting 
phenomenology appears, including, e.g., rheotaxis 
\cite{Pine2013,Uspal2015b}, cross-stream rheotaxis \cite{Uspal2018b}, 
and gravitaxis \cite{Ebbens2013,Stark2011,Bechinger2014}. On the other 
hand, theoretical studies of self-phoresis of active particles in the vicinity of 
confining surfaces suffer from the fact that even for conceptually simple 
models of spherical active particles \cite{Golestanian2005,Golestanian2007} it 
is difficult to analytically solve the equations describing the motion. Therefore, 
either approximate far-field analyses \cite{Liverpool2015,Liverpool2016} or 
numerical methods (or a combination of the two) 
\cite{Uspal2015a,Uspal2015b,Uspal2016,Simmchen2016,Howse2015,Popescu2017b}) have 
been employed in order to obtain the corresponding solutions. In a few cases 
formal analytical solutions can be obtained in the form of series representations
\cite{Popescu2011EPL,Koplik2016,Koplik2016b,Reigh2015}. Since, however, the 
corresponding coefficients must be determined numerically, an intuitive 
understanding of the result is impeded.

In view of the aforementioned exact mapping (in unbounded space) to squirmer 
models and in view of the wealth of knowledge concerning the behavior of squirmers in 
confinement, it is therefore not surprising that further analogies with 
pushers or pullers have been made. For example, such analogies have been used 
in order to 
interpret an attractive or repulsive character of the effective interaction 
between an active particle and a wall \cite{Howse2015} or a larger inert 
particle \cite{Brown2016b} as potentially discriminating between distinct mechanisms 
of motility. However, these analogies should be considered cautiously. In 
contrast to squirmers with a prescribed slip, which is independent of 
the configuration (i.e., distance and orientation of the particle relative to 
the wall) \cite{Lauga2006,Spagnolie2012,Pedley2006,Ishimoto2013},  for active 
particles the disturbance of the distributions of densities of chemical 
species (e.g., due to a confining surface or the presence 
of other particles) leads to changes in the phoretic slip. Consequently, active 
particles exhibit complex hydrodynamic interactions which are modulated by these 
disturbances of the chemical fields 
\cite{Uspal2015a,Uspal2015b,Simmchen2016,Michelin2015,Popescu2011EPL,
Liverpool2015}. 

Here we employ a basic model of self-diffusiophoresis of chemically active 
particles with several choices for activity functions in order to 
illustrate the aforementioned mapping \cite{Majee2013,Michelin2014,Liverpool2016} 
onto effective squirmer models and the corresponding hydrodynamic flows in 
unbounded space. By turning to the conceptually simple, but physically 
insightful case of a central collision between two active particles, we 
demonstrate that, even in the far field, an active particle and its 
corresponding effective squirmer model may exhibit qualitatively different 
effective interactions. 

\section{\label{model_squirm} Model of chemically active spherical colloids 
and its mapping onto a squirmer}

As the model for an active particle we use the one introduced in 
Refs. \cite{Golestanian2005,Golestanian2007}. This model, which has been 
analyzed further in, e.g., Refs. 
\cite{Howse2007,Golestanian2007,Golestanian2012,Popescu2010,Seifert2012a,Koplik2013,
Michelin2014}, is conceptually simple but nonetheless captures the relevant 
phenomenology observed in experimental studies 
\cite{Howse2007,Golestanian2012,Simmchen2016}. The model is succinctly summarized 
below.

\subsection{\label{model_def} Model of active particles}

\begin{figure}[!htb]
    \centering
    \includegraphics[width=.78\columnwidth]{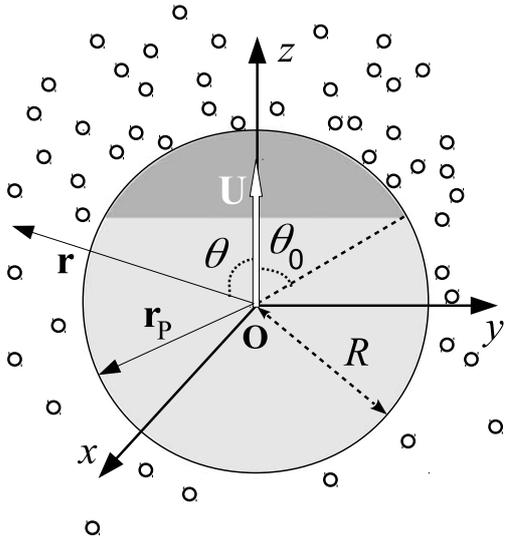}
     \caption{
\label{fig1}
Schematic illustration of a chemically active spherical particle of radius 
$R$ immersed in 
an unbounded solution and moving with velocity $\mathbf{U}$ (white thick arrow). 
The two parts of the surface are spherical caps (dark gray and light 
gray areas, respectively) delimited by the latitudinal circle with 
polar angle $\theta_0$, and have different properties in terms of 
``chemical activity''. For instance, they can release or annihilate a solute 
molecule (small white circles) which diffuses in 
the surrounding solution, or they are chemically inert (i.e., neither releasing 
nor annihilating solute). Due to the distinct ``chemical activity'' of the 
two parts, a gradient in the solute density around the particle builds up 
(schematically indicated by the non-uniform distribution of the small white 
circles). The origin $O$ of the fixed (laboratory) system of coordinates is chosen 
such that it coincides with the instantaneous position of the center of the particle, 
and the $Oz$ axis is aligned with the axis of symmetry; $r$, $\theta$, and 
$\phi$ (not shown) denote the spherical coordinates defined in the usual way in 
the $Oxyz$ frame of reference; $\mathbf{r}_P$ is a point on the surface of 
the particle.
}
\end{figure}

The ``activity''  of a particle is represented as sources (or sinks) of a 
molecular solute which diffuses in the surrounding solution, taken as 
an incompressible Newtonian liquid of viscosity $\mu$ (see Fig. \ref{fig1}). 

The simplest realization of sources and sinks is the so-called ``constant-flux'' 
boundary condition \cite{Golestanian2007,Brady2011}, i.e., at each point 
$\mathbf{r}_P$ at the surface of the particle the normal component of the solute 
current takes a prescribed, time-independent value given by
${\cal K}(\mathbf{r}_P) := Q f(\mathbf{r}_P)$ (with units m$^{-2} \times$ 
s$^{-1}$), where $Q > 0$ is a prefactor which takes care of the 
dimensionality so that $f(\mathbf{r}_P)$ is dimensionless. The latter will 
be referred to as the ``activity function'' and accounts for the sign of 
${\cal K}(\mathbf{r}_P)$ corresponding to ``production'' ($f > 0$) 
or ``annihilation'' ($f < 0$) of solute at $\mathbf{r}_P$. The more complex 
case of a reaction with first order kinetics, i.e., ${\cal K}(\mathbf{r}_P)$ 
being proportional to the local density of a ``fuel'' species at $\mathbf{r}_P$, 
will be considered separately in, c.f., Sec. \ref{first_order_kin}.

The motion of the particle and the diffusion of the molecular solute are assumed 
to be such that the P{\'eclet} number of the solute and the Reynolds number of 
the flow are very small, such that the number density of the 
solute relaxes towards the steady state distribution $c(\mathbf{r})$ much 
faster than the characteristic time scale of the motion of the particle (e.g., the 
time needed for the particle to pass a distance equal to its radius). The 
interaction of the solute 
molecules (in excess to the one of a solvent molecule \cite{Mazur_book,Julicher2009}) 
with the surface of the particle is encoded into a phoretic mobility coefficient 
$b(\mathbf{r}_P) := b_0 g(\mathbf{r}_P)$ (with the units m$^5$/s), such 
that $g(\mathbf{r}_P)$ is dimensionless and $b_0 > 0$ is a characteristic 
value (e.g., $b_0 = (4\pi R^2)^{-1} \int d^2 \mathbf{r}_P |b(\mathbf{r}_P)|$). 
The mobility coefficient describes the phoretic slip boundary condition 
\cite{Anderson1989,Derjaguin1966}, 
\begin{equation}
\label{eq:phor_slip_connect}
\mathbf{u}_s (\mathbf{r}_P) = -b_0 g(\mathbf{r}_P) \nabla_{||} c(\mathbf{r}_P)\,,
\end{equation}
for the hydrodynamic field $\mathbf{u}(\mathbf{r})$ of the surrounding solution. 
The phoretic mobility coefficient can be either positive or negative, depending 
on the attractive or repulsive character of the excess interaction of the 
solute with the surface. The sign of $b(\mathbf{r}_P)$ will be accounted for 
by $g(\mathbf{r}_P)$. Typical experimental realizations 
\cite{Paxton2004,Howse2007,Bechinger2013b,Bechinger2016} are such that the  
surface (or the whole volume) of the particle consists of two parts composed 
of different materials but preserving axial symmetry. Accordingly, the 
surface of the model particle is divided into two spherical caps (the poles of 
which define the symmetry axis) corresponding to an opening polar angle $\theta_0$ 
(see Fig. \ref{fig1}). The activity function $f$ as well as, 
in general, the phoretic mobility $g$ differ over the two caps. 

The translational and angular velocities of the particle follow from the 
requirement of zero net force and torque on the particle, consistent with the 
case of overdamped motion. 

\subsection{\label{math_form} Active particle in an unbounded fluid}

The dynamics of this model particle is governed by boundary-value problems for the 
chemical and hydrodynamic fields $c(\mathbf{r})$ and $\mathbf{u}(\mathbf{r})$, 
respectively, and by the force balance on the particle. Accordingly, in an 
unbounded, quiescent fluid, one has:
\newline
\textbullet~\textit{Laplace equation for} $c(\mathbf{r})$:
\begin{subequations}
 \label{eq:bvp_for_c_thin}
\begin{equation}
\label{eq:diff_c_thin} 
\nabla^2 c(\mathbf{r}) = 0\,,
 \end{equation}
\textit{with the boundary conditions (BCs)}
\begin{equation}
\label{eq:bcs_for_c_part_thin} 
- D \left.\left[\mathbf{n} \cdot 
\nabla c(\mathbf{r})\right]\right|_{\mathbf{r} = \mathbf{r}_P} 
= Q f(\mathbf{r}_P)\,,
\end{equation}
where $D$ is the diffusion constant of the solute molecules, and
\begin{equation}
\label{eq:bcs_for_c_infty_thin} 
c(|\mathbf{r}| \to \infty) \to c_\infty\,.
\end{equation}
\end{subequations}
\noindent\textbullet~\textit{incompressible Stokes equations for} 
$\mathbf{u}(\mathbf{r})$:
\begin{subequations}
\label{eq:bvp_for_u_thin}
\begin{equation}
 \label{eq:Stokes_thin}
 \nabla \cdot \hat {\boldsymbol{\sigma}} = 0\,,~~~~ 
 \nabla \cdot \mathbf{u}(\mathbf{r}) = 0\,,
\end{equation}
\textit{with the BCs}
\begin{equation}
 \label{eq:bc_for_u_part_thin}
 \mathbf{u}(\mathbf{r}_P) 
= \mathbf{U}+\mathbf{u}_s (\mathbf{r}_P)\,,
\end{equation}
\begin{equation}
 \label{eq:bc_for_u_infty_thin}
 \mathbf{u}(|\mathbf{r}| \to \infty) = 0\,.
\end{equation}
\end{subequations}
In these equations, $\mathbf{u}_s (\mathbf{r}_P)$ is given by Eq. 
(\ref{eq:phor_slip_connect}), 
\begin{equation}
  \label{eq:stress}
  \hat{\boldsymbol{\sigma}} := - p \hat{\mathbf{I}} + \mu 
  \left[\nabla \mathbf{u} + \left(\nabla \mathbf{u} \right)^T \right]\,
\end{equation} 
denotes the stress tensor (for a Newtonian fluid of viscosity $\mu$) 
with the pressure $p(\mathbf{r})$. $( )^T$ indicates a 
transposed quantity; 
we use the convention that in the absence of an explicitly indicated operation 
two adjacent vectors (or vector operators) denote the tensor (dyadic) product. 

\noindent\textbullet~\textit{vanishing net force on the particle}\footnote{For the 
axisymmetric systems considered in Sects. \ref{model_squirm} - \ref{first_order_kin} 
the motion involves only translation along the axis of symmetry; thus here only 
the component of the force balance equation along this axis has to be considered.}:
\begin{equation}
\label{eq:zero_F_T_thin}
 \mathbf{F}_{ext} + \int\limits_{|\mathbf{r}| = R} dS\,  
\hat{\boldsymbol{\sigma}} \cdot \mathbf{n}  = 0\,,
\end{equation}
where $\mathbf{F}_{ext}$ denotes the external force on the particle.

In the following, we set $c_\infty = 0$ without loss of generality. This 
amounts to introducing $c(\mathbf{r}) \rightarrow \tilde c(\mathbf{r}) = 
c(\mathbf{r}) - c_\infty$ as the deviation from the ``bulk'' value $c_\infty$, 
which leaves Eqs. (\ref{eq:phor_slip_connect}) and (\ref{eq:bvp_for_c_thin}) 
unchanged. The model is therefore completely specified by providing the geometrical 
parameter $\theta_0$, the activity function $f(\theta)$, the phoretic mobility 
function $g(\theta)$, and the external forces acting on the spherical particle.

Before proceeding with the formal solution of 
Eqs. (\ref{eq:phor_slip_connect})-(\ref{eq:zero_F_T_thin}), dimensional analysis 
allows one to introduce the following quantities: \newline
(i) from Eq. (\ref{eq:bcs_for_c_part_thin}), a characteristic number density
\begin{subequations}
\label{def:char_scales}
\begin{equation}
 \label{eq:def_C0}
 C_0 := \frac {Q R}{D}\,;
\end{equation}
(ii) from Eqs. (\ref{eq:phor_slip_connect}) and (\ref{eq:def_C0}), a 
characteristic velocity 
\begin{equation}
 \label{eq:def_U0}
 U_0 := \frac{b_0 C_0}{R} = \frac{Q b_0}{D}\,;
\end{equation}
(iii) and, from Eqs. (\ref{eq:stress}) and (\ref{eq:zero_F_T_thin}), a 
characteristic force 
\begin{equation}
 \label{eq:def_p0_F0}
 F_0 := 6 \pi \mu R U_0\,,
\end{equation}
\end{subequations}
respectively. These provide the scales for the corresponding dimensional 
quantities.

\subsection{\label{squirm} Squirmer representation}

By expanding $f(\theta)$ in terms of Legendre polynomials,
\begin{subequations}
\label{eq:def_f}
\begin{equation}
 \label{eq:exp_f}
 f(\theta) = \sum\limits_{n \geq 0} f_n P_n(\cos\theta)\,,
\end{equation}
where
\begin{equation}
 \label{eq:f_coef}
 f_n = (n+1/2) \int\limits_0^\pi d\theta \, \sin \theta\,f(\theta) P_n(\cos\theta)\,,
\end{equation}
\end{subequations}
the solution of the diffusion problem (Eqs. (\ref{eq:bvp_for_c_thin})(a)-(c)) 
can be expressed in terms of a multipole expansion \cite{Golestanian2007}:
\begin{equation}
 \label{eq:c_series}
 \frac{c(r,\theta)}{C_0} = \sum\limits_{n \geq 0} \frac{f_n}{n+1} 
\left(\frac{R}{r}\right)^{n+1} P_n(\cos\theta)\,.
\end{equation}
The individual terms allow for clear physical interpretations: the first 
one (monopole)
corresponds to a source or sink (net release or annihilation), the second one 
corresponds to a dipole (fore-aft asymmetry in release or annihilation), etc. 
Combining Eqs. (\ref{eq:c_series}), (\ref{eq:phor_slip_connect}), and 
(\ref{Pn1_vs_Pn}), and defining \cite{Lighthill1952,Blake1971}
\begin{equation}
 \label{eq:def_Vn}
 V_n(\cos \theta) = \frac{2}{n (n+1)} P_n^1 (\cos\theta)\,,
\end{equation}
where $P_n^1(\cos\theta)$ denotes the associated Legendre function of degree $n$ 
and order 1 \cite{abram}, renders the phoretic 
slip \cite{Majee2013,Michelin2014,Liverpool2016}
\begin{eqnarray}
 \label{eq:phor_slip_squirm}
 \frac{\mathbf{u}_s}{U_0} &=& 
 -\left[ g(\theta) \sum\limits_{n \geq 1} \frac{f_n}{n+1} P_n^1 (\cos\theta) \right] 
\mathbf{e}_\theta \nonumber\\
 &:= & 
 \left(\sum\limits_{n \geq 1} B_n V_n(\cos\theta)\right) \mathbf{e}_\theta\,,
\end{eqnarray}
with 
\begin{equation}
 \label{eq:Bn_coef}
 B_n = - \left(n + \frac{1}{2}\right)
 \sum\limits_{k \geq 1} \frac{f_k}{k+1} {\cal I}_{n,k} \,,
\end{equation}
and 
\begin{equation}
 \label{eq:def_Ink}
{\cal I}_{n,k} := \dfrac{1}{2} \int\limits_{0}^{\pi} d \theta\,\sin\theta\, g(\theta) 
P_k^1 (\cos\theta) P_n^1 (\cos\theta)\,.
\end{equation}
Note that, with the exception of $f_0$, all coefficients $f_k$ in the 
expansion 
of the activity function contribute to each of the coefficients $B_n$ with weights 
${\cal I}_{n,k}$ determined by the variation of the phoretic mobility encoded 
in $g(\theta)$.

As discussed in Refs. \cite{Majee2013,Michelin2014,Liverpool2016} the hydrodynamic 
problem defined by Eq. (\ref{eq:bvp_for_u_thin}), together with the expression 
for the phoretic slip  (Eq. (\ref{eq:phor_slip_squirm})) and the force-free 
condition (Eq. (\ref{eq:zero_F_T_thin}) with $\mathbf{F}_{ext} = 0$) 
is mathematically identical to a ``squirmer''  model \cite{Blake1971,Lighthill1952}. 
For reasons given below, it is advantageous to explicitly 
account for an external force $\mathbf{F}_{ext} = F \mathbf{e}_z$, which preserves 
the axial symmetry of the system.\footnote{Obviously, the solution for the squirmer 
exposed to an external force $\mathbf{F}_{ext}$ can be straightforwardly 
formulated by adding to the flow corresponding to a force-free  squirmer the known 
flow field of a no-slip sphere driven by $\mathbf{F}_{ext}$ \cite{HaBr73}.} 

Following standard procedure, by using the general results  derived by 
Brenner \cite{HaBr73} for the flow around a sphere and the corresponding 
hydrodynamic force exerted on the sphere, we arrive at the following expressions 
for the flow field $\mathbf{u}(r,\theta) = 
u_r(r,\theta) \mathbf{e}_r + u_\theta(r,\theta) \mathbf{e}_\theta$ in the 
fixed laboratory frame:
\begin{subequations}
 \label{eq:flow_lab_syst}
 \begin{eqnarray}
  \label{eq:u_radial}
&&\frac{u_r(r,\theta)}{U_0} = \frac{1}{2} \left[3 \left(\frac{R}{r}\right) - 
\left(\frac{R}{r}\right)^3 \right] \left(\frac{F}{F_0}\right) 
P_1(\cos\theta) \hspace*{1.cm}\nonumber\\
&&\hspace*{0.5cm} - \frac{2}{3} B_1 \left(\frac{R}{r}\right)^3 P_1(\cos\theta)\\ 
&&\hspace*{0.5cm} +\sum\limits_{n \geq 2} \left[ \left(\frac{R}{r}\right)^n - 
\left(\frac{R}{r}\right)^{n+2} \right] B_n P_n(\cos\theta)\,,\nonumber
 \end{eqnarray}
 \begin{eqnarray}
  \label{eq:u_tan}
&&\frac{ u_\theta(r,\theta)}{U_0} = \frac{1}{4} \left[- 3 \left(\frac{R}{r}\right) + 
\left(\frac{R}{r}\right)^3 \right] \left(\frac{F}{F_0}\right) V_1(\cos\theta)\nonumber\\
&&\hspace*{0.5cm} + \frac{1}{3} B_1 \left(\frac{R}{r}\right)^3 V_1(\cos\theta)\\
&&\hspace*{0.5cm} - \frac{1}{2} \sum\limits_{n \geq 2} 
\left[(n-2)\left(\frac{R}{r}\right)^n - 
n \left(\frac{R}{r}\right)^{n+2} \right] B_n V_n(\cos\theta)\,.\nonumber
 \end{eqnarray} 
\end{subequations}
The velocity $U:= \mathbf{U} \cdot \mathbf{e}_z$ of the particle is given by 
\begin{equation}
 \label{U_part}
 \frac{U}{U_0} = \frac{F}{F_0} - \frac{2}{3} B_1\,.
\end{equation}
(If needed, the flow field $\mathbf{u}^C(r,\theta)$, in a coordinate 
system aligned with $Oxyz$ and co-moving with the particle, is 
straightforwardly obtained as $\mathbf{u}^C(\mathbf{r}) = 
\mathbf{u}(\mathbf{r}) - \mathbf{U}$.)

Equation (\ref{eq:flow_lab_syst}) identifies the contributions to the flow due 
to the external force (see the first line in Eqs. (\ref{eq:flow_lab_syst})(a) 
and (b)) and due to the phoretic slip -- or self-motility -- 
(i.e., the remaining terms). This reflects the linearity of the Stokes 
equations. Similarly, the expression for the velocity of the particle (Eq. 
(\ref{U_part})) reflects 
a contribution due to the external force (i.e., the first term on the right 
hand side (RHS)) and one due to self-propulsion. It is well established that 
the latter depends only on the first ``squirmer mode'' $B_1$ \cite{Lighthill1952}. 
There are two set-ups of particular interest (see also Ref. \cite{Majee2013}): 
(a) a particle moving in the absence of external forces ($F = 0$, force free (f)),
with velocity $U = U^{(f)}$, and (b) a ``stalling'' (st) configuration, i.e., 
an active particle which is immobilized ($U = 0$) due to an external force 
$F = F_{st}$ acting along the symmetry axis. In the first case, 
with $F = 0$, Eq. (\ref{U_part}) renders 
\begin{equation}
 \label{U_free}
 U^{(f)}/U_0 = - \frac{2}{3} B_1\,.
\end{equation}
In the second case, with $U = 0$, Eq. (\ref{U_part}) renders
\begin{equation}
 \label{F_stall}
 F_{st}/F_0 = \frac{2}{3} B_1\,.
\end{equation}
By combining the two relations, one finds \cite{Majee2013,Gleb2017}
\begin{equation}
\label{stallF_Uf}
F_{st} = - 6 \pi \mu R U^{(f)}\,, 
\end{equation}
which is a deceptively simple expression, in particular in view of its 
exact resemblance to the Stokes formula for a \textit{dragged} spherical 
particle\footnote{This provides a straightforward rationale for the relation 
of the force measurement with the free-particle velocity measurement reported in 
Ref. \cite{Ma2015}.}.

These results are relevant for experimental studies involving active particles.
For example, it is very difficult to measure directly, by three-dimensional 
tracking a moving active particle \cite{Ebbens2013,Golestanian2014}, the velocity 
$U^{(f)}$ of force-free motion in an unbounded fluid. (See also the 
measurements of the flow around a swimming micro-organism reported in Ref. 
\cite{Goldstein2010}.) On the other hand, if it is possible to realize a 
stall-force experiment by trapping an active particle far away from boundaries 
while minimally interfering 
with the mechanism of activity, i.e., without affecting the coefficient $B_1$, 
Eq. (\ref{stallF_Uf}) provides the value $U^{(f)}$ from the measured 
stall force; the set-up in Ref. \cite{Ma2015} could provide such an example.

\section{\label{C_and_flow} Solute distribution and flow around 
active particles with constant-flux activity in unbounded space}

We apply the results derived in the previous section in order to study 
how the choice of a constant-flux activity function, as well as variations 
in the phoretic mobility over the surface of the particle, influence the 
structure of the flow around such model active particles suspended in an 
unbounded fluid. Three model activities, which lend themselves for 
experimental realizations of active particles 
\cite{Golestanian2007,Howse2007,Liverpool2016,Posner2017}, will be considered: 
(i) $f^{(pi)}(\theta)$ describing a spatially uniform \textit{p}roduction of 
solute over one part of the surface, the other part being chemically \textit{i}nert; 
(ii) $f^{(pa)}(\theta)$ describing a uniform \textit{p}roduction of 
solute over one part of the surface and uniform \textit{a}nnihilation (in general 
at a different rate) of the solute over the other part, such that there is no 
net production of solute by the particle; and (iii) $f^{(vi)}(\theta)$ 
describing a spatially \textit{v}arying production of solute over one 
part of the surface (presumably reflecting a certain systematic dependence of 
catalytic properties on the thickness of the coating for very thin films of 
catalysts \cite{Ebbens2013}), the other part being chemically \textit{i}nert. 
We shall discuss separately the case in which the phoretic mobility is position 
independent, focusing on the influence of $\theta_0$ on the resulting flow, and 
the case of a position dependent phoretic mobility which takes distinct values 
on the two parts of the surface. In the latter case, for which closed-form 
formulas cannot be derived, we restrict the discussion to the case of Janus 
colloids, i.e., $\theta_0 = \pi/2$ (which experimentally is the most common 
case).

We employ the terminology of squirmers in order to discuss the force-free 
far-field flow of the corresponding model active particle. For a motile, force-free 
particle (i.e., $B_1 \neq 0$), the second squirming mode $B_2$, which is 
related to the magnitude of the flow due to a stresslet 
\cite{Lighthill1952,Blake1971,Ishimoto2013}, provides that contribution to the 
flow with the slowest decay $\sim r^{-2}$ (see Eq. (\ref{eq:flow_lab_syst})). 
Therefore, the parameter $S := - (B_2/|B_1|)$ (which we shall denote as 
``squirmer parameter'')\footnote{The use of the absolute value $|B_1|$ and of 
the minus sign has to be included in the definition of $S$ 
in order to maintain consistency with the usual sign convention in the 
squirmer literature, in which the direction $Oz$ is chosen to be the same as that 
of the velocity $\mathbf{U}^{(f)}$, i.e., $U^{(f)} > 0$. Since we have fixed 
the direction $Oz$ independently of the direction of motion, we have thus 
allowed for both positive and negative values of $U^{(f)}$. Accordingly, if 
our calculation leads to $U^{(f)} < 0$, in order to facilitate the comparison with 
the squirmer language 
one should change the direction of the $z$ axis: $\mathbf{e}_z \to \mathbf{e}'_z = 
-\mathbf{e}_z$, i.e., $\theta \rightarrow \theta' = \pi-\theta$ and $\mathbf{e}_\theta 
\rightarrow \mathbf{e}_{\theta'} = - \mathbf{e}_{\theta}$. Since upon mapping 
$\theta \rightarrow \theta'$ the polynomials 
$P_n^1(\cos \theta)$ acquires a factor of $(-1)^{n+1}$, one infers that the 
coefficients $B_n$ with odd indices change sign upon this transformation, while 
the ones with even indices remain unchanged. This explains the need for the 
use of the absolute value $|B_1|$ of $B_1$ and for a minus sign in the definition 
of $S$, while there is no such factor needed in the definition of $S'$.} has been 
used in the 
studies of squirmers in order to distinguish between pusher ($S < 0$), puller 
($S > 0$), and neutral ($B_2 = 0$) squirmers. If $B_2 = 0$, the slowest decaying 
contribution to the flow is the one proportional to $r^{-3}$, which is the flow 
due to either 
a source-dipole or a force quadrupole \cite{Ishimoto2013,Michelin2014}. The 
squirming modes contributing to it are $B_1$ and $B_3$. For the models studied 
here, it does not occur that both coefficients $B_2$ and $B_3$ are 
vanishing simultaneously. Therefore the dependences of $S := -(B_2/|B_1|)$ 
and $S' = (B_3/B_1)$ on the parameters of the model suffice to characterize 
the far-field hydrodynamic flow.

\subsection{\label{const_b} Position independent phoretic mobility $b(\mathbf{r}_P)$}

We chose $g(\theta) = -1$, which corresponds to a repulsive effective 
interaction between solute molecules and the surface of the particle. (The case 
$g(\theta) = +1$ can be obtained from the results presented here by simply changing 
the sign of the velocities for both the particle and the flow.) In this case, 
the weights ${\cal I}_{n,k}$ (Eq. (\ref{eq:def_Ink})) take the simple form 
(see also Eq. (\ref{Pn1_norm})) 
\begin{equation}
{\cal  I}_{n,k} = -\frac{n (n+1)}{2n + 1} \delta_{n,k}\,,
\end{equation}
and each of the coefficients $B_n$ depends solely on the corresponding coefficient 
$f_n$ with the same index $n$:
\begin{equation}
 \label{Bn_unif_b}
 B_n = \frac{n f_n}{2}\,.
\end{equation}
As discussed in Ref. \cite{Golestanian2007}, the straightforward implication of 
Eqs. (\ref{U_free}) and (\ref{Bn_unif_b}) is that a spherical particle with 
uniform properties in terms of chemical activity -- i.e., only the amplitude $f_0$ 
is nonzero -- can neither exhibit self-motility (for which $U^{(f)} 
\neq 0$) nor induce flow in an unbounded fluid because in this case all 
coefficients $B_n$ vanish.
\begin{figure}[!tbh]
    \centering
   \includegraphics[width=.95\columnwidth]{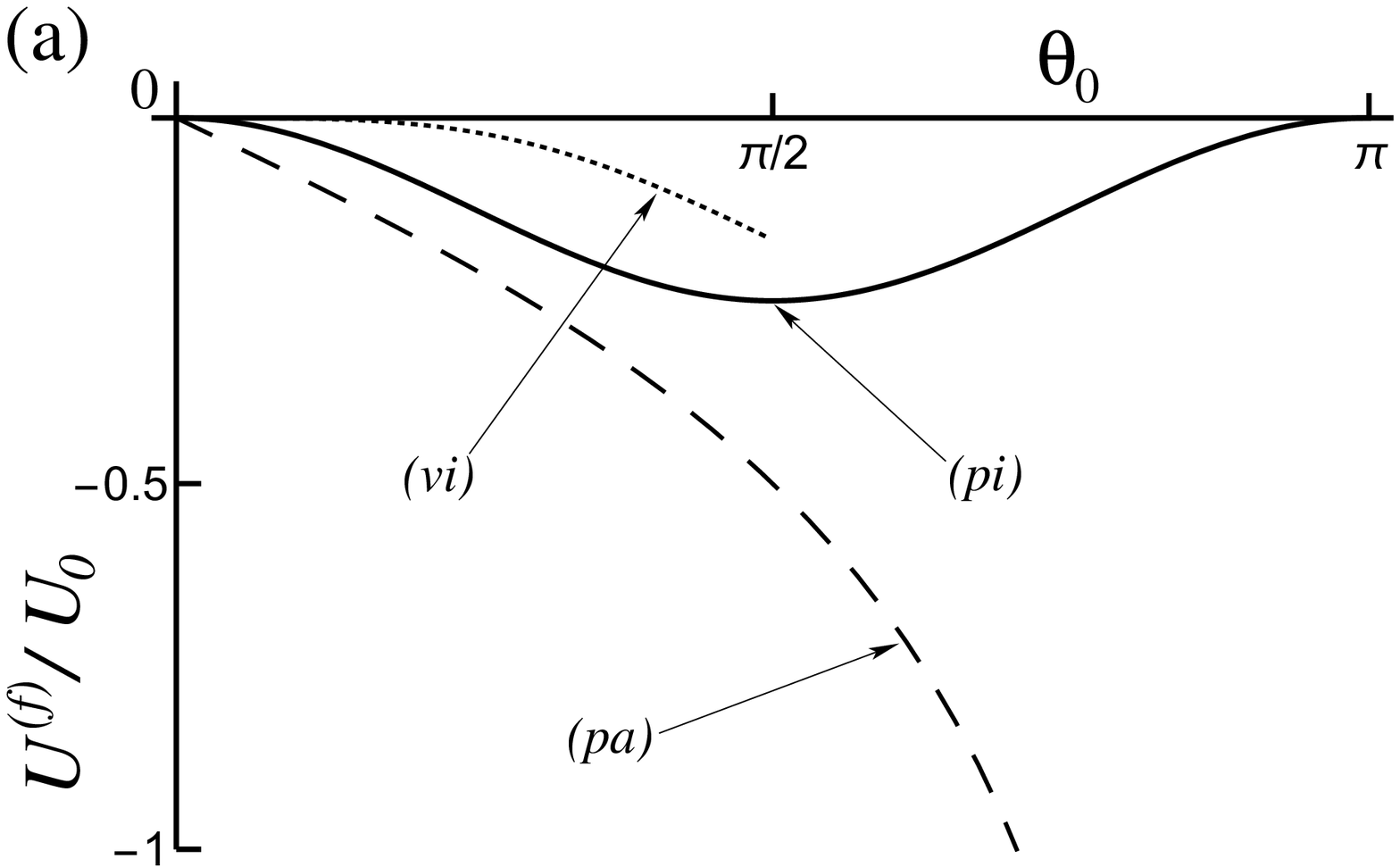}
   \vspace*{0.5cm}\hfill\\
   \includegraphics[width=.95\columnwidth]{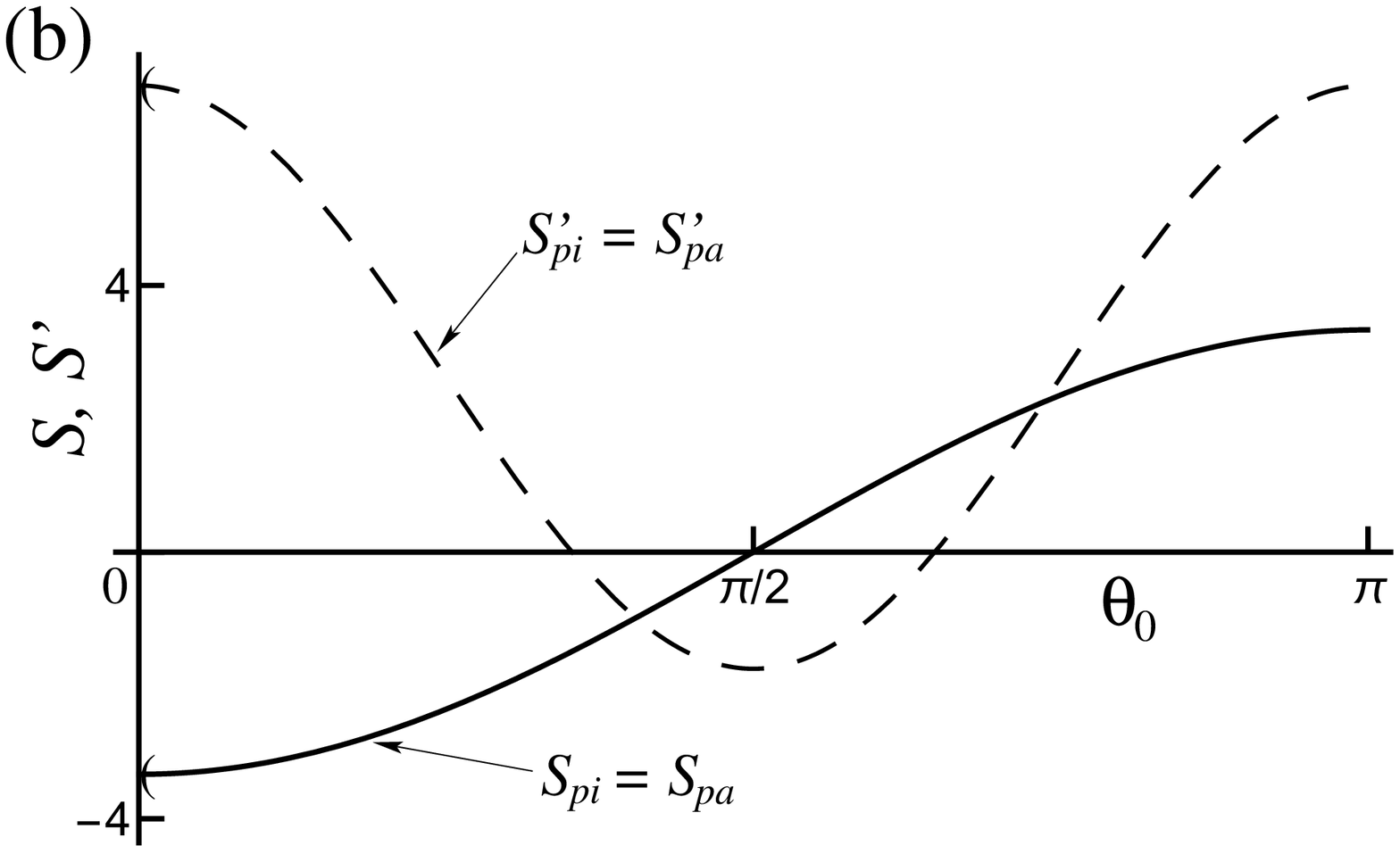}   
   \vspace*{0.5cm}\hfill\\
   \includegraphics[width=.95\columnwidth]{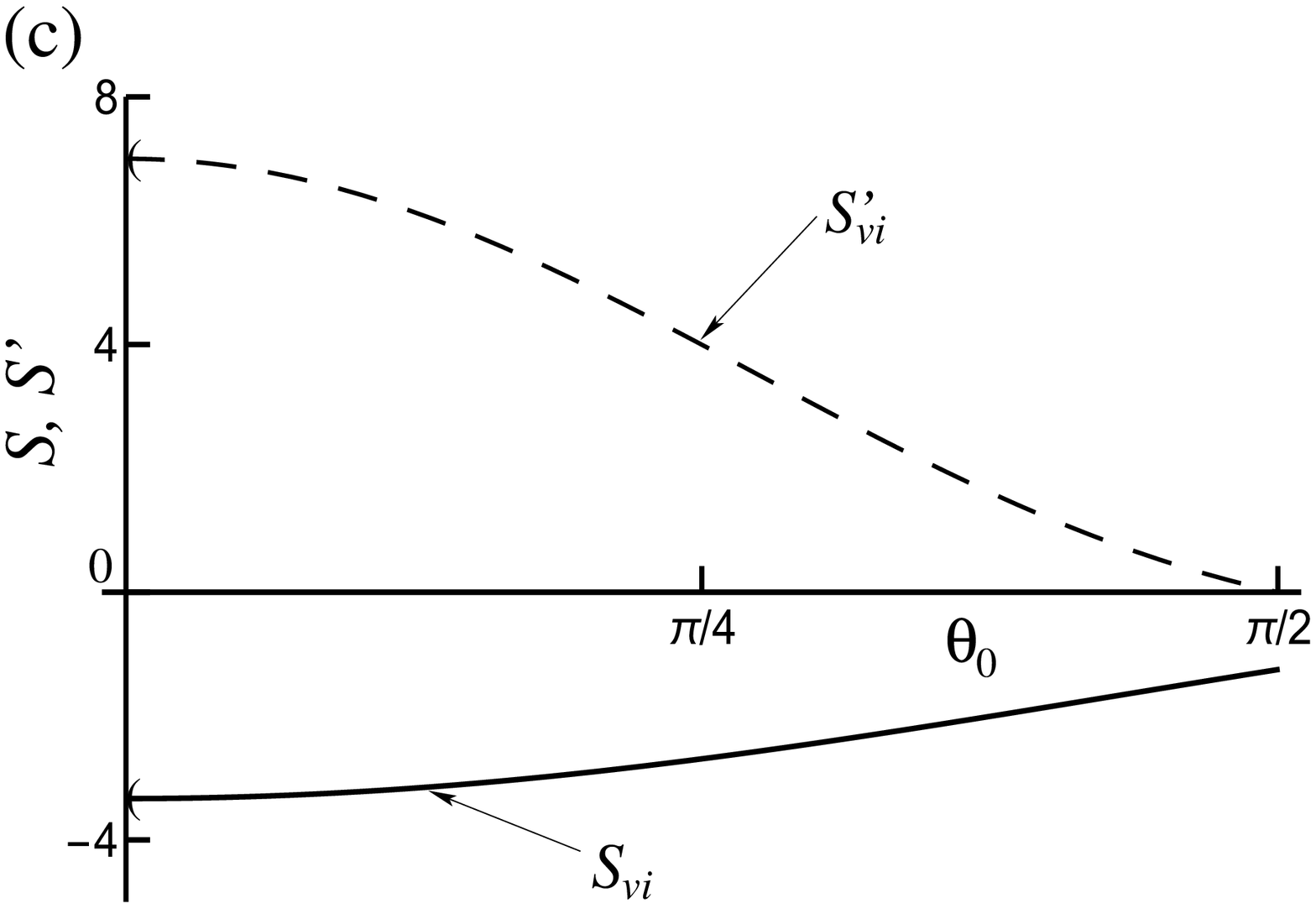}
     \caption{
\label{fig2}
(a) The velocity $U^{(f)}$ of a force free particle as a function of the opening 
angle $\theta_0$ for three model activities (see the main text) and for a 
position-independent, negative phoretic mobility. The curve for $(vi)$ ends at 
$\theta_0 = \pi/2$ because that model is defined for $\theta_0 \leq \pi/2$ (see 
Eq. (\ref{f_vi}) in the main text). (b), (c) The squirmer 
parameters $S = -(B_2/B_1)$ (note that $B_1 > 0$, panel (a)) and $S' = B_3/B_1$ as 
functions of $\theta_0$ for the model activities (see the main text) $(pa)$ 
and $(pi)$ (panel (b)) and $(vi)$ (panel (c)) for a position-independent, 
negative phoretic mobility. The open interval marks in (b) and (c) remind that 
$S$ and $S'$ are not defined at $\theta_0 = 0$, where $B_1$ vanishes, but their 
limits as $\theta_0 \to 0$ exist.
}
\end{figure}

\subsubsection{Particle with position-independent activity over a spherical cap 
and being inert over the rest of the surface}

The activity function corresponding to this case is given by
\begin{equation}
 \label{f_pi}
 f^{(pi)}(\theta) = 
 \begin{cases}
1, ~ & 0 \leq \theta < \theta_0\,,\\ 
0, ~ & \theta_0 < \theta \leq \pi\,,
\end{cases}
\end{equation}
with $0 < \theta_0 < \pi$. The cases of a completely inert ($\theta_0 = 0$) or 
an entirely active ($\theta_0 = \pi$) particle, 
which -- owing to the spherical symmetry -- are not motile in an 
unbounded fluid, are excluded from the discussion here. 
The corresponding amplitudes $f_n$ are (see Eq. (\ref{subint_integr_Pn}))
\begin{equation}
 \label{fn_pi}
 f_n^{(pi)}(\theta_0) = 
 \begin{cases}
 \dfrac{1}{2} (1 - \cos \theta_0) > 0\,, & n = 0\,,\\
  -\dfrac{n+1/2}{n (n+1)} \, \sin \theta_0 P_n^1(\cos\theta_0)\,, & 
 n \geq 1\,,
 \end{cases}
\end{equation}
and the velocity corresponding to force-free motion is given by
\begin{equation}
\label{Uf_pi}
 \frac{U^{(f)}_{pi}}{U_0} = -\frac{\sin\theta_0^2}{4} < 0\,,
\end{equation}
i.e., as expected \cite{Golestanian2007}, the motion is in the negative $z$ 
direction (away from the active cap), irrespective of the value of $\theta_0$ 
(see also Fig. \ref{fig2}(a)). The dependence of the velocity $U^{(f)}_{pi}$ 
on $\theta_0$, shown by the solid line in Fig. \ref{fig2}(a), exhibits the 
expected symmetry with respect to $\theta_0 = \pi/2$ 
\cite{Golestanian2007,Popescu2010}.

\begin{figure}[!ht]
\centering
\includegraphics[width=.72\columnwidth]{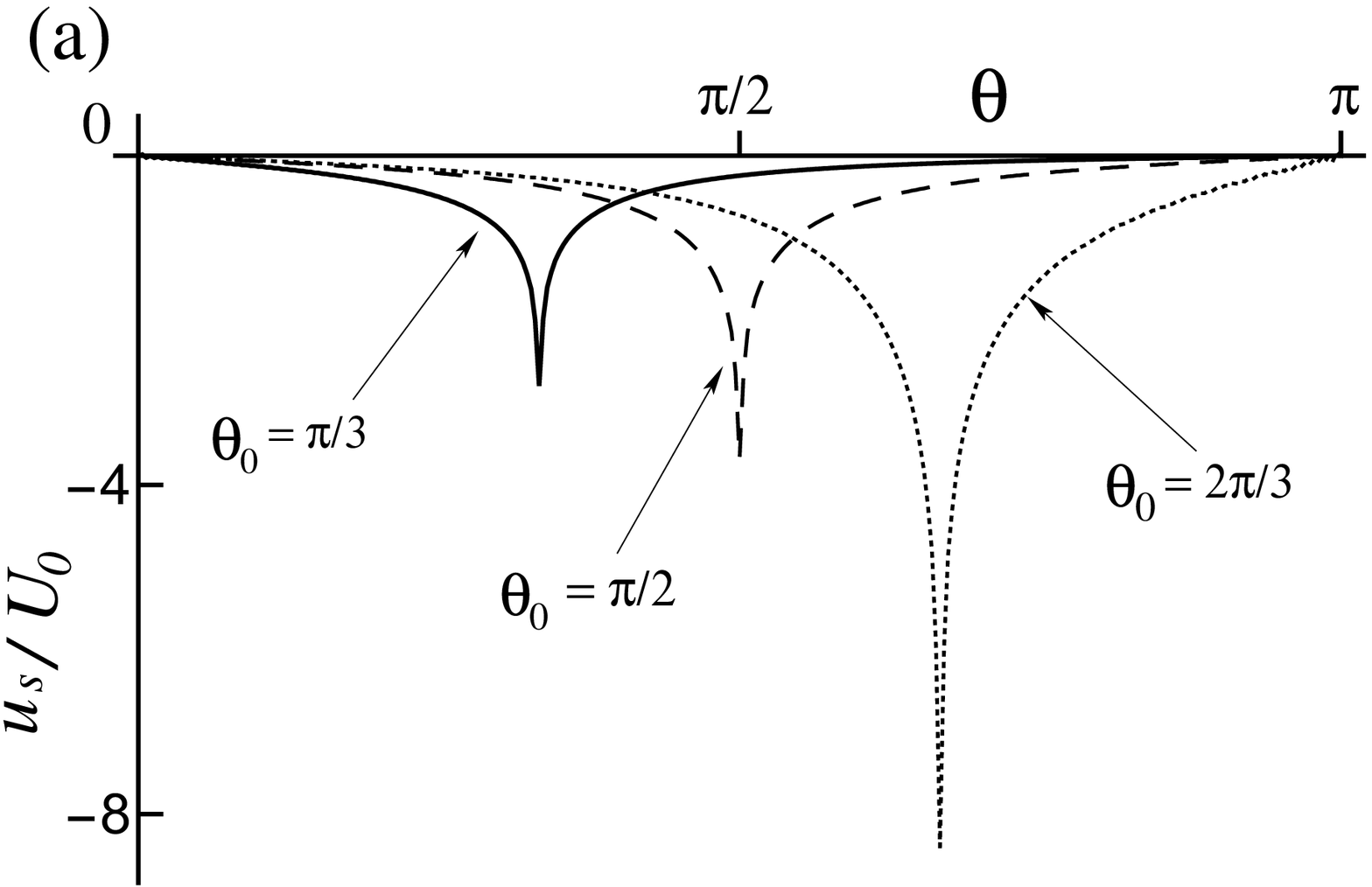}\hspace*{1.cm}
\vspace*{0.12cm}\hfill\\
\includegraphics[width=.6\columnwidth]{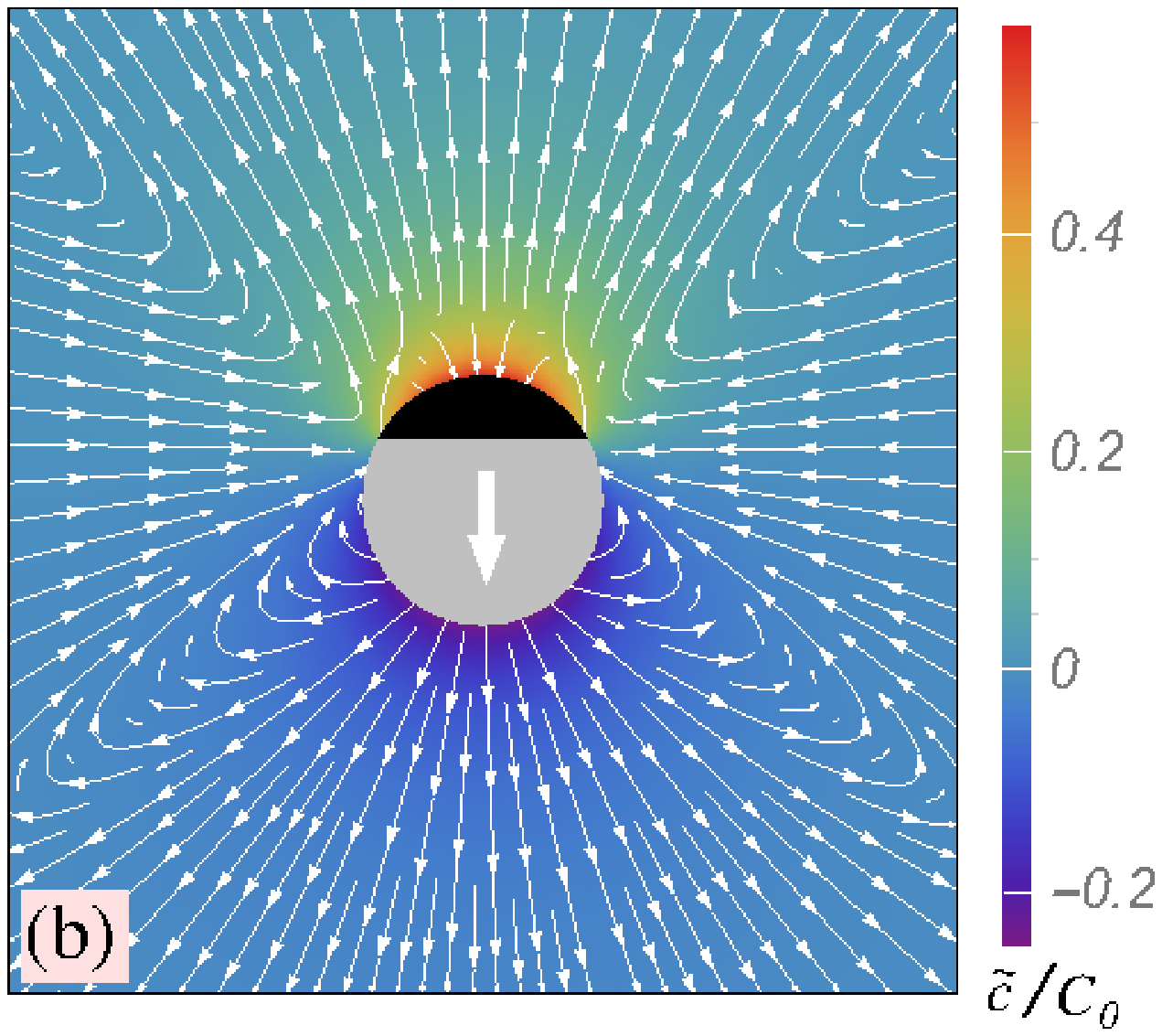}
\vspace*{0.1cm}\hfill\\
\includegraphics[width=.6\columnwidth]{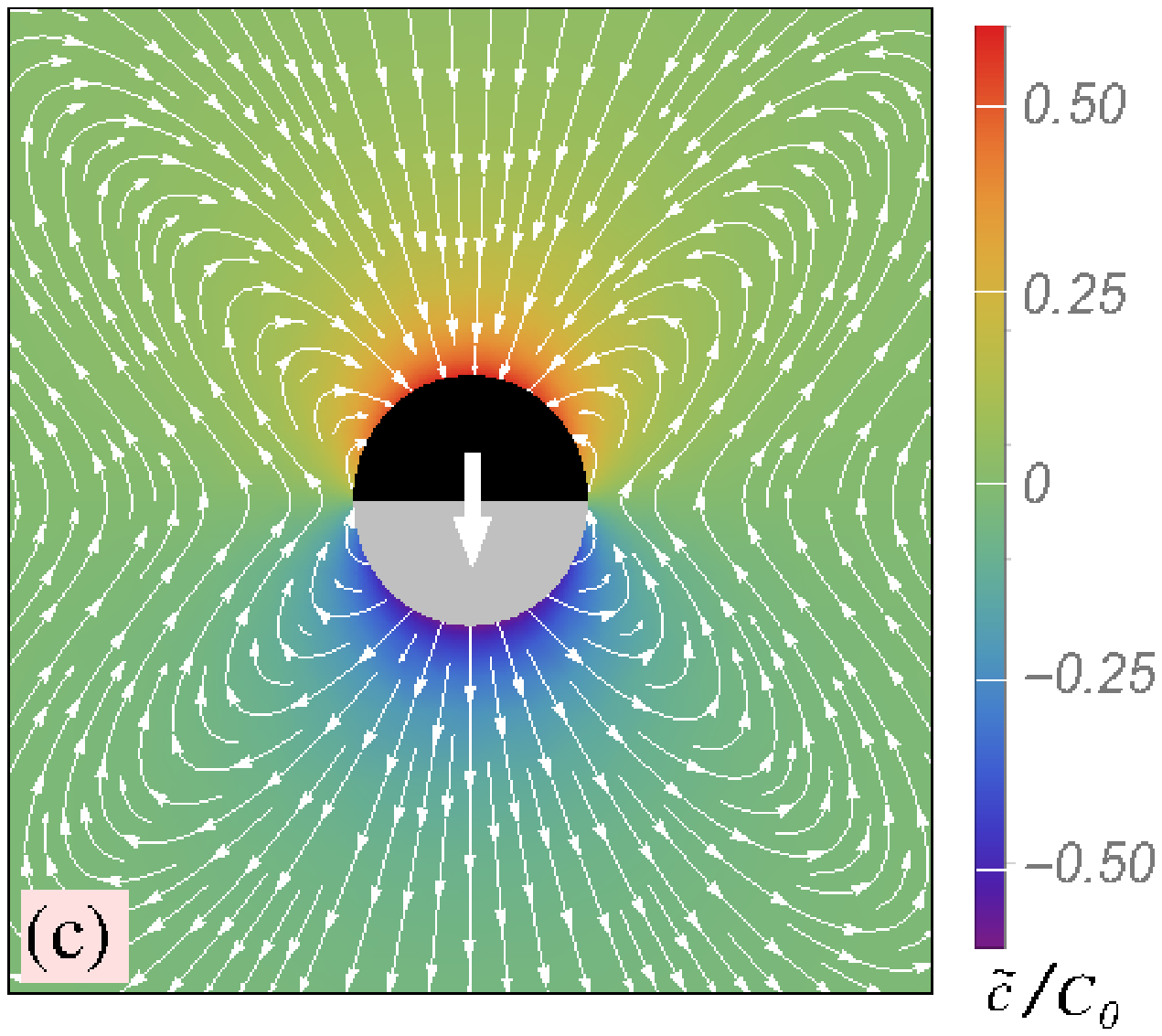}
\vspace*{0.1cm}\hfill\\
\includegraphics[width=.6\columnwidth]{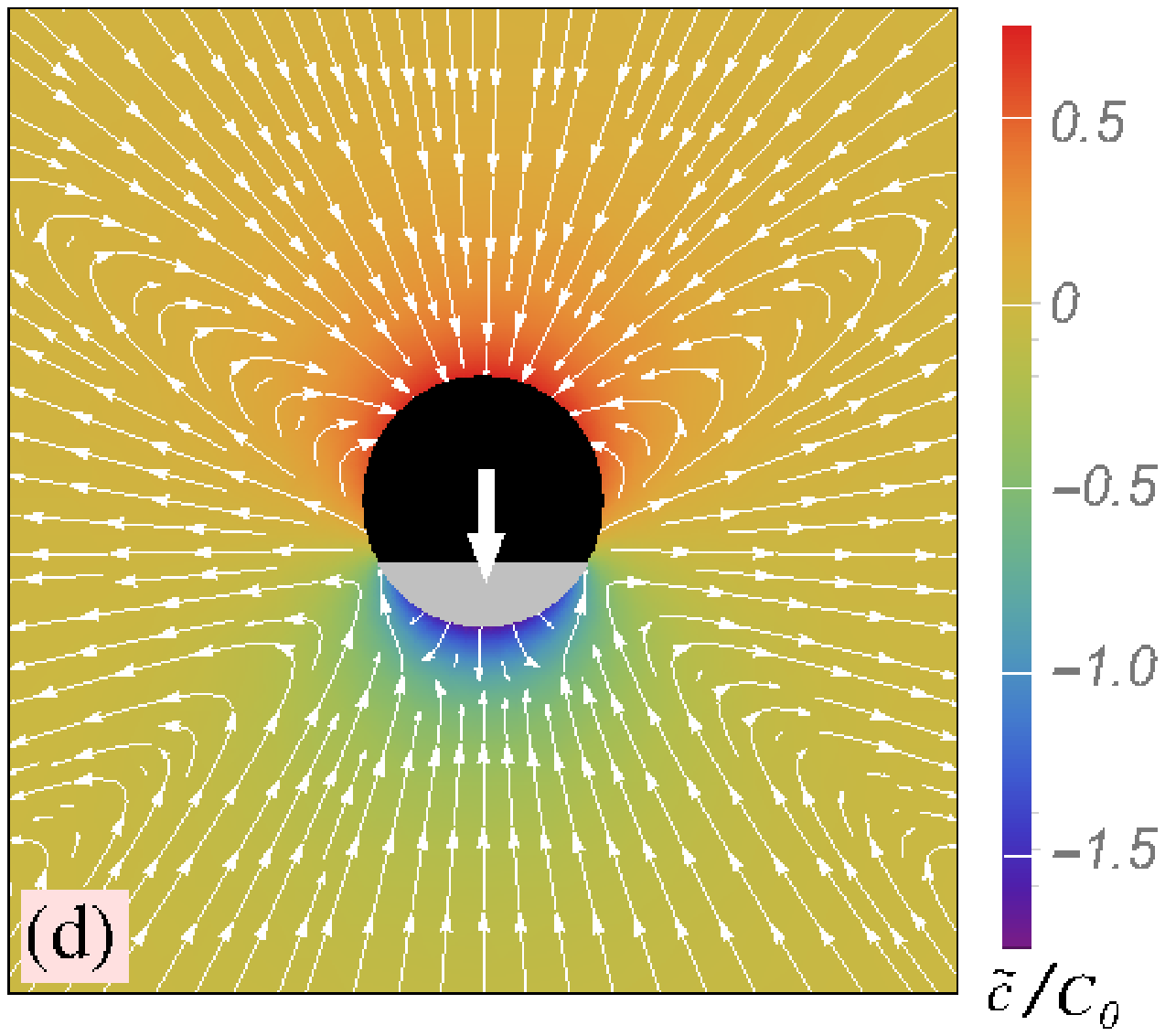}
\caption{
\label{fig3}
(a) Phoretic slip $u_s:= \mathbf{u}_s \cdot \mathbf{e}_\theta$ for model 
$(pa)$ as a function of the angular position $\theta$ 
[Eq. (\ref{eq:phor_slip_squirm}), truncated at the first 300 terms] 
for opening angles $\theta_0 = \pi/3$, $\pi/2$, and $2 \pi/3$, respectively. 
(b)-(d) The force-free flow field in the laboratory system (lines) (Eq. 
(\ref{eq:flow_lab_syst})) and the distribution of solute  (color coded) relative 
to the bulk density (Eq. (\ref{eq:c_series})) for model $(pa)$ and opening 
angles $\theta_0 = \pi/3$, $\pi/2$, and $2 \pi/3$, respectively. For both 
the density and the flow the series are truncated at the first 50 terms. 
The thick white arrows show the direction of the motion of the particle. 
In (a)-(d) the phoretic mobility is position independent and negative. 
}
\end{figure}

Since $B_1(\theta_0) \geq 0$, the parameters $S$ and $S'$ corresponding to this 
model activity function are given by
\begin{subequations}
 \label{S_S'_pi}
 \begin{equation}
  \label{S_pi}
  S_{pi} := -\frac{B_2}{B_1} = - \frac{10}{3} \cos\theta_0\,,
 \end{equation}
and
 \begin{equation}
  \label{S'_pi}
  S'_{pi} : = \frac{B_3}{B_1} = \frac{7}{4} \left(5 \cos^2\theta_0 - 1\right)\,,
 \end{equation}
\end{subequations}
and are shown in Fig. \ref{fig2}(b). One notices that $S_{pi}(\theta_0)$ is 
negative for $\theta_0 < \pi/2$ and that it changes sign at $\theta_0 = \pi/2$.  
(At that point, $S'_{pi}$ is non-zero and negative (see Fig. \ref{fig2}(b)), 
in agreement with the observation that for the models we consider the two 
coefficients do not vanish simultaneously.) Therefore, the 
far-field flows in this model correspond to those of a pusher ($\theta_0 < \pi/2$), 
neutral ($\theta_0 = \pi/2$), and 
puller ($\theta_0 > \pi/2$), respectively (see also, c.f., Fig. \ref{fig3}). 
Thus by varying $\theta_0$ this model can exhibit the whole spectrum (puller, 
pusher, neutral) of squirmer behaviors. In view of the results reported in 
Ref. \cite{Sano2016}, this could be advantageously exploited in, e.g., 
studies of the collective behavior of chemically active particles.

This model has been extensively studied (see, e.g., Refs. 
\cite{Golestanian2007,Popescu2010,Michelin2014,deGraaf2015,Posner2017,
Reigh2015}) and representative plots of the solute distribution (see, e.g., 
Ref. \cite{Michelin2014}), the phoretic slip distribution, and the 
flow field (see, e.g., 
Refs. \cite{Majee2013,Reigh2015,deGraaf2015,Zahra2016}) corresponding to this 
model can be found in the literature. Furthermore, it turns out that the phoretic slip 
distribution and the hydrodynamic flow exhibit patterns similar to the ones 
corresponding to model $(pa)$ (see next subsection), and therefore they will 
be discussed there.

\subsubsection{Particle with position-independent production or annihilation 
activity, respectively, over the two spherical caps}

The activity function corresponding to this case is given by
\begin{equation}
 \label{f_pa}
 f^{(pa)}(\theta) = 
 \begin{cases}
1, ~ & 0 \leq \theta < \theta_0\,,\\ 
q, ~ & \theta_0 < \theta \leq \pi\,.
\end{cases}
\end{equation}
By imposing that there is no net production or annihilation of solute, the 
value of the  parameter $q$ is fixed to 
\begin{equation}
 \label{eq:def_q}
 q = -\frac{1-\cos\theta_0}{1+\cos\theta_0} < 0\,.
\end{equation}
The parameter $0 < \theta_0 < \pi$, i.e., the cases of in which the whole surface 
is either producing or annihilating, respectively, are excluded from the 
discussion here. (Moreover, the limit $\theta_0 \to \pi$ is unphysical 
because in that case there is a point-like sink with a diverging 
rate of annihilation.) The corresponding amplitudes $f_n$ are given by (see Eq. 
(\ref{subint_integr_Pn}))
\begin{equation}
 \label{fn_pa}
 f_n^{(pa)}(\theta_0) = 
 \begin{cases}
 0\,, & n = 0 \,,\\
 -\dfrac{2 n + 1}{n (n+1)} \dfrac{P_n^1(\cos\theta_0)}{1 + \cos \theta_0}\,, 
 & n \geq 1\,.
 \end{cases}
\end{equation}
As expected, due to the requirement that there is no net 
production or annihilation, the amplitude $f_0$ of the monopole term vanishes. 
For three values of $\theta_0$, the number density $\tilde c(\mathbf{r})$ of solute 
in excess of the bulk density is shown as color code in Figs. \ref{fig3} (b)-(d). 
In all cases there is a region of excess density (red color) around the cap which 
releases solute and a depletion region 
(deep blue up to violet color) around the cap which annihilates solute. The 
size of these regions as well as the magnitude of the excess or the depletion 
(see the range of the color bars at the right of the corresponding panels) 
increases upon increasing the size of the release area (i.e., upon 
increasing $\theta_0$), while the dipolar structure of the solute distribution 
becomes more pronounced.  

The velocity corresponding to force-free motion is given by
\begin{equation}
\label{Uf_pa}
  \frac{U^{(f)}_{pa}}{U_0} = -\frac{1}{2} \tan\left(\frac{\theta_0}{2}\right) < 0\,,
\end{equation}
i.e., as for the model $(pi)$, also in this case the motion is in the 
direction of negative $z$ (i.e., away from the active cap), irrespective 
of the value of $\theta_0$ (see also Fig. \ref{fig2}(a)). However, the dependence on 
$\theta_0$ is different due to the change in the activity over the lower cap: the 
gradients of the solute number density along the surface are enhanced 
(see Fig. \ref{fig3}), and, as a consequence, in this model the magnitude of 
the force-free velocity is larger than that in the model $(pi)$ and 
exhibits a monotonic increase with $\theta_0$. As $\theta_0$ 
approaches $\pi$, the absolute value of the peak in the phoretic slip 
distribution increases and diverges for $\theta_0 \to \pi$. As discussed above, 
in that limit the latter is an unphysical feature due to the model being 
ill-defined  with a diverging rate of annihilation.

By comparing Eqs. (\ref{fn_pa}) and (\ref{fn_pi}), one concludes that for a 
given $\theta_0$ the parameters $S_{pa}$ and $S'_{pa}$ (and, in general, all 
ratios $B_n/B_1$, $n \geq 2$) take the same values as those corresponding 
to the model $(pi)$ analyzed in the previous subsection, i.e.,
\begin{equation}
 \label{S_S'_pa}
  S_{pa} = S_{pi} \,,~~ S'_{pa} = S'_{pi} \,.
\end{equation}
Therefore, as noticed above, the flow field and the phoretic slip distribution 
have the same characteristics and appearances as the ones corresponding to the 
case $(pi)$; they differ only in magnitude by a $\theta_0$-dependent velocity 
scale factor $B_1^{(pi)}/B_1^{(pa)} = \sin\theta_0 \cos^2(\theta_0/2)$. 
Accordingly, one can conclude that the two models, although physically 
different with respect to the mechanism and the character of their chemical 
activity, exhibit, up to a velocity scale factor, similar hydrodynamic fields 
associated with their motion in an unbounded fluid.  

As shown in Fig. \ref{fig3}(a), the phoretic slip over the surface of the particle 
is negative everywhere (i.e., it points into the direction of 
$-\mathbf{e}_\theta$, and thus towards the region with higher density of solute) 
Accordingly, the particle moves into the opposite direction 
(see the thick white arrows), in line with the sign in Eq. (\ref{Uf_pa}). 
(We note that here and below the series representation of the phoretic slip 
in Eq. (\ref{eq:phor_slip_squirm}) has been truncated at the first 300 
terms.) The magnitude of the phoretic slip varies non-monotonically with the 
angular position $\theta$ along the surface and, as discussed in, e.g., 
Ref. \cite{deGraaf2015}, it has a sharply peaked maximum (but, despite of 
the appearance, there is neither a divergence nor a cusp) at $\theta_0$, where 
the discontinuity in the activity function is located. 

The flow fields in the laboratory system are shown in Figs. \ref{fig3}(b)-(d) 
for three values of the parameter $\theta_0$ selected such that, according to the 
discussion above and in the previous subsection of the squirmer parameter $S$, 
the far-field behavior corresponds to a pusher ($\theta_0 < \pi/2$), 
a neutral ($\theta_0 = \pi/2$), and a puller ($\theta_0 > \pi/2$) 
squirmer, respectively. (Note that these values correlate with the phoretic slip 
distribution (Fig. \ref{fig3}(a)) being significantly peaked at the hemisphere with 
the active pole, at the equator, and at the hemisphere with the inert pole, 
respectively.) The presence of a stagnation point for $\theta_0 < \pi/2$ (Fig. 
\ref{fig3}(b)) and $\theta_0 > \pi/2$ (Fig. \ref{fig3}(d)), respectively, and its 
location behind or ahead of the particle, respectively, are indeed consistent 
with the pusher and puller characteristics. The clear fore-aft symmetry of the 
streamlines in Fig. \ref{fig3}(c), for which $\theta_0 = \pi/2$ and thus $S = 0$, 
is expected for a neutral squirmer. In all three cases there are strong 
deviations of the shape of the streamlines from the expected far-field ones. In 
the case of the neutral squirmer, the formation of a ``saddle''- or 
``butterfly''-like feature at $\theta \simeq \pi/2$ is particularly noteworthy. 
This signals that for $r/R \simeq 4$ the squirming modes with $n \geq 3$ still 
make significant contributions to the flow around the active particle.

Finally, we note that $\tilde c(\mathbf{r})$, as shown in Figs. \ref{fig3}(b)-(d),  
underscores that in order for the model $(pa)$ to be well defined, i.e., the 
solute number density $c(\mathbf{r})$ to be non-negative everywhere, 
the background number density $c_\infty$ must be sufficiently large. This is a 
result of the assumption that the annihilation reaction has a constant rate 
independent of, rather than being proportional to, the local density of 
solute molecules. Although this assumption reduces the significance of the 
results for experimental studies, it has the merit of providing the means for 
straightforwardly building a conceptually clear example that two different models 
of activity can lead to effective squirmers which exhibit identical behaviors 
(in the sense of identical coefficients $B_n$ for $n \geq 2$). Furthermore, 
the generalization to activity functions corresponding to reactions with 
first-order kinetics will be discussed in Sec. \ref{first_order_kin}.

\subsubsection{Particle with position-dependent production over a spherical cap 
and being inert over the rest of the surface}
\begin{figure}[!htb]
    \centering
   \includegraphics[width=.8\columnwidth]{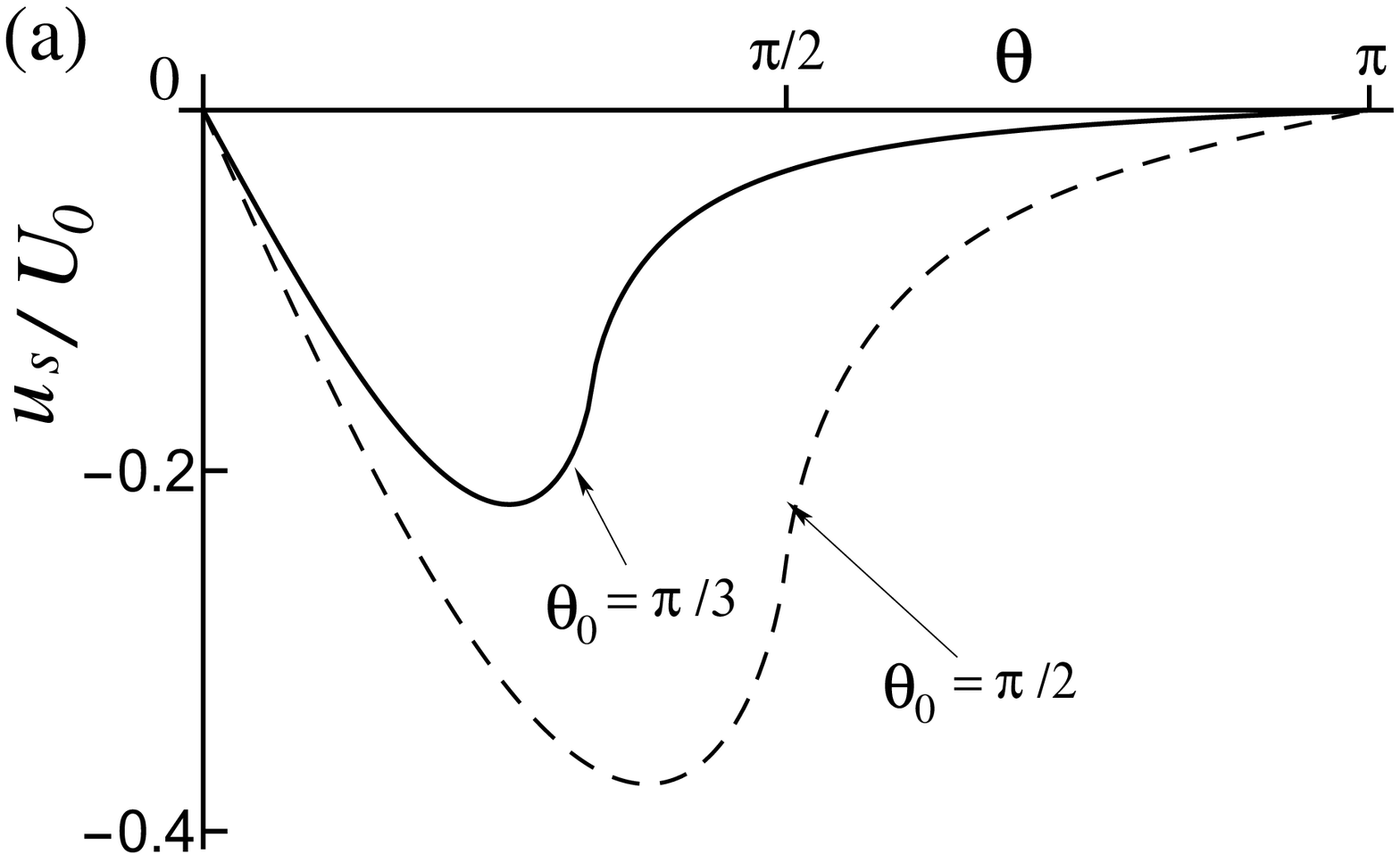}\hspace*{1.cm}
   \vspace*{0.5cm}\hfill\\
   \includegraphics[width=.61\columnwidth]{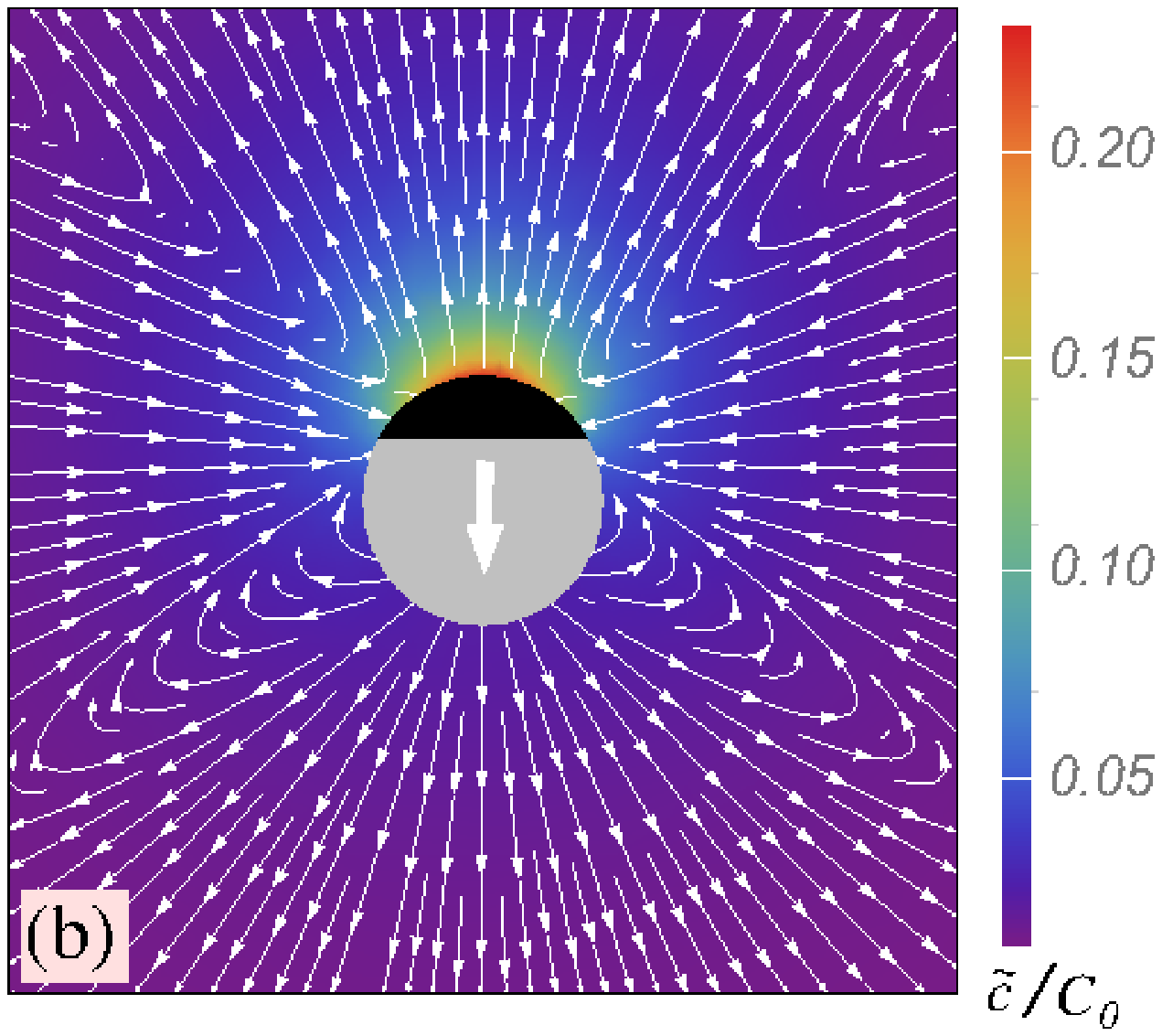}   
   \vspace*{0.5cm}\hfill\\
   \includegraphics[width=.61\columnwidth]{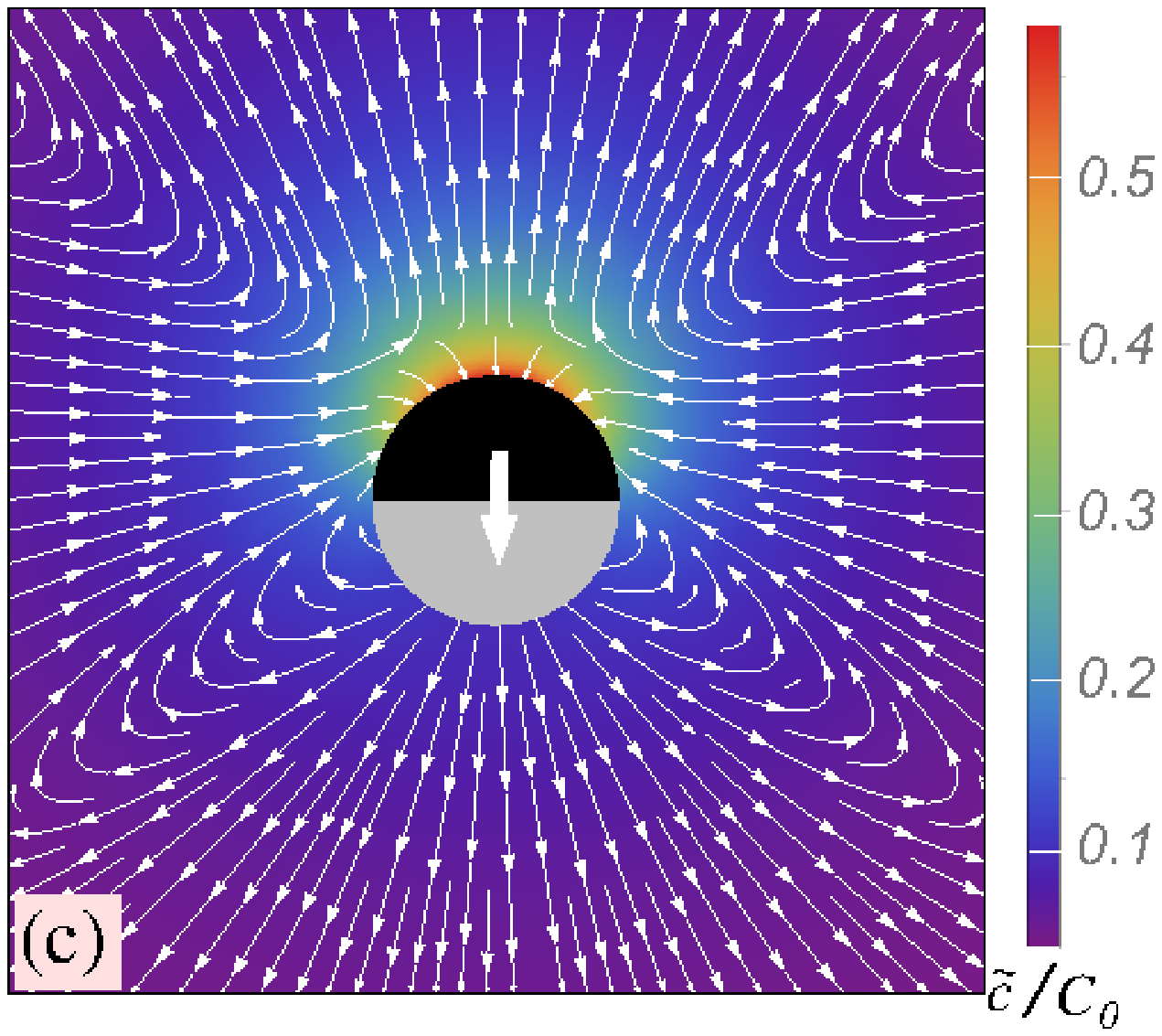}
     \caption{
\label{fig4}
(a) Phoretic slip for model $(vi)$ as a function of position $\theta$ 
[Eq. (\ref{eq:phor_slip_squirm}), truncated at the first 300 terms] 
for opening angles $\theta_0 = \pi/3$ and $\pi/2$, respectively. 
(b), (c) The force-free flow field in the laboratory system (lines, Eqs. 
(\ref{eq:flow_lab_syst})), and the distribution of solute  (color coded) 
relative to the bulk density (Eq. (\ref{eq:c_series})) for model $(vi)$ 
and opening angles $\theta_0 = \pi/3$ and $\pi/2$, respectively. For both 
density and flow the series are truncated at the first 50 terms. The 
thick white arrows show the direction of the motion for the particle. 
In (a)-(c) the phoretic mobility is position independent and negative. 
}
\end{figure}
The activity function corresponding to this case is chosen to be of the form 
\begin{equation}
 \label{f_vi}
 f^{(vi)}(\theta) = 
 \begin{cases}
\cos\theta - \cos\theta_0, ~ & 0 \leq \theta < \theta_0~(\leq \pi/2)\,,\\ 
0, ~ & \theta_0 \leq \theta \leq \pi\,.
\end{cases}
\end{equation}
This choice is motivated by Refs. \cite{Ebbens2013,Brown2014}, in which it is 
argued, based on experimental evidence, that there is a possible 
dependence of the activity on the thickness of the catalyst coating. This 
thickness varies on the surface of the sphere from a maximum at the pole 
towards a minimum (i.e., no catalyst) upon approaching the equator. For 
reasons of simplicity, the parameter $\theta_0$ is constrained to the typical range 
$0 < \theta_0 \leq \pi/2$ employed in experimental studies. The constant term 
$\cos\theta_0$ ensures that the activity function is continuous at $\theta_0$.

By using Eqs. (\ref{subint_integr_Pn}) and (\ref{subint_integr_Pn_Pm}), 
the corresponding amplitudes $f_n$ (see Eq. (\ref{eq:def_f})) can be 
expressed as
\begin{eqnarray}
 \label{fn_vi}
 &&f_n^{(vi)}(\omega_0) 
 = \left(n + \frac{1}{2}\right) \left(1-\omega_0^2\right)^{1/2} \\
 &\times& \left[ \frac{P_1(\omega_0) P_n^1(\omega_0)}{n (n+1)}  - 
 \frac{P_n(\omega_0) P_1^1(\omega_0) - P_1(\omega_0) P_n^1(\omega_0)}
 {2 - n (n+1)}\right]\,,\nonumber
 \end{eqnarray}
where $\omega_0 := \cos(\theta_0)$.

The velocity corresponding to force-free motion is given by
\begin{equation}
\label{Uf_vi}
  \frac{U^{(f)}_{vi}}{U_0} = -\frac{1}{12}\, 
  \left[2+ \cos\theta_0 \left(\cos^2\theta_0-3\right)\right] \,.
\end{equation}
For $0 < \theta_0 \leq \pi/2$ one has $U^{(f)}_{vi}/U_0 < 0$ (see 
Fig. \ref{fig2}(a)). As expected, the motion is in the negative $z$ 
direction (i.e., away from the active cap), which is consistent with 
the phoretic slip pointing towards the $-\mathbf{e}_\theta$ 
direction (see Fig. \ref{fig4}(a)). The smoothly decreasing production 
rate over the surface leads to smaller gradients in the solute number 
density and, accordingly, to visibly reduced velocities in comparison to 
those in the previous two models. This also leads to the removal of 
the sharp peaks, as observed in the other models, in the distribution of the 
phoretic slip around the surface of the particle (see Fig. \ref{fig4}(a)).

From Eq. (\ref{fn_vi}) one obtains the parameters $S$ and $S'$ as 
\begin{subequations}
 \label{S_S'_vi}
 \begin{equation}
  \label{S_vi}
  S_{vi} : = - \frac{B_2}{|B_1|} = -\frac{5}{2} \,
  \dfrac{\sin^4\theta_0}{2+ \cos\theta_0 \left(\cos^2\theta_0-3\right)}
 \end{equation}
and
 \begin{equation}
  \label{S'_vi}
  S'_{vi} : = \frac{B_3}{B_1} = \frac{21}{4}\, 
  \dfrac{\cos\theta_0\,\sin^4\theta_0}{2+ \cos\theta_0 
  \left(\cos^2\theta_0-3\right)}\,,
 \end{equation}
\end{subequations}
respectively. Their dependence on $\theta_0$ is shown in Fig. \ref{fig2}(c). As for 
the other models, $S_{vi}(\theta_0) < 0$ for $\theta_0 < \pi/2$; but it 
remains negative also at $\theta_0 = \pi/2$. Therefore, within the whole range 
of $\theta_0$ the far-field flows in this model correspond to those of a 
pusher. This is illustrated in Figs. \ref{fig4}(b)-(c), where we show the 
flows for $\theta_0 = \pi/3$ (b) and $\theta_0 = \pi/2$ (c).

\subsection{\label{pos_dep_mob} Position dependent phoretic 
mobility $b(\mathbf{r}_P)$}

As noted in the beginning of Sec. \ref{C_and_flow}, if the phoretic mobility 
varies over the surface it is not possible, in general, to obtain simple 
expressions -- such as, e.g., Eq. (\ref{Bn_unif_b}) -- for the coefficients $B_n$. 
Therefore we continue the discussion of the model activity functions 
under additional constraints in order to reduce the number of free parameters. 
We shall focus on the case $\theta_0 = \pi/2$, which is a typical value 
in experimental studies, and we shall consider only models with the phoretic 
mobility described by a piecewise constant function:
\begin{equation}
\label{eq:two_val_g}
g(\theta) = 
\begin{cases}
- 1, & 0 \leq \theta < \theta_0 = \pi/2,\\
- \gamma, & \pi/2=\theta_0 < \theta \leq \pi.
\end{cases}
\end{equation}
This corresponds to a negative phoretic mobility $- b_0$ over the upper cap (see 
Fig. \ref{fig1}) and a different value,  $- \gamma \, b_0$, over the lower cap. 
This choice is motivated by the typical realizations of such particles, in which 
the two parts of the particle consist of two distinct materials (such as the 
Au-Pt rods employed in the experiments reported in Ref. \cite{Paxton2004}); 
alternatively, a part of their surface is coated by a different material. This 
is, e.g., the case for the particles employed in Ref. \cite{Simmchen2016} for which 
one part is silica (inactive) while the other part is covered by Pt catalyst 
(active). We note that $\gamma = 1$ corresponds to models with position 
independent phoretic mobility, as studied in the previous subsection. 

With this choice for $g(\theta)$, the weights ${\cal I}_{n,k}$ (Eq. (\ref{eq:def_Ink})) 
take the form 
\begin{eqnarray}
\label{Ink_twoval} 
{\cal I}_{n,k} &=& - \frac{1+ (-1)^{n+k} \,\gamma}{2}  
\int\limits_0^1 dx P_n^1(x) P_k^1(x) \nonumber\\
&=:& - \frac{1+ (-1)^{n+k} \,\gamma}{2} J_{n,k}\,,~~n,k \geq 1\,.
\end{eqnarray}
Although an insightful, closed form expression for the integrals 
$J_{n,k}$ defined above is not available, certain simplifications of the calculations 
below are possible by noticing that if $n+k$ is an even number one has
\begin{eqnarray}
\label{Ink_twoval_even} 
J_{n,k}  & := & \int\limits_{0}^1 dx P_n^1(x) P_k^1(x) = 
\frac{1}{2} \int\limits_{-1}^1 dx P_n^1(x) P_k^1(x) \nonumber\\
& = & \frac{n (n+1)}{2n +1} \delta_{n,k}\,,~~\mathrm{for}~ 
n,k \geq 1\,,~ n+k~\mathrm{even}.~~~
\end{eqnarray}

Furthermore, as discussed in the previous section, in the models $(pa)$ 
(production and annihilation) and $(pi)$ (production and inert) 
the coefficients $f_n$  with $n \geq 1$, and therefore the coefficients $B_n$, 
differ only by a constant factor independent of $n$ (e.g., for $\theta_0 = \pi/2$, 
the coefficients $f^{(pi)}_n$ are twice as large as the coefficients 
$f^{(pa)}_n$). Consequently, the ensuing hydrodynamic flows they induce have the 
same structure and differ solely in terms of a velocity scale, irrespective of the 
specific dependence of the phoretic mobility on the position at the surface. Thus 
in the remaining part of this subsection we study only the models $(pa)$ and $(vi)$.

\subsubsection{Production and annihilation (pa) activity function}

\begin{figure}[!t]
    \centering
  \includegraphics[width=.95\columnwidth]{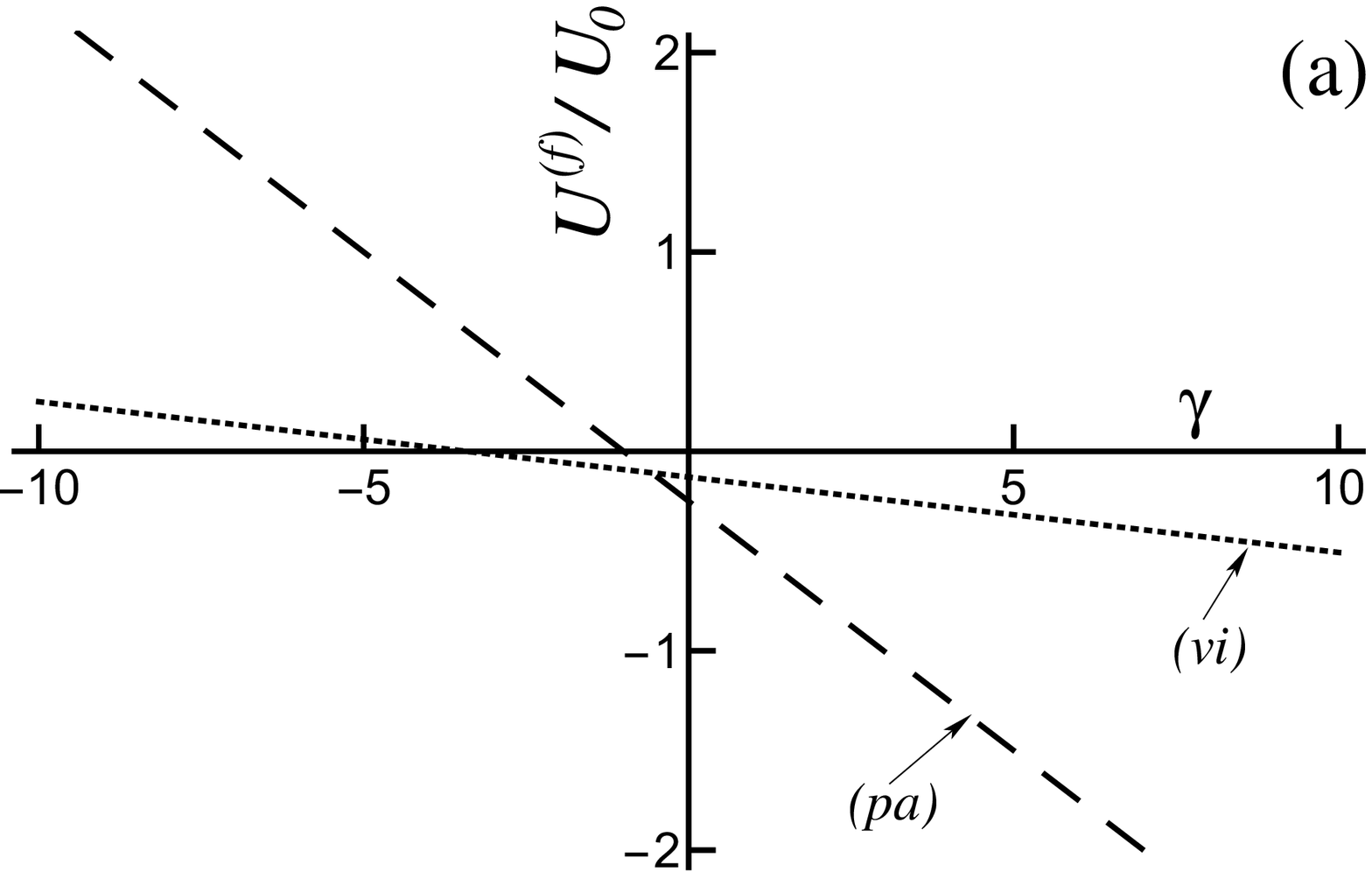}
   \vspace*{0.5cm}\hfill\\
   \includegraphics[width=.95\columnwidth]{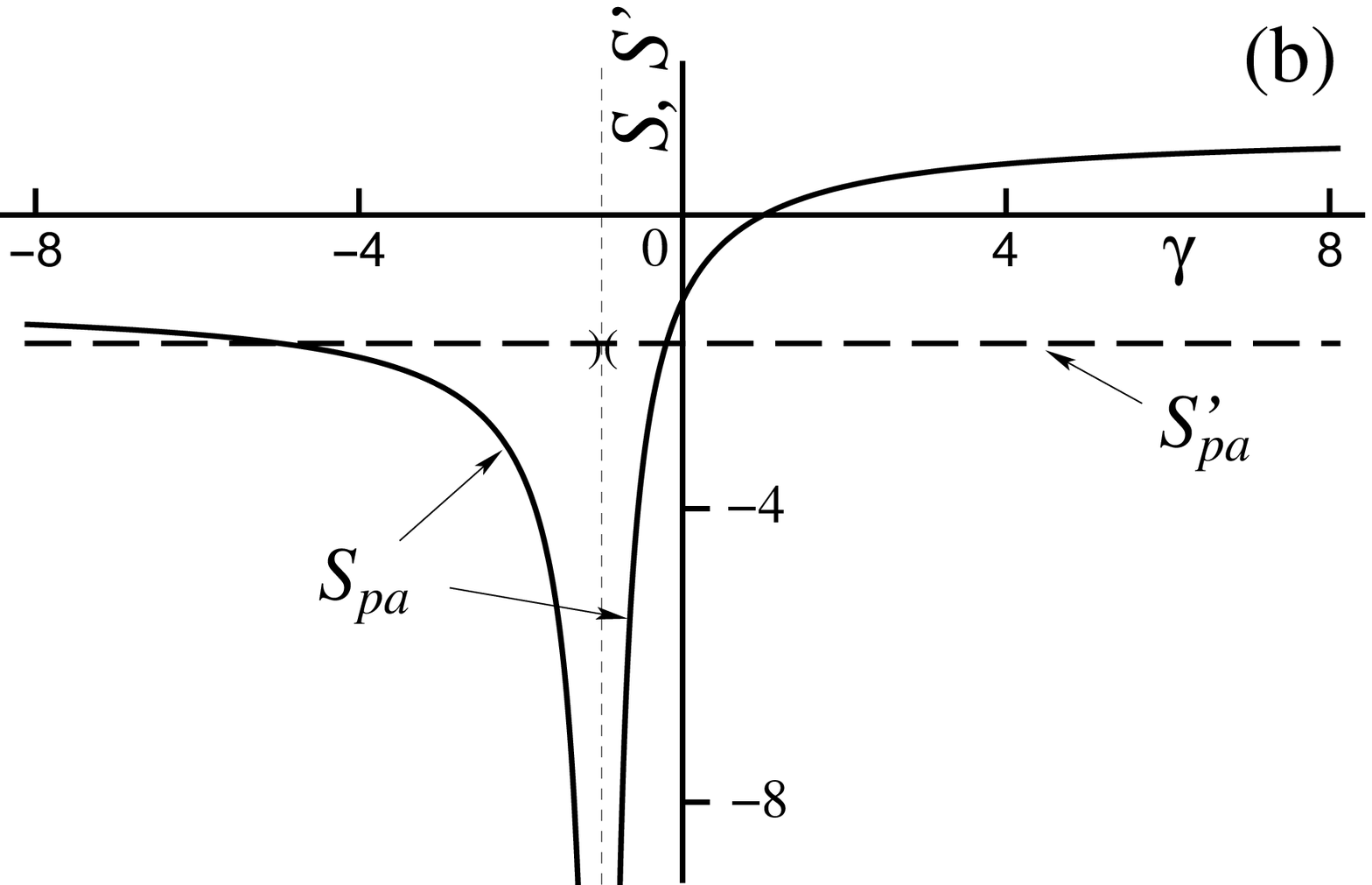}   
    \vspace*{0.5cm}\hfill\\
    \includegraphics[width=.95\columnwidth]{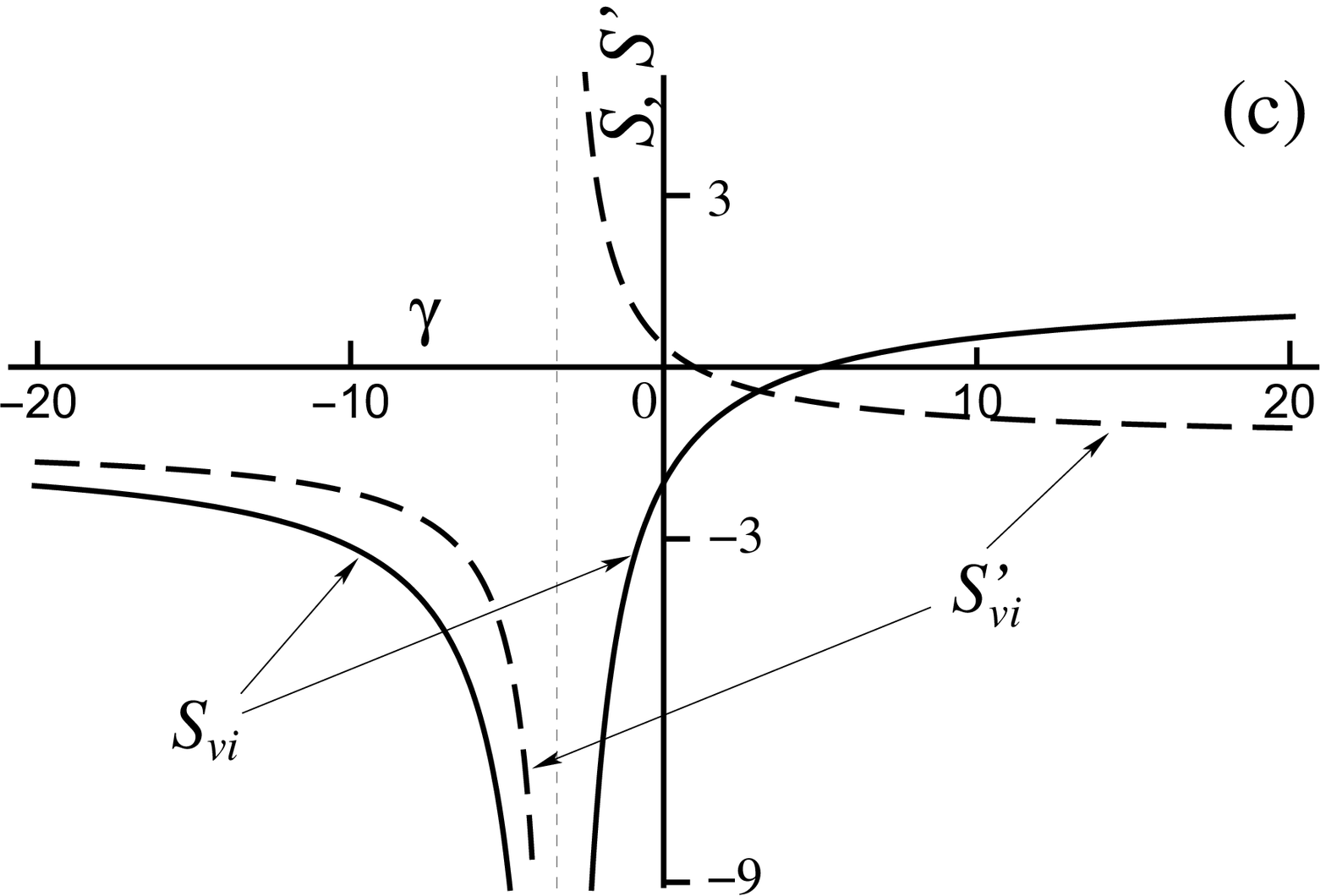}
     \caption{
\label{fig5}
(a) The velocity $U^{(f)}$ of a force-free spherical Janus particle 
($\theta_0 = \pi/2$) as a function of the ratio $\gamma$ of the phoretic mobilities 
over the two hemispheres, for the model activities (see the main text) $(pa)$ 
(dashed line) and $(vi)$ (dotted line). (b), (c) The squirmer parameters 
$S = -(B_2/|B_1|)$ and $S' = B_3/B_1$ as functions of the ratio $\gamma$ for 
the model activities $(pa)$ 
(panel (b)) and $(vi)$ (panel (c)). In all three panels (a)-(c) the phoretic 
mobility over the upper hemisphere ($\theta < \pi/2$, see Fig. \ref{fig1}), 
where the solute is released into solution, is negative. The left and right 
open interval signs in panel (b) indicate that the function $S'_{pa}$ is not 
defined at $\gamma = -1$ (where $B_1$ vanishes), but the left and right limits 
$S'_{pa} (\gamma \to -1)$ do exist.  
}
\end{figure}

In this case and for $\theta_0 = \pi/2$ only the coefficients $f_n^{(pa)}$ 
with an odd index $n$ are nonzero (see Eq. (\ref{fn_pa})). Consequently, from 
Eqs. (\ref{eq:Bn_coef}), (\ref{Ink_twoval}), and (\ref{Ink_twoval_even}) one concludes 
that for this model the coefficients $B_n$ are given by 
\begin{equation}
  \label{Bn_pa_twoval}
  B_n = \left[1 + (-1)^{n+1} \gamma \right] \chi_n\,,~~n \geq 1 \,,
\end{equation}
where
\begin{equation}
\label{chi_def}
 \chi_n = \frac{1}{4} \times 
 \begin{cases}
 n f_n^{(pa)}\,, &~n~\textrm{odd}\,,\\
 (2 n +1) \displaystyle{\sum\limits_{k~\mathrm odd}} J_{n,k} \,
 \dfrac{f_k^{(pa)}}{k+1}\,, &~n~\textrm{even}\,.
 \end{cases}
\end{equation}
Therefore, by combining Eqs. (\ref{U_free}), (\ref{fn_pa}), (\ref{Bn_pa_twoval}), and 
(\ref{chi_def}) one arrives at the following simple expression for the force-free 
velocity in an unbounded fluid:
\begin{equation}
 \label{Uf_pa_twoval}
 \frac{U^{(f)}}{U_0} = - \frac{1+\gamma}{6} f_1^{(pa)} = - \frac{1+\gamma}{4}\,.
\end{equation}
This expression shows that, as discussed in Ref. \cite{Golestanian2007}, even 
if the phoretic mobility varies over the surface, i.e., $\gamma \neq 1$, a spatial 
variation in the activity, i.e., $f_1^{(pa)} \neq 0$, remains a necessary condition 
for self-motility ($U^{(f)} \neq 0$). One also finds that for $\gamma =  -1$ 
(i.e., if the phoretic mobility over that hemisphere, where the solute 
is released, is equal in magnitude but opposite in sign to the one over 
the hemisphere where the solute is annihilated) the velocity 
vanishes. Nevertheless, the particle still induces a hydrodynamic flow. In 
the language of squirmers, this situation corresponds to a ``shaker'' 
($S \to \pm \infty$). For $\gamma < -1$, the velocity reverses sign, and 
the force-free active particle moves in the positive $z$ direction, i.e., towards 
that cap where the solute is produced (see Fig. \ref{fig5}(a)).

\begin{figure*}[!ht]
\centering
\includegraphics[width=.8\columnwidth]{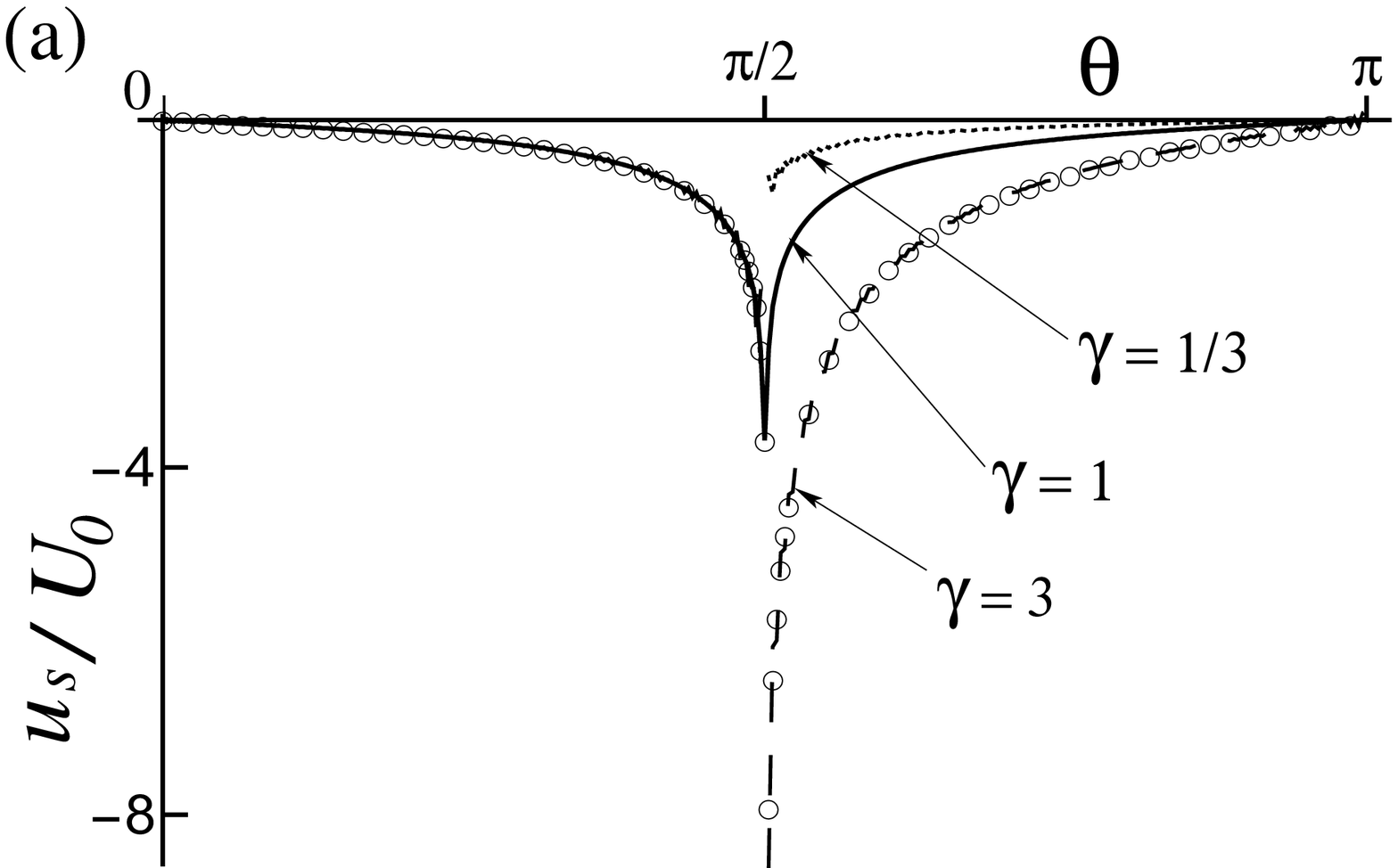}\hspace*{1.cm}%
\includegraphics[width=.8\columnwidth]{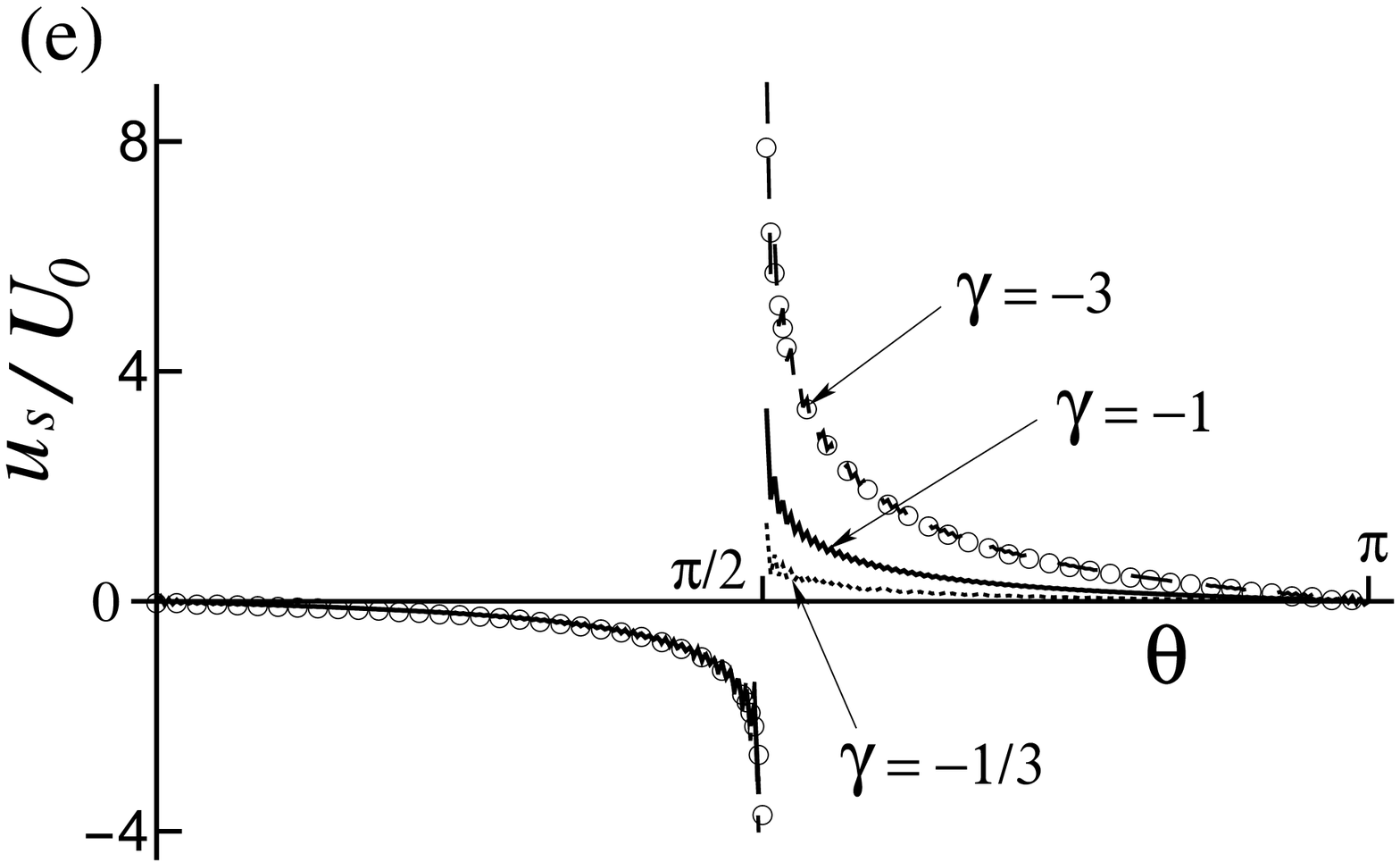}\hspace*{1.cm}\hfill\\
\vspace*{0.1cm}\hfill\\
\includegraphics[width=.66\columnwidth]{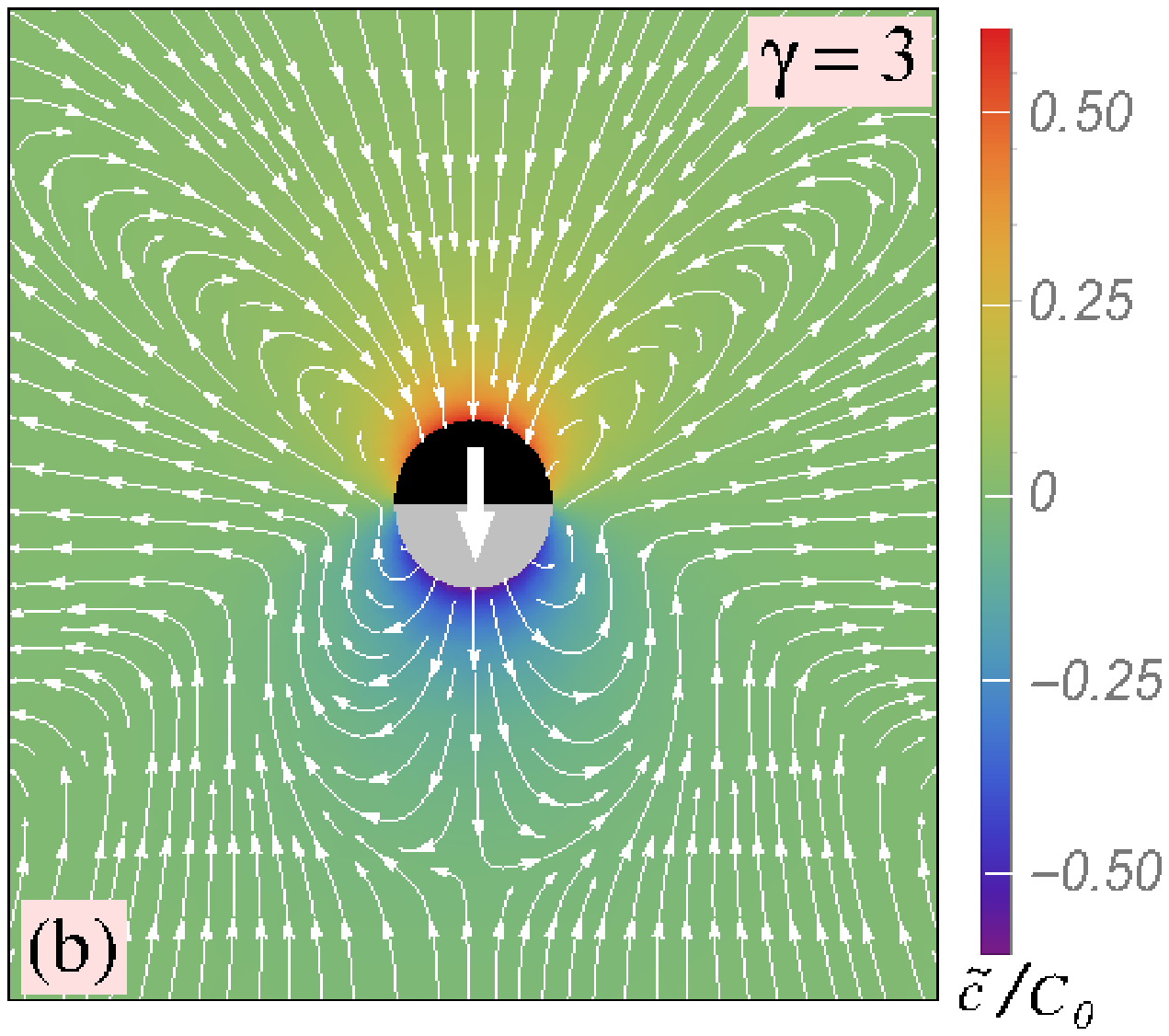}\hspace*{2.cm}%
\includegraphics[width=.66\columnwidth]{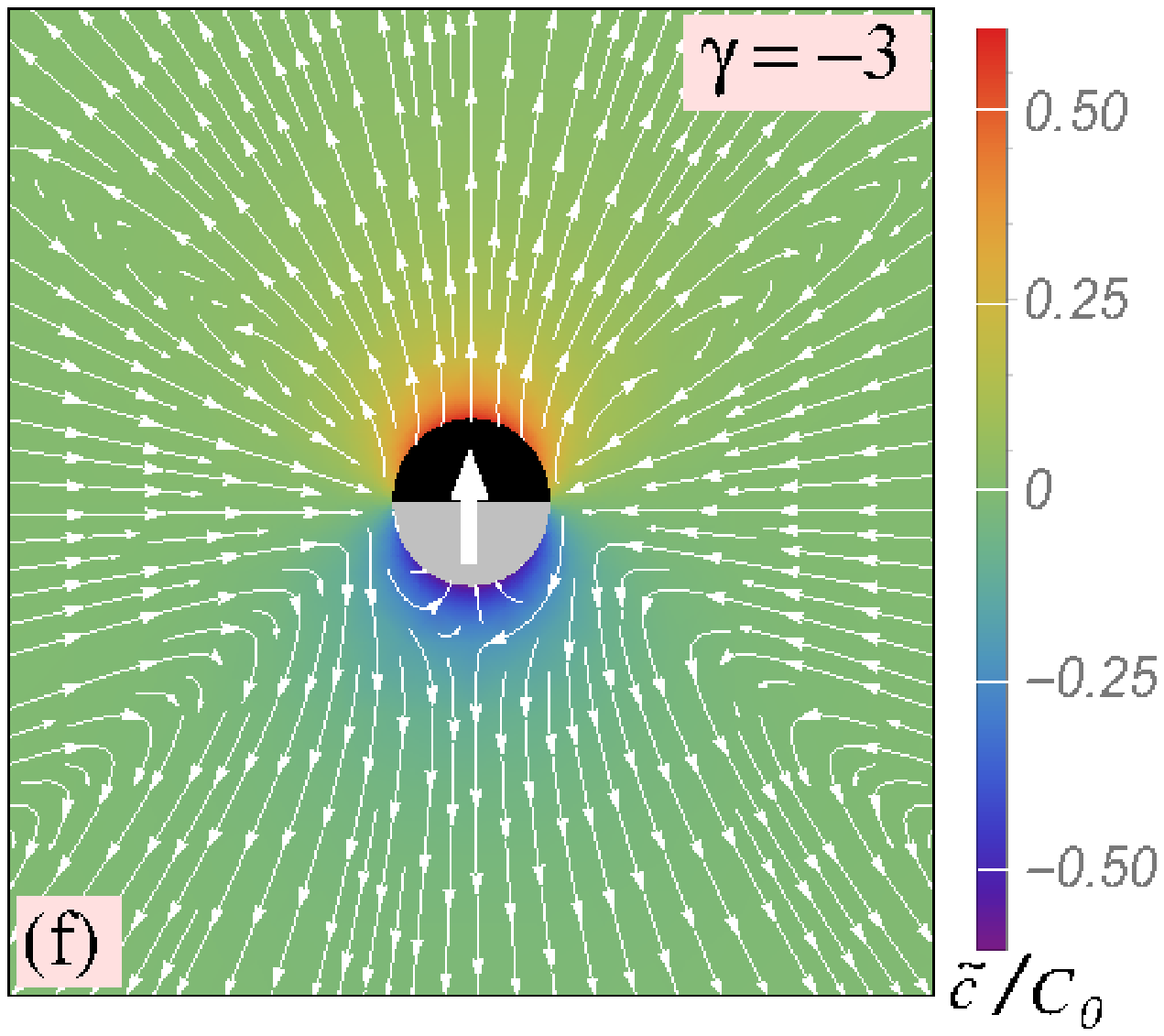}\hfill\\
\vspace*{0.1cm}\hfill\\
\includegraphics[width=.66\columnwidth]{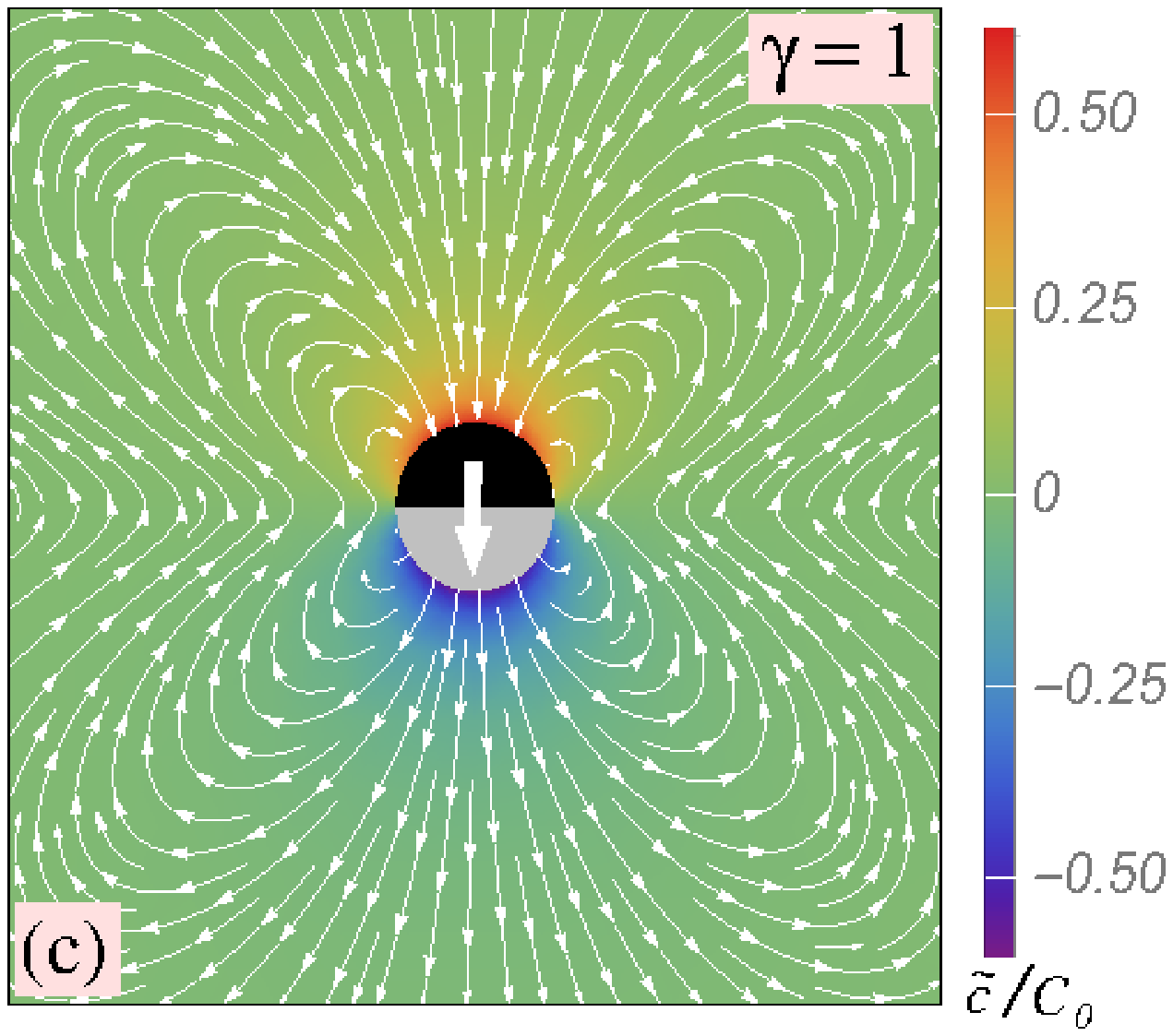}\hspace*{2.cm}%
\includegraphics[width=.66\columnwidth]{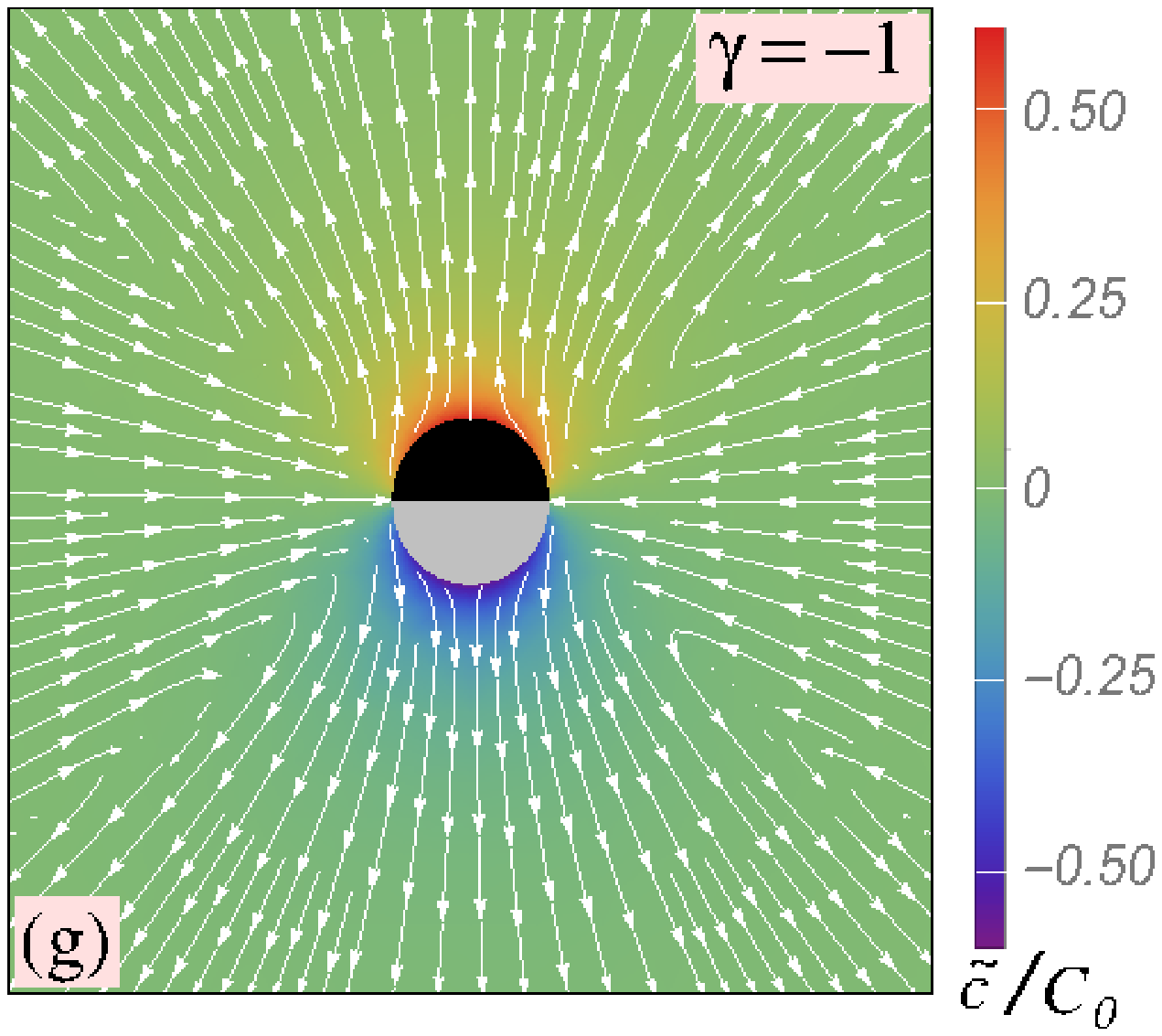}\hfill\\
\vspace*{0.1cm}\hfill\\
\includegraphics[width=.66\columnwidth]{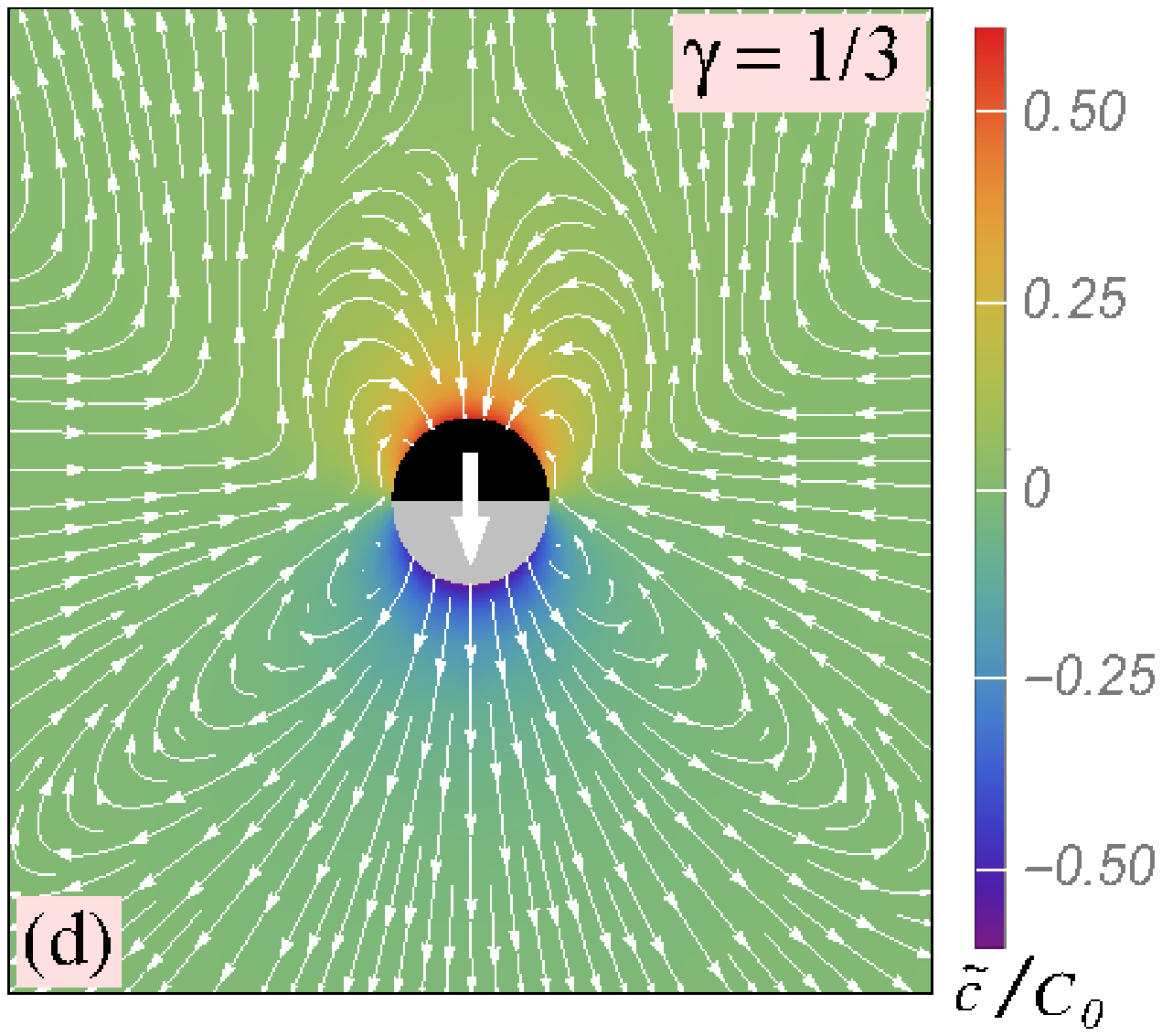}\hspace*{2.cm}%
\includegraphics[width=.66\columnwidth]{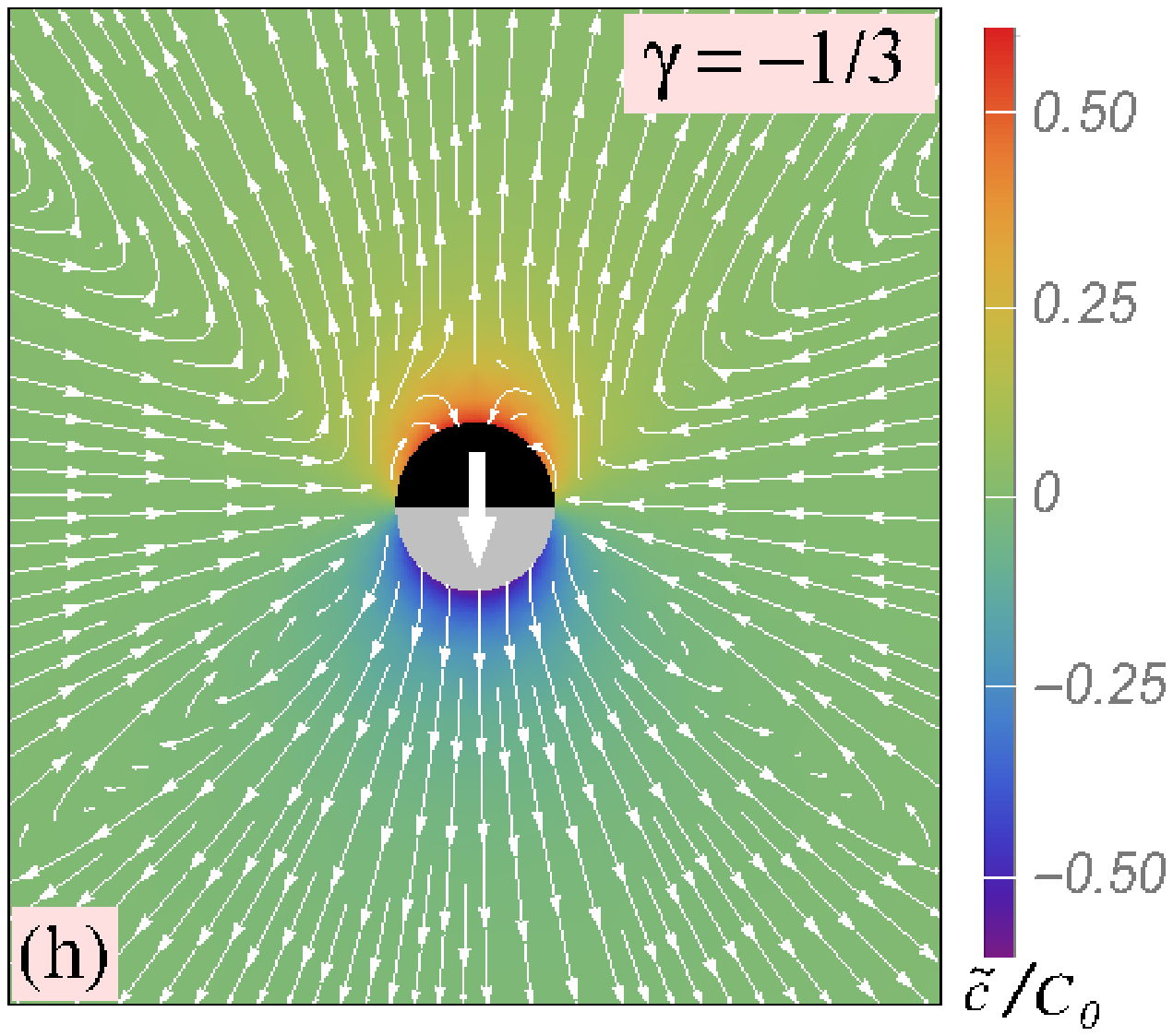}\hfill
\caption{
\label{fig6}
(a), (e): Phoretic slip  as a function of position $\theta$ 
[Eq. (\ref{eq:phor_slip_squirm}), truncated at the first 
300 terms] for model $(pa)$ with a binary-valued phoretic mobility 
(Eq. (\ref{eq:two_val_g})) and for values $\gamma > 0$ (panel (a)) and 
$\gamma < 0$ (panel (e)). The open circles show the phoretic slip obtained 
by multiplying with a factor $\gamma = \pm 3$, respectively, the branch 
$\theta > \pi/2$ of the corresponding $(pa)$ result at $\theta_0 = \pi/2$ 
from Sec. \ref{const_b}. The solid curve ($\gamma = 1$) in (a) corresponds 
to a particle with the same phoretic mobility $-b_0$ over the whole surface and 
thus reproduces (up to numerical accuracy, see the main text) the curve 
$\theta_0 = \pi/2$ in Fig. \ref{fig3}(a). For $\theta < \pi/2$, in each 
panel all three curves coincide. Note that for $\gamma \neq 1$ there is 
a discontinuity at $\theta = \pi/2$ owing to the binary-valued phoretic mobility 
(see Eq. (\ref{eq:two_val_g}) and the main text). (b)-(d) and (f)-(h): 
The flow field in the laboratory system (lines, Eq. (\ref{eq:flow_lab_syst}) 
and the number density of solute (color coded) relative to the bulk 
density (Eq. (\ref{eq:c_series})) for model $(pa)$ with a binary-valued 
phoretic mobility function (Eq. (\ref{eq:two_val_g})) and for values 
$\gamma > 0$ (panels (b)-(d)) and $\gamma < 0$ (panels (f)-(h)), respectively. 
For both the density and the flow the series are truncated at the first 
50 terms. The thick white arrows show the direction of the motion of 
the particle. In all cases the phoretic mobility is negative over the 
hemisphere where the solute is released.
}
\end{figure*}
The mode $B_2$, which enters into the definition of the squirmer parameter $S$, 
cannot be determined analytically in closed form; on the other hand, Eqs. 
(\ref{Bn_pa_twoval}) and (\ref{chi_def}) do render an expression of closed 
form for the mode $B_3$. By numerically evaluating the coefficient $\chi_2$ in 
Eq. (\ref{chi_def}), we obtain $\chi_2 \simeq -0.44$. For $n \leq 300$, the series 
entering in the definition of the coefficients $B_n$ have been evaluated by 
keeping the first $k \leq 200$ terms in Eq. (\ref{chi_def}); this ensured 
convergence of the truncation in all cases. With respect to the evaluation of the 
slip velocity over the surface of the particle, the discontinuity of the phoretic 
slip (due to the binary-valued phoretic mobility (Eq. (\ref{eq:two_val_g}))) at 
$\theta_0 = \pi/2$ cannot be captured  accurately by a truncated series representation 
(Eq. (\ref{eq:phor_slip_squirm})) even if one keeps up to $n = 300$ 
coefficients $B_n$. However, the result appears to be reasonably accurate, 
as shown in Figs. \ref{fig6}(a) and (e), except near the discontinuity at 
$\theta_0 = \pi/2$, where the curves are still slightly noisy. A 
cross-check is provided by the observation that for the phoretic mobility defined 
in Eq. (\ref{eq:two_val_g}), the slip distribution can be obtained from the one 
at $\theta_0 = \pi/2$ shown in Fig. \ref{fig3}(a) by leaving the branch 
$\theta < \pi/2$ unchanged while multiplying the branch $\theta > \pi/2$ 
with the factor $\gamma$. For $\gamma = \pm 3$, the result of this procedure 
is shown in Figs. \ref{fig6}(a) and (e) by symbols (circles); it 
compares very well with the results obtained from Eq. (\ref{eq:phor_slip_squirm}) 
(solid lines). 

\begin{figure*}[!ht]
\centering
\includegraphics[width=.8\columnwidth]{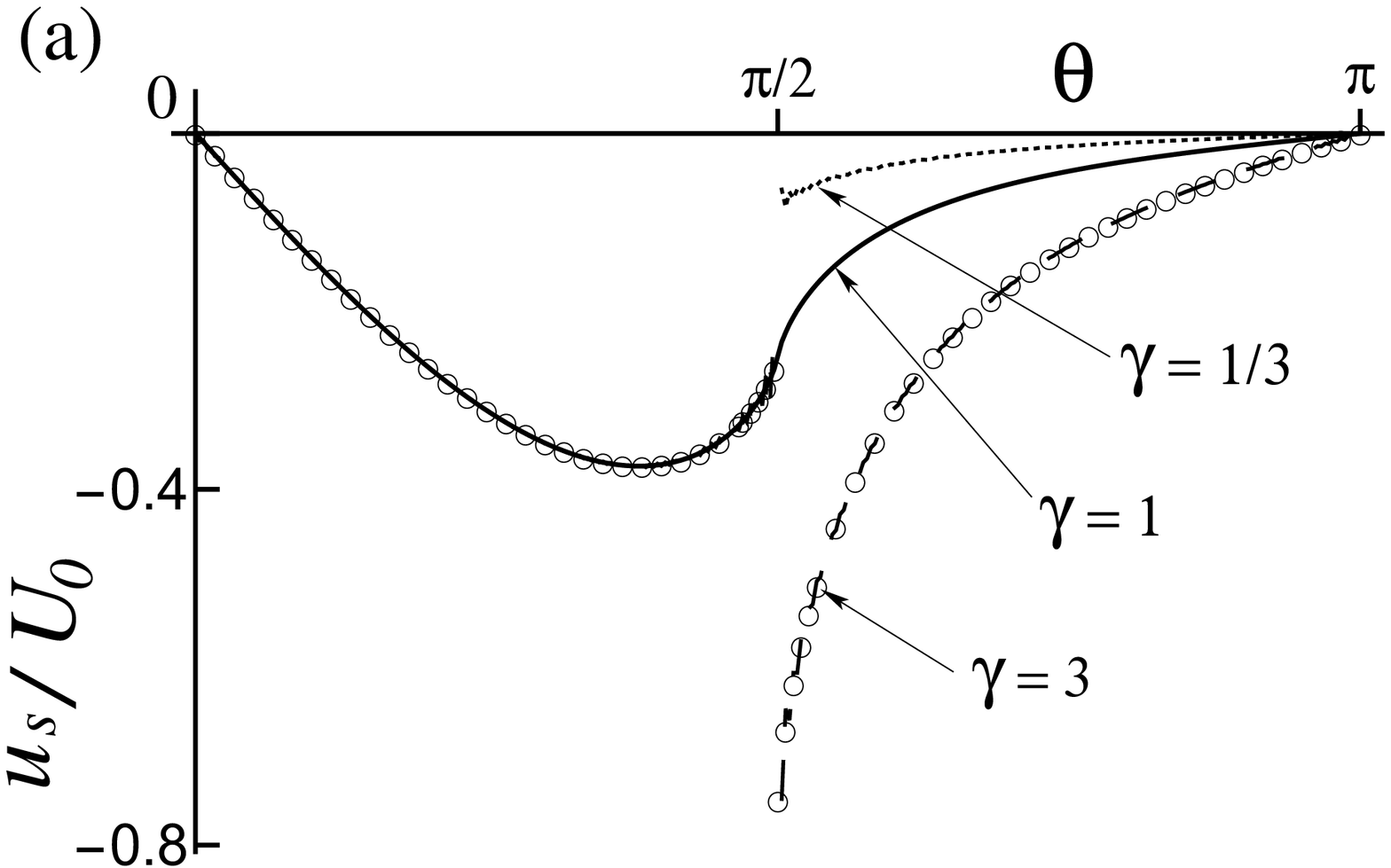}\hspace*{1.cm}%
\includegraphics[width=.8\columnwidth]{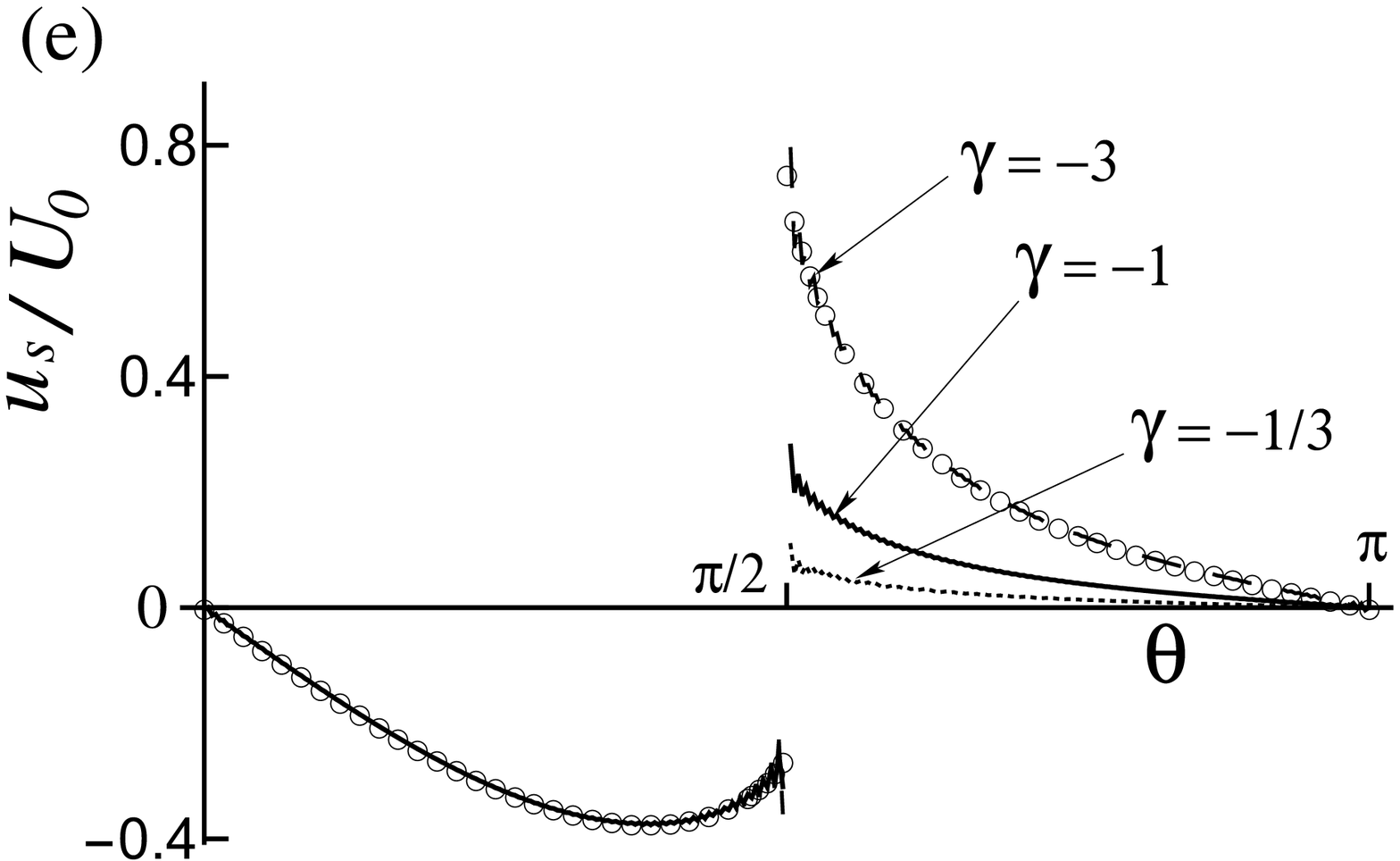}\hspace*{1.cm}\hfill\\
\vspace*{0.1cm}\hfill\\
\includegraphics[width=.66\columnwidth]{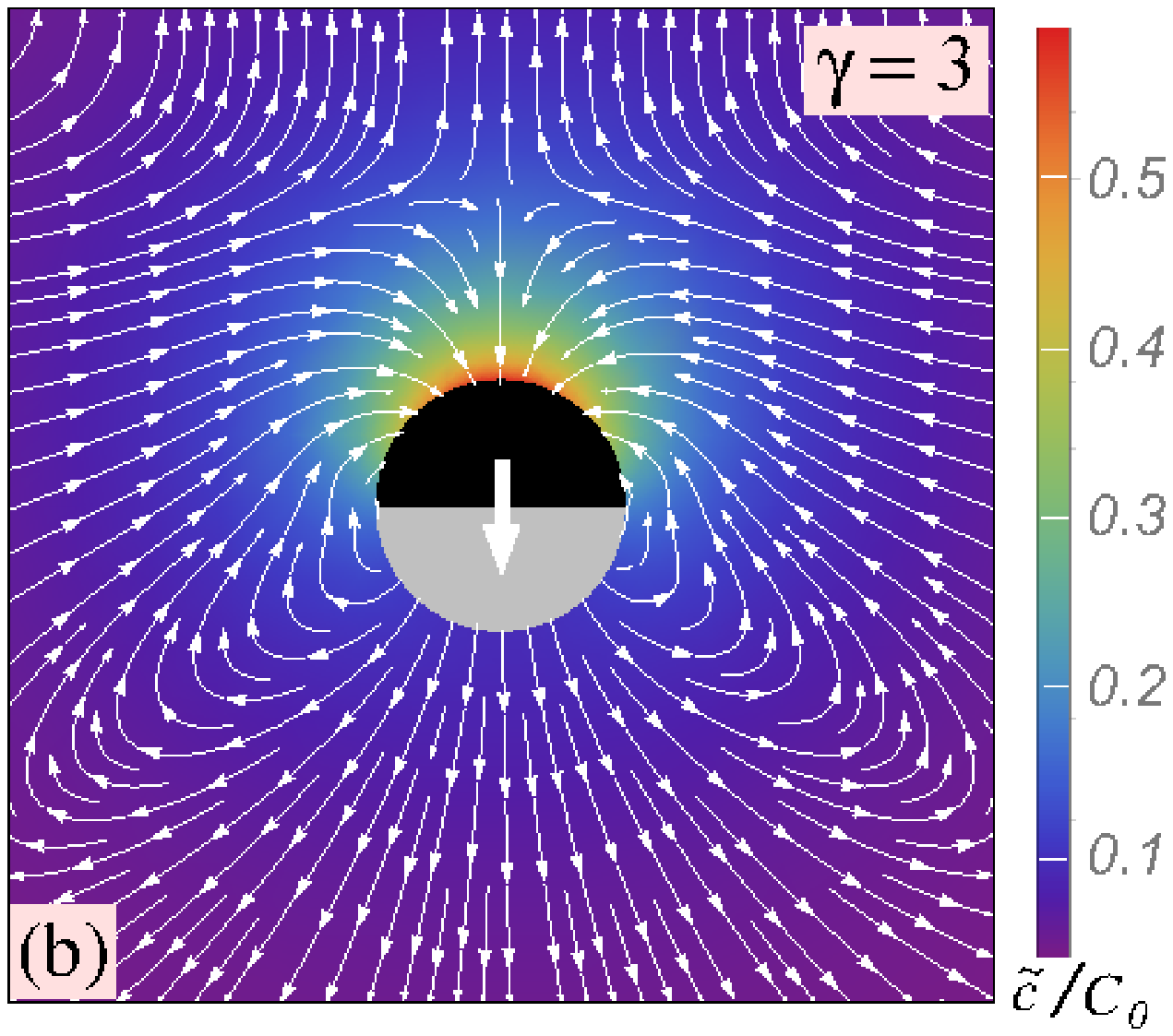}\hspace*{2.cm}%
\includegraphics[width=.66\columnwidth]{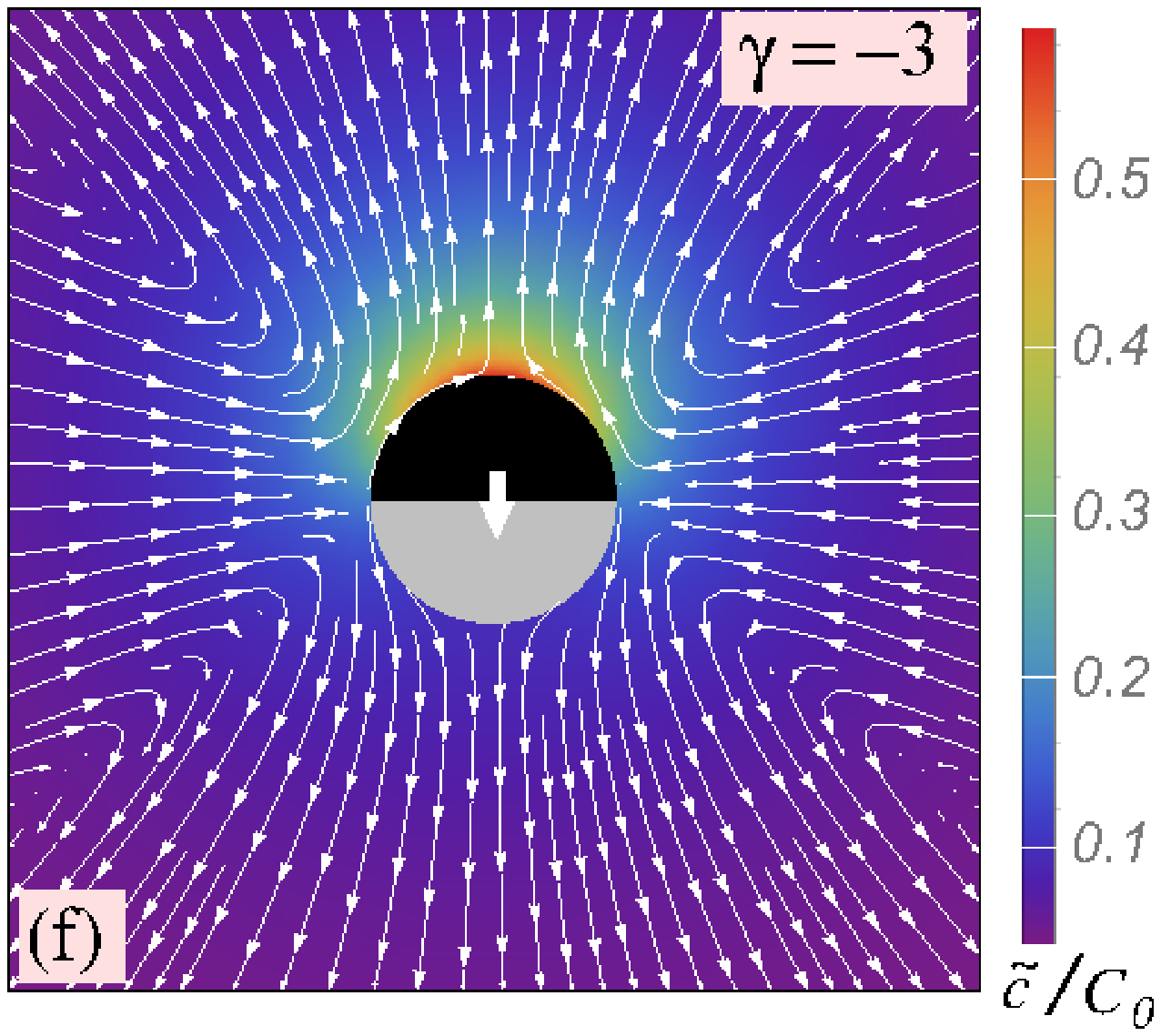}\hfill\\
\vspace*{0.1cm}\hfill\\
\includegraphics[width=.66\columnwidth]{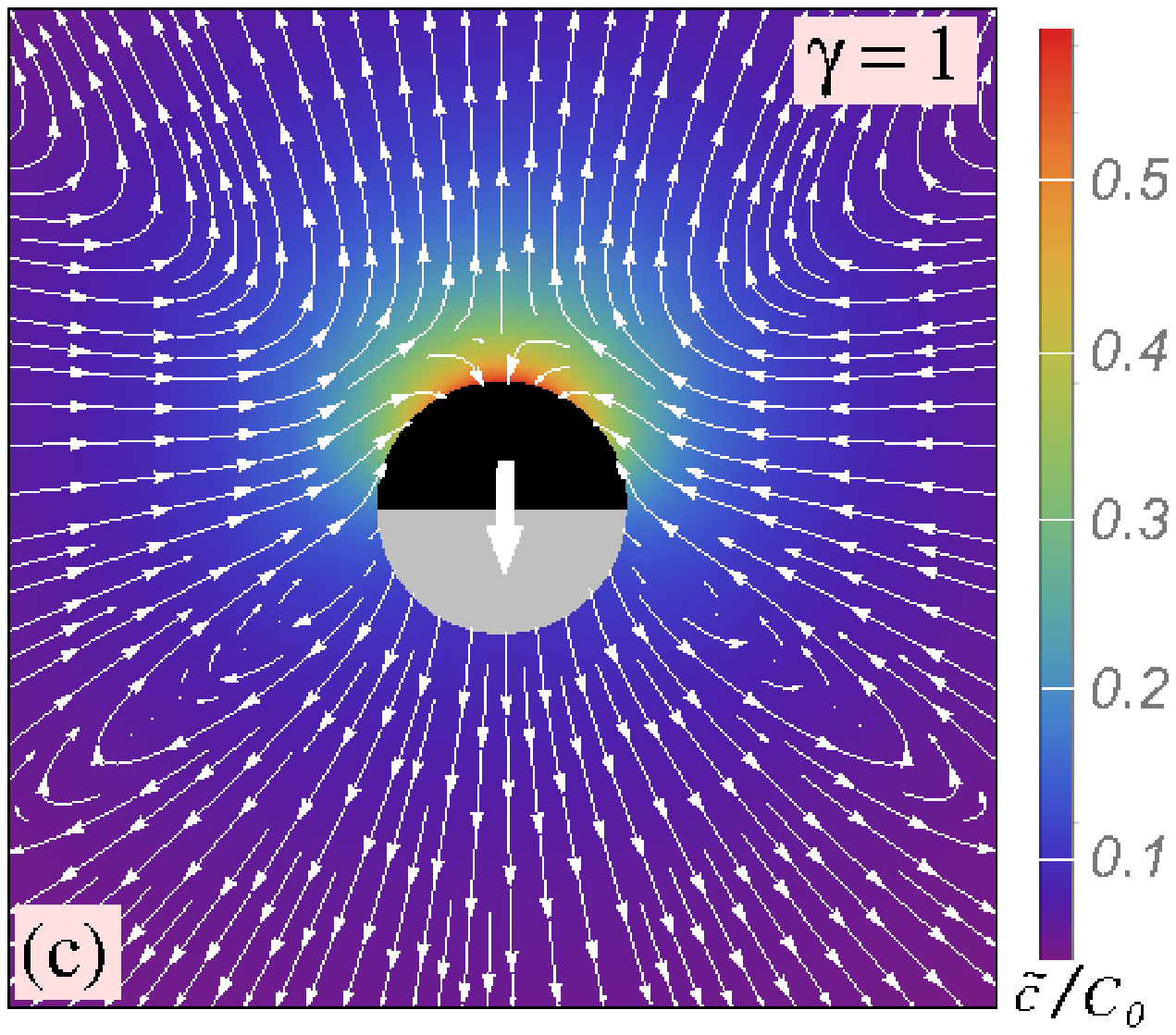}\hspace*{2.cm}%
\includegraphics[width=.66\columnwidth]{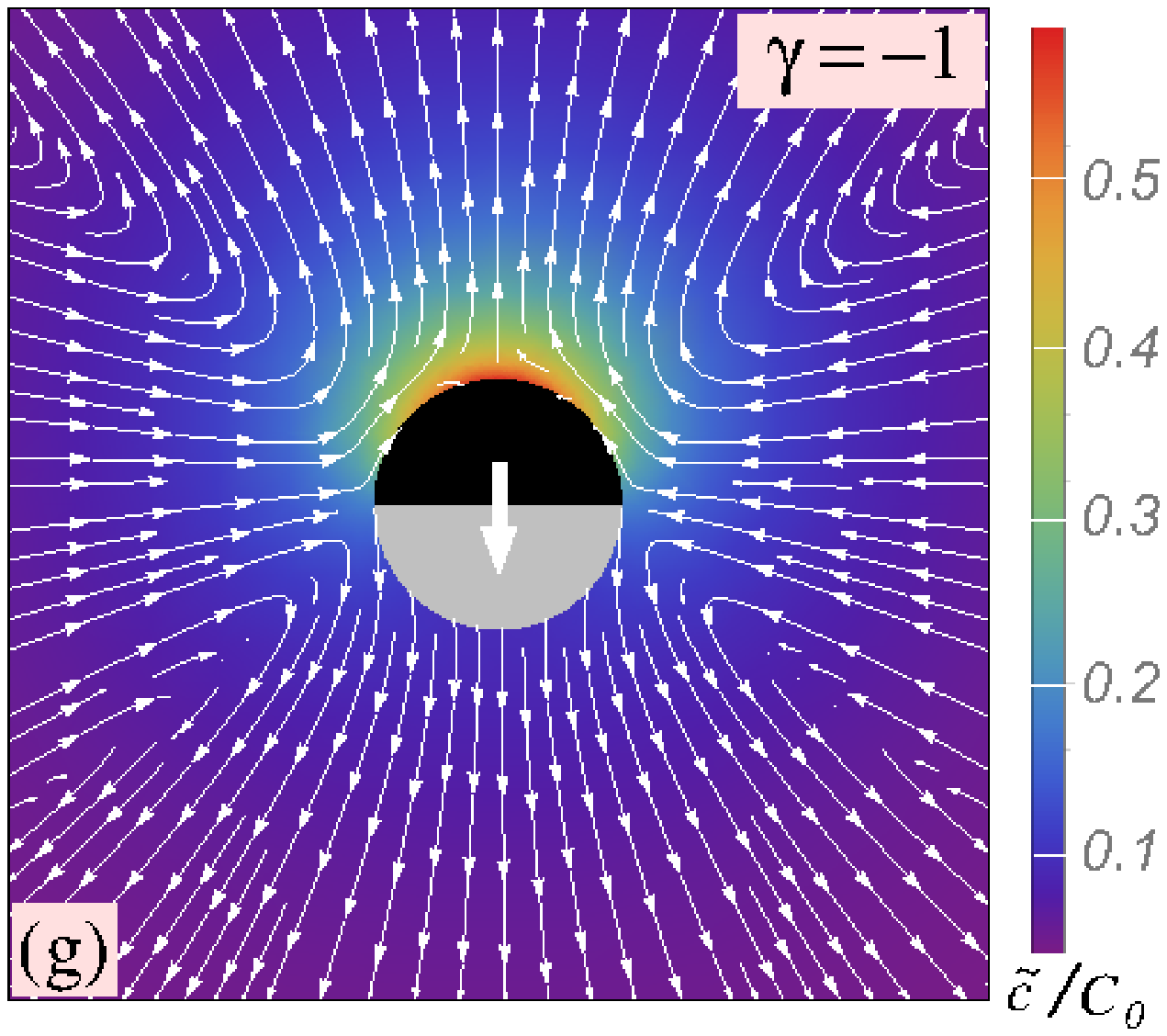}\hfill\\
\vspace*{0.1cm}\hfill\\
\includegraphics[width=.66\columnwidth]{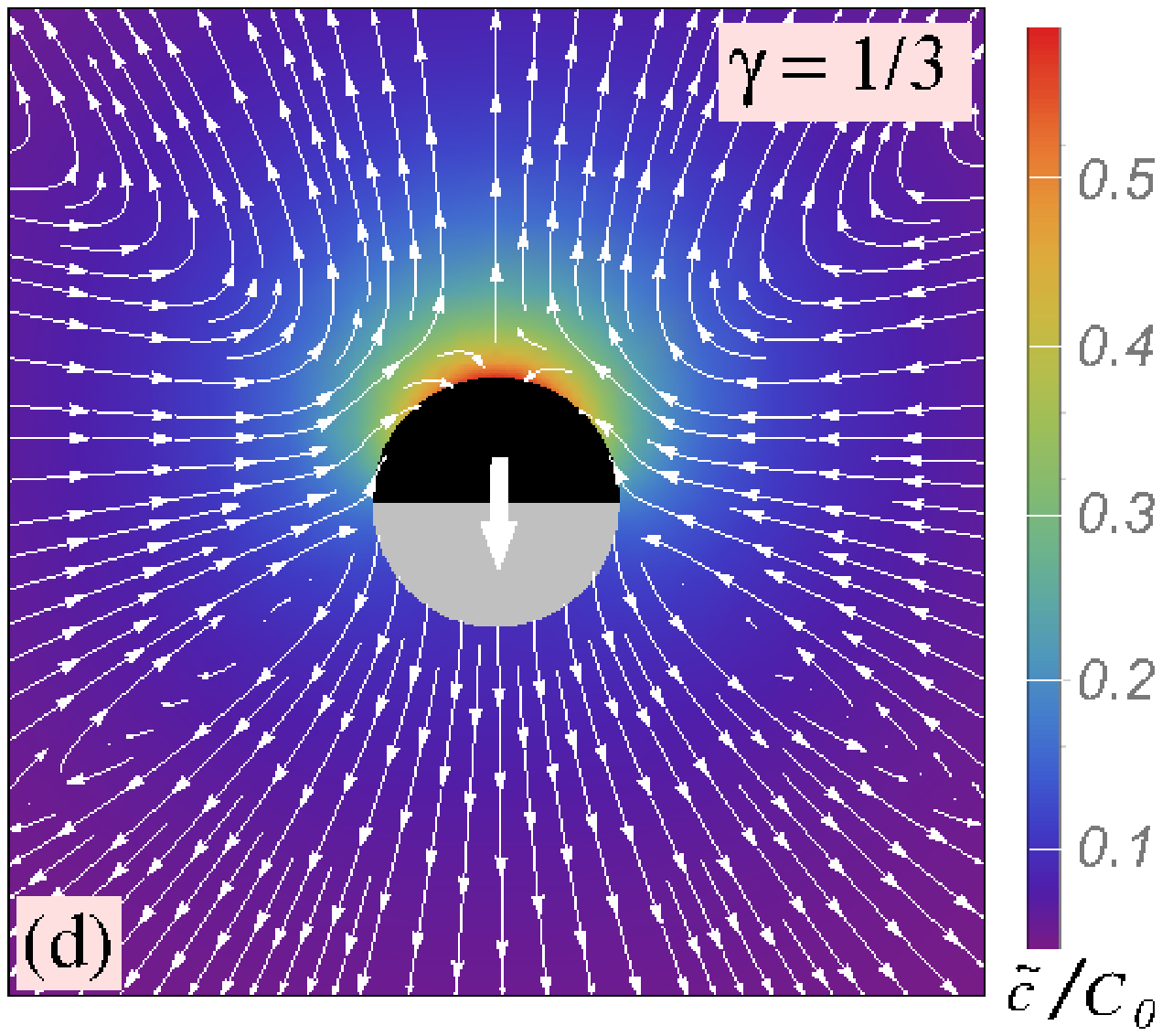}\hspace*{2.cm}%
\includegraphics[width=.66\columnwidth]{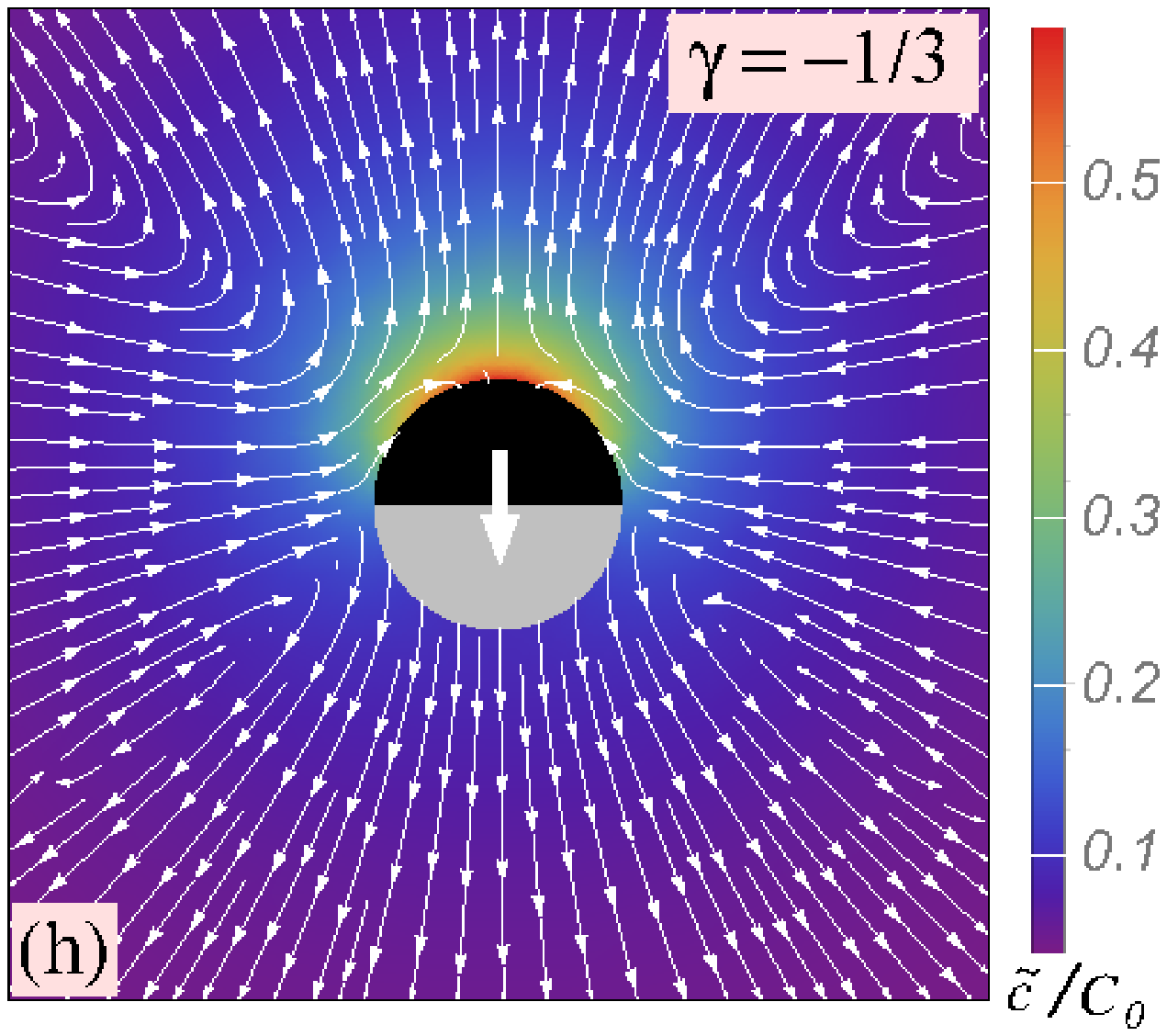}\hfill
\caption{
\label{fig7}
(a), (e): Phoretic slip  as a function of position $\theta$ 
[Eq. (\ref{eq:phor_slip_squirm}), truncated at the first 
300 terms] for model $(vi)$ with a binary-valued phoretic mobility 
(Eq. (\ref{eq:two_val_g})) and for values $\gamma > 0$ (panel (a)) and 
$\gamma < 0$ (panel (e)). The open circles show the phoretic slip obtained 
by multiplying with a factor $\gamma = \pm 3$ the branch $\theta > \pi/2$ of 
the corresponding result for $(vi)$ at $\theta_0 = \pi/2$ from Sec. \ref{const_b}. 
For $\theta < \pi/2$, in each panel all three curves coincide. Note 
that for $\gamma \neq 1$ there is a discontinuity at $\theta = \pi/2$ owing to 
the binary-valued phoretic mobility (see Eq. (\ref{eq:two_val_g}) and the main 
text). (b)-(d) and (f)-(h): The flow field in the laboratory 
system (lines, Eqs. (\ref{eq:flow_lab_syst})) and the number density of 
solute (color coded) relative to the bulk density (Eq. (\ref{eq:c_series})) 
for model $(vi)$ with a binary-valued phoretic mobility function 
(Eq. (\ref{eq:two_val_g})) and for values $\gamma > 0$ (panels (b)-(d)) and 
$\gamma < 0$ (panels (f)-(h)), respectively. For both the density and the 
flow the series are truncated at the first 50 terms. The thick white arrows 
show the direction of the motion of the particle; the shorter arrow in panel 
(f) is a reminder that at $\gamma = -3$ the behavior is very similar to that of 
a shaker (which corresponds to  $\gamma \simeq -3.4$). In all cases the phoretic 
mobility is negative over the hemisphere where the solute is released.
}
\end{figure*}
The parameters $S$ and $S'$ are thus given by
\begin{subequations}
 \label{S_S'_pa_twoval}
 \begin{equation}
\label{S_pa_twoval}
 S_{pa} := -\frac{B_2}{|B_1|} \simeq 1.16 \times 
 \frac{\gamma - 1}{|\gamma + 1|}\,,\textrm{ 
for } \gamma \neq -1\,,
\end{equation}
\begin{equation}
\label{S'_pa_twoval}
S'_{pa} := B_3/B_1 = -\frac{7}{4}\,,\textrm{ for } \gamma \neq -1\,;
\end{equation}
\end{subequations}
they are shown in Fig. \ref{fig5}(b). For $\gamma > 1$ one has $S_{pa} > 
0$ and the far-field flow is that of a puller (Fig. \ref{fig6}(b), see the 
location of the stagnation point in front of the particle). For 
$\gamma < 1$, $\gamma \neq -1$, $S_{pa}$ takes the opposite sign, 
$S_{pa} < 0$, which is reflected by a far-field flow corresponding to a pusher 
(Figs. \ref{fig6}(d), (f), and (h)). For $\gamma = 1$, one has $S_{pa} =  0$ 
and the particle behaves as a neutral squirmer (Fig. \ref{fig6}(c)). For 
$\gamma \to -1$, for which the velocity 
vanishes, the squirmer parameter $S_{pa}$ diverges ($S_{pa} \to - \infty$) 
and, as mentioned above, the particle behaves as a shaker; the corresponding 
flow field is illustrated in Fig. \ref{fig6}(g)). The change in sign of 
$S_{pa}$ upon decreasing $\gamma$ occurs ahead of the change in sign of 
the velocity. Accordingly, in the range $-1 < \gamma < 1$ the particle maintains 
the motion away from the ``producing'' cap but changes its hydrodynamic signature 
from a puller (as it is for $\gamma > 1$) to a pusher, which 
implies that the stagnation point is now located behind the particle. 
Finally, we note that, as can be read off from Eq. (\ref{S_pa_twoval}), in the 
limit of a very large ratio of mobilities, i.e., $\gamma \to \pm \infty$, 
the squirmer parameter $S_{pa}$ attains the limiting values $S_{pa} \to \pm 1$.

\subsubsection{Variable production and inert (vi) activity function}

The analysis of model $(vi)$ proceeds along very similar lines. For 
$\theta_0 = \pi/2$, all coefficients $f_n^{(vi)}$ of odd index $n$, with the 
exception of $f_1^{(vi)}$,  vanish  (see Eq. (\ref{fn_pa})). Consequently, 
Eqs. (\ref{eq:Bn_coef})-(\ref{Ink_twoval_even}) render the coefficients
\begin{equation}
  \label{Bn_vi_twoval}
  B_n = \left[1 + (-1)^{n+1} \gamma \right] \xi_n 
  + \left[1 + (-1)^{n} \gamma \right] \nu_n\,,~n \geq 1\,,
\end{equation}
where
\begin{subequations}
\label{xi_and_nu}
 \begin{equation}
\label{xi_def} 
 \xi_n = \frac{n+1/2}{4} \,f_1^{(vi)} \, J_{n,1}
 \end{equation}
and
\begin{equation}
\label{nu_def}
 \nu_n = \frac{1}{4} \times 
 \begin{cases}
 n f_n^{(vi)}\,, &~n~\textrm{even}\,,\\
 (2 n +1) \displaystyle{\sum\limits_{k>0,~\mathrm even}} J_{n,k} \,
 \dfrac{f_k^{(vi)}}{k+1}\,, &~n~\textrm{odd}\,.
 \end{cases}
\end{equation}
\end{subequations}
Thus for this model the coefficients $B_n$ with $n$ odd have to be evaluated 
numerically by suitably truncating -- typically after 200 terms -- the series 
entering into the definition of $\nu_n$.

By combining Eqs. (\ref{U_free}), (\ref{fn_vi}), (\ref{Bn_vi_twoval}), 
and (\ref{xi_and_nu}) one arrives at the following expression for the 
force-free velocity in an unbounded fluid:
\begin{equation}
 \label{Uf_vi_twoval}
 \frac{U^{(f)}}{U_0} \simeq -\frac{2}{3} \left[0.07 \,(1 - \gamma) + \frac{1}{8}\, 
 (1 + \gamma)\right]\,;
\end{equation}
this is shown as a dotted line in Fig. \ref{fig5}(a). This behavior is 
very similar to the one of model $(pa)$ discussed above: for 
$\gamma > \gamma_c \simeq -3.4$ the velocity is negative (i.e., the particle 
moves away from the active cap), while for $\gamma < \gamma_c$ the particle moves 
in the positive $z$ direction, i.e., towards the active cap. At 
$\gamma = \gamma_c$ the velocity vanishes and the hydrodynamic flow field 
induced by the particle resembles that of a shaker -- very similar to the 
flow illustrated in Fig. \ref{fig7}(f) for $\gamma = -3$.

As with the model corresponding to Fig. \ref{fig6}, upon evaluating the slip 
velocity over the surface of the particle (Eq. (\ref{eq:phor_slip_squirm})) the 
discontinuity (due to the bi-valued phoretic mobility (Eq. (\ref{eq:two_val_g}))) 
at $\theta_0 = \pi/2$ cannot be captured accurately by a truncated series 
representation, even if keeping up to $n = 300$ coefficients $B_n$. However, the 
result of such a truncation turns out to be reasonably accurate, as shown in 
Figs. \ref{fig7}(a) and (e), except near $\theta_0 = \pi/2$, where some noise in the 
curves remains 
visible. Furthermore, as for the previously discussed model, the cross-check with 
the slip distribution obtained from the one in Fig. \ref{fig4}(a) at $\theta_0 = \pi/2$  
by multiplying the branch $\theta > \pi/2$ with the factor $\gamma$ (shown by 
open circles in Figs. \ref{fig7}(a) and (e) for $\gamma = \pm 3$) is satisfactory.

The parameters $S$ and $S'$ are given by
\begin{eqnarray}
 \label{S_S'_vi_twoval}
 S_{vi} &:=& -\frac{B_2}{|B_1|} \simeq 
 \frac{- 0.4 + 0.08 \,\gamma}{|0.2 + 0.06 \,\gamma|}\,,\textrm{ 
for } \gamma \neq \gamma_c \,,\nonumber \\
S'_{vi} &:=& B_3/B_1 \simeq \, \frac{0.07 \,(1-\gamma)}{0.2 + 0.06 \,\gamma}\,,
\textrm{ for } \gamma \neq \gamma_c\,;
\end{eqnarray}
these are shown in Fig. \ref{fig5}(c). As a function of $\gamma$, 
$S_{vi}$ exhibits qualitatively the same behavior as $S_{pa}$ (compare with 
the solid curve in Fig. \ref{fig5}(b)), with the only difference that 
the zero, here at $\gamma_0 \simeq 5$, is shifted to larger positive values 
of $\gamma$ while the position $\gamma_c$ of the singularity is shifted further 
to negative values $\gamma$; the limits for $\gamma \to \pm \infty$ are 
somewhat larger, $S_{vi} \to \pm 1.33$ (see Eq. (\ref{S_S'_vi_twoval})), 
than those in model $(pa)$. Thus, 
for values $-3 \leq \gamma \leq 3$ -- as used for the examples shown in 
Figs. \ref{fig7}(a)-(h) -- the velocity of the particle is negative and the 
far-field hydrodynamic flow has the characteristics of a pusher. Finally, we note 
that in contrast to the case of model $(pa)$, the parameter $S'_{vi}$ varies 
with $\gamma$ (compare Fig. \ref{fig5}(b)); at the point $\gamma_0$ where 
$S_{vi}$ vanishes, it has a negative value $S'_{vi}(\gamma_0) \simeq -0.6$ 
(Fig. \ref{fig5}(c)).

\section{\label{first_order_kin} Number densities and flow 
around active particles exhibiting first order reaction kinetics in unbounded space}

We proceed by discussing similar mappings onto a squirmer model for more complex 
chemical 
activities of the particle, while keeping the spherical shape with axial symmetry 
unchanged, 
i.e., as before we consider a spherical colloid with a spherical cap covered by a 
catalyst. The chemical 
activity model investigated in this section depends on the local density 
of fuel molecules $A$ at the surface of the particle, and it consists of a 
catalyst-promoted chemical conversion:
\begin{equation}
 \label{kin_eq}
 A + \mathrm{catalyst}~{\rightarrow}~B + \mathrm{catalyst}\,.
\end{equation} 

As for the cases discussed in the previous section, we shall focus on the dynamics 
in steady state. We further assume that diffusion of the molecular 
species $A$ and $B$ is sufficiently fast so that advection by the flow is 
negligible compared to the transport by diffusion. Therefore, considering the case of a 
dilute solution and treating the species $A$ and $B$ as forming non-interacting, 
ideal gases, the diffusion boundary-value problem defined by 
Eq. (\ref{eq:bvp_for_c_thin}) is replaced by two problems for the number 
densities $c_{A,B}(\mathbf{r})$ of the fuel ($A$) and the product ($B$) molecular 
species \cite{Golestanian2012,Michelin2014,Kapral2007,Kapral2013}:
\begin{subequations}
 \label{eq:bvp_for_A_and_B}
\begin{equation}
\label{eq:diff_A_and_B} 
\nabla^2 c_{A,B}(\mathbf{r}) = 0\,.
 \end{equation}
They are subject to the BCs
\begin{equation}
\label{eq:bcs_for_c_A_and_B} 
- D_{A,B} \left.\left[\mathbf{n} \cdot 
\nabla c_{A,B}(\mathbf{r})\right]\right|_{\mathbf{r} = \mathbf{r}_P} 
= \mp {\cal K}(\mathbf{r}_P)\,,
\end{equation}
\begin{equation}
\label{eq:c_A_and_B_infty} 
c_{A,B}(|\mathbf{r}| \to \infty) \to C_{A,B}^{(\infty)}\,.
\end{equation}
\end{subequations}
The minus sign applies for species $A$, and the activity function 
${\cal K}(\mathbf{r}_P)$, with $\mathbf{r}_P$ on the surface of 
the particle, is given by
\begin{equation}
\label{eq:act_func_A_and_B}
{\cal K}(\mathbf{r}_P) = 
\begin{cases}
& \kappa \, c_A(\mathbf{r}_P)\,,~\mathrm{for}~\mathbf{r}_P \in ~\mathrm{catalyst}\,,\\
& 0\,, ~\mathrm{otherwise}\,.
\end{cases}
\end{equation}
Moreover, we assume that $\kappa$ (with units m/s) is constant over the catalyst 
covered area, i.e., $\kappa$ is a constant independent of the position $\mathbf{r}_P$ 
within the catalytic patch. For a steady state with $c_{A,B} > 0$ to be stable it is 
necessary that $C_{A}^{(\infty)} > 0$; on the other hand, Eq. (\ref{eq:bvp_for_A_and_B}) 
reveals that $C_{B}^{(\infty)}$ will enter into the final result only as an additive 
constant. Thus $C_{B}^{(\infty)}$ is irrelevant for the motion of the particle (driven 
by gradients of the densities) and for the hydrodynamic flow of the solution; in 
the following we set $C_{B}^{(\infty)} = 0$. 

The form (Eq. (\ref{eq:act_func_A_and_B})) of the right hand side of Eq. 
(\ref{eq:bcs_for_c_A_and_B}) for the species $A$ and $B$ defines the reaction as 
exhibiting a first order chemical kinetics: the rate of consumption of $A$ 
molecules (and the rate of production of $B$ molecules, respectively) by 
the catalytic chemical reaction is proportional to the local number density $c_A$ 
of the fuel molecules. As implied by Eq. (\ref{eq:bcs_for_c_A_and_B}), at 
the catalyst covered region of the particle the chemical reaction  acts as a sink 
term for the flux of $A$ molecules, and as a source term of the same magnitude for 
the flux of $B$ molecules. As noted in Refs. \cite{Brady2008,Golestanian2012}, 
this implies that the quantity 
\begin{equation}
 \label{sum_def}
 {\cal N} (\mathbf{r}) := D_A c_A(\mathbf{r}) + D_B c_B(\mathbf{r})
\end{equation}
is spatially constant.\footnote{From Eq. (\ref{eq:bvp_for_A_and_B}) 
it follows that ${\cal N}$ obeys the Laplace equation subject to a 
homogeneous Neumann BC (i.e., vanishing normal derivative) on the sphere and subject to 
the BC of a constant value ($D_A C_A^{(\infty)}$) at infinity. The solution of this 
boundary value problem is a constant.} 
Therefore, it is sufficient to solve one of the boundary value problems 
in Eq. (\ref{eq:bvp_for_A_and_B}), e.g., the one for $c_B(\mathbf{r})$. The 
other density is obtained from ${\cal N} (\mathbf{r}) = D_A C_A^{(\infty)}$ 
(recall that $C_B^{(\infty)} = 0$) as
\begin{subequations}
\begin{equation}
 \label{dens_A_from_conserv}
c_A(\mathbf{r}) = C_A^{(\infty)} - \frac{D_B}{D_A} c_B(\mathbf{r}) \,.
\end{equation}

In order to reduce the number of free parameters, we restrict the subsequent  
discussion to the particular case in which the species $A$ and $B$ have similar 
diffusion constants $D_A \simeq D_B =: D$. In this case, 
Eq. (\ref{dens_A_from_conserv}) takes the simpler form 
\begin{equation}
\label{dens_A_same_diff}
c_A(\mathbf{r}) = C_A^{(\infty)} - c_B(\mathbf{r})\,. 
\end{equation}
\end{subequations}
Furthermore, for the same reason of reducing the number of free parameters, we 
consider only the case of a particle half-covered by catalyst, i.e., we fix 
$\theta_0 = \pi/2$, which experimentally is the most relevant case. 
Additionally, we assume that the phoretic mobility $b_A$ of the fuel species 
$A$ vanishes. This facilitates a straightforward comparison with the 
cases discussed in the previous sections. Moreover, under the constraint in Eq. 
(\ref{dens_A_from_conserv}), the generalization to the case $b_A \neq 0$ merely 
amounts to a redefinition of the parameter $b_B$ (see, e.g., Ref. 
\cite{Golestanian2012}). As it will become clear in the following, a generalization of 
the calculations to arbitrary values for the diffusion constants and the coverage 
$\theta_0$ is straightforward but involves significantly more cumbersome algebra.

The relative importance of the transport by diffusion compared to the production 
(and annihilation) of molecular species through the chemical reaction is characterized by 
the dimensionless Damk{\"o}hler number\footnote{For distinct values 
of the diffusion coefficients of the two species, it is customary to define the 
Damk{\"o}hler number in terms of the reactant (i.e., fuel) species.}
\begin{equation}
 \label{eq:Da_def}
 Da = \frac{\kappa R}{D}\,;
\end{equation}
in the chemical kinetics literature, the limits $Da \ll 1$ and $Da \gg 1$ are known 
under the physically intuitive names of ``reaction limited'' and ``diffusion limited'' 
regimes, respectively. In the following we shall use 
\begin{equation}
\label{eq:def_barC0}
 {\bar C}_0 = Da \, C_A^{(\infty)}
\end{equation}
as a characteristic number density. This choice is motivated by the fact that, 
as we shall show below, in the ``reaction limited'' regime ${\bar C}_0$ 
turns into the expression for the characteristic number density $C_0$ 
defined by Eq. (\ref{eq:def_C0}). 

Introducing $n_B(\mathbf{r}): = c_B(\mathbf{r})/{\bar C}_0$ and recalling 
$D_A = D_B$, Eqs. (\ref{eq:bvp_for_A_and_B}), (\ref{dens_A_same_diff}), 
(\ref{eq:Da_def}), and (\ref{eq:def_barC0}) render the following boundary-value problem for 
$n_B(\mathbf{r})$:
\begin{subequations}
 \label{eq:bvp_for_nB}
 \begin{equation}
 \label{eq:Lap_nB}
 \nabla^2 n_B(\mathbf{r}) = 0\,.
 \end{equation}
The BC on the particle surface ($r = R$) is
\begin{equation}
  \label{eq:BC_part_nB}
  - R \, \partial_r n_B(R,\theta) = 
  \begin{cases}
   1 - Da \,\times\, n_B(R,\theta)\,,&~0 \leq \theta \leq \pi/2\,,\\
   0\,,& \pi/2 < \theta \leq \pi\,,
  \end{cases}
 \end{equation}
and the BC at infinity is
 \begin{equation}
 \label{eq:BC_infty_nB}
 n_B(r \to \infty) \to 0\,.
 \end{equation}
 \end{subequations}
Before we proceed, we note that in the limit $Da \ll 1$ one has $1 - Da \, 
 n_B(\mathbf{r}_P) \simeq 1$; thus Eq. (\ref{eq:bvp_for_nB}) indeed takes 
 the same form as the one describing the model $(pi)$ in Sec. \ref{C_and_flow}.1.1 
 if one identifies $Da\,\times\,C_A^{(\infty)} = C_0 = const$. Furthermore, 
 from  Eq. (\ref{dens_A_same_diff}) it follows that in the same limit $Da \ll 1$ 
 one has $n_A(\mathbf{r}) \simeq 1$; thus indeed $\kappa c_A(\mathbf{r}_P) \simeq 
\kappa C_A^{(\infty)}$ can be consistently identified with the production rate $Q$ 
in model $(pi)$. Similarly, in the opposite limit $Da \gg 1$, in order 
to have $Da \,\times \, n_B < 1$ (see the first line of the BC in 
Eq. (\ref{eq:BC_part_nB})) 
$n_B$ must satisfy $n_B(R, \theta \leq \pi/2) \ll 1$. By combining this new 
BC on the catalytic hemisphere with that of vanishing normal derivative on the lower 
hemisphere and with the BC in Eq. (\ref{eq:BC_infty_nB}), it follows that in this case 
 $n_B(\mathbf{r})$ is vanishingly small everywhere, i.e., 
 $n_B (\mathbf{r}) \ll 1$ \cite{Gleb2017} (which is consistent with a similar 
 conclusion obtained by considering Eq. (\ref{dens_A_same_diff}) in the limit 
 $Da \gg 1$); this implies that the particle remains at rest and that there is no 
 flow of the solution (see also Ref. \cite{Gleb2017}).
  
Since the limiting cases $Da \gg 1$ and $Da \ll 1$ are either trivial (the 
former one) or has been already analyzed in the previous sections (the latter one), 
in the following we shall focus on the case $Da \sim {\cal O}(1)$. As in the previous 
sections, the solution can be conveniently expressed in terms of a multipole 
series expansion:
\begin{equation}
 \label{eq:nB_series}
 n_B(r,\theta) = \sum\limits_{n \geq 0} \frac{f_n^{(kin)}}{n+1} 
 \left(\frac{R}{r}\right)^{n+1} P_n(\cos\theta)\,.
\end{equation}
The coefficients $f_n^{(kin)}$ (the superscript referring to the reaction 
exhibiting a first-order \textit{kin}etics) are determined by plugging 
Eq. (\ref{eq:nB_series}) into the BC in Eq. (\ref{eq:BC_part_nB}). This 
renders
\begin{eqnarray}
 \label{eq:coef_exp_nB}
 &&\sum_{k \geq 0} f_k^{(kin)} P_k (\cos \theta)  \\ 
 & = & \begin{cases}
  1 - Da \displaystyle{\sum\limits_{k \geq 0} \frac{f_n^{(kin)}}{k+1}} 
  P_k (\cos \theta) \,,&~0 \leq 
  \theta \leq \pi/2\,,\\
   0\,,& \pi/2 < \theta \leq \pi\,. \nonumber
 \end{cases}
\end{eqnarray}
Multiplying both sides with $(\sin \theta )\, P_n(\cos\theta)$, integrating 
the first and the second lines of the RHS over their corresponding intervals, 
adding the results, using the orthogonality of the Legendre polynomials 
(Eq. (\ref{Pn_norm})), and defining (see also Appendix \ref{formulas})
\begin{subequations}
 \label{eq:def_aux_Hfunc}
 \begin{equation}
 {\cal H}_n(\theta_0) 
 := \int\limits_{0}^{\theta_0} \,d\theta \sin \theta \,P_n(\cos \theta)
 \end{equation}
 and
 \begin{equation}
 H_{n,m}(\theta_0) := \int\limits_{0}^{\theta_0} \,d\theta \sin \theta \,
 P_n(\cos \theta) \, P_m(\cos \theta)\,,
 \end{equation}
\end{subequations}
one obtains the following infinite system of linear equations for the coefficients 
$f_n^{(kin)}$:
\begin{eqnarray}
 \label{eq:sys_coef_nB}
 \frac{2}{2 n + 1}\, f_n^{(kin)} &+& 
 Da \,\sum_{k \geq 0}\frac{H_{n,k}(\pi/2)}{k+1} f_k^{(kin)} \nonumber\\ 
 &=& {\cal H}_n(\pi/2)\,,~ n = 0, 1, \dots~~~\,.
\end{eqnarray}

Such systems of equations (equivalently, Eq. (\ref{eq:coef_exp_nB})) are known 
as dual series problems \cite{Collins1960}. They are often encountered in mixed 
boundary value problems \cite{Sneddon_book} in the context of, e.g., 
electrostatics of hemispherical conducting shells \cite{Collins1960} or the 
calculation of steric factors in the chemical kinetics literature 
\cite{Deutch1978,Shoup1981,Shoup1982,Traytak1994,Traytak1995a,Traytak1995b,
Traytak1995c,Traytak2007,Gleb2017}. There are cases in which the solution of such 
equations can be obtained analytically (see, e.g., Ref. \cite{Majee2013} for an 
example in the context of active colloids and Ref. \cite{Sneddon_book} for a 
rather exhaustive list). However, these cases are the exception rather than 
the rule. Therefore, in general only numerical or approximate solutions are 
possible (see, e.g., Refs. \cite{Traytak1994,Traytak1995b}). Here we follow 
the latter approach and numerically solve the system of equations by 
truncating it at a sufficiently large order $N$, i.e., we set 
$f_{n > N}^{(kin)} = 0$. We have used the cutoff $N = 300$, for which the 
linear system of equations (\ref{eq:sys_coef_nB}) can be solved without requiring 
particular technical efforts, for values of $Da$ within the broad range 
$10^{-4} \leq Da \leq 10^3$. Finally, we remark that Eq. (\ref{eq:sys_coef_nB}) can be 
straightforwardly generalized to the case of a cap with opening angle 
$\theta_0 \neq \pi/2$ by replacing $\pi/2 \to \theta_0$ in the arguments of the 
functions ${\cal H}_n$ and $H_{n,k}$.
\begin{figure}[!htb]
    \centering
   \includegraphics[width=.99\columnwidth]{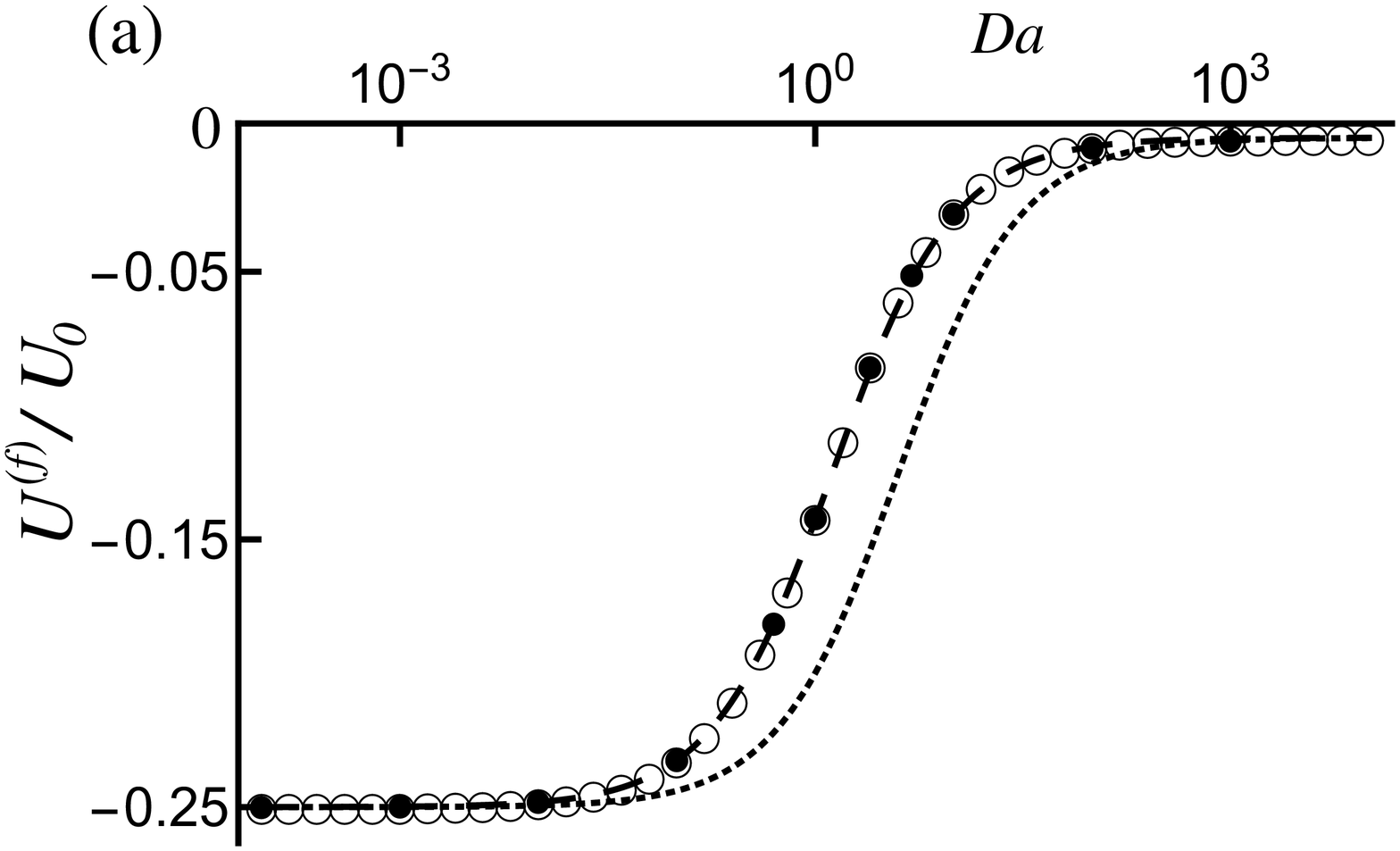}
   \vspace*{1.cm}\hfill\\
    \includegraphics[width=.95\columnwidth]{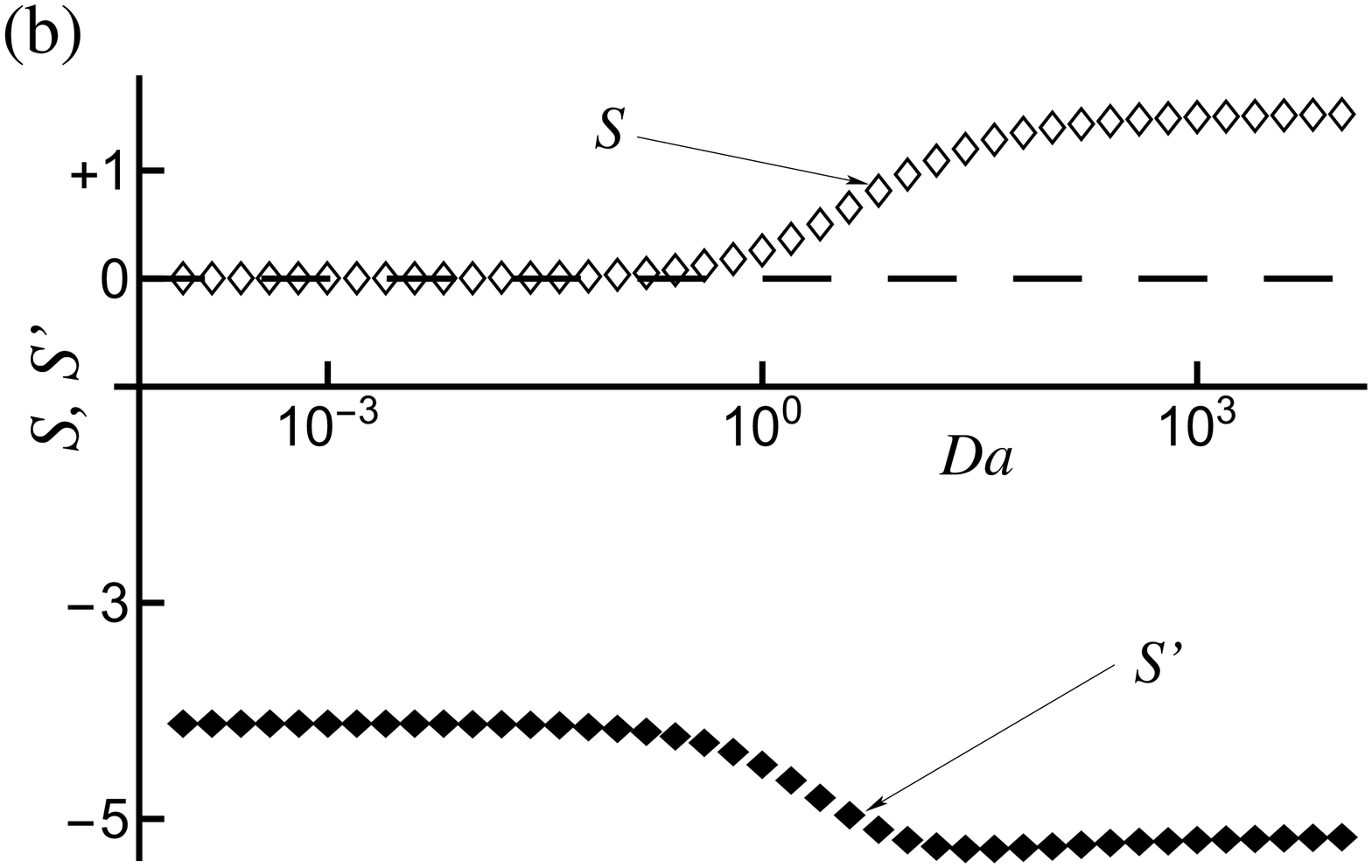}   
     \caption{
\label{fig8}
(a) The velocity $U^{(f)}$, in units of $U_0$ (see the main text), of a force-free 
Janus particle ($\theta_0 = \pi/2$) as a function of the Damk{\"o}hler number $Da$. 
The open symbols correspond to the solution of Eq. (\ref{eq:sys_coef_nB}), the 
filled symbols are the results of BEM numerical solutions, the dotted line is the 
approximate solution $-1/(4 + Da)$ obtained by disregarding the off-diagonal terms 
in Eq. (\ref{eq:sys_coef_nB}) (i.e., by setting $H_{n,k} = H_{n,n} \, \delta_{n,k}$), 
while the dashed line corresponds to the heuristic guess (fit) $- 1/(4 + 3 Da)$.
(b) The squirmer parameters $S = -(B_2/B_1)$ (note that $B_1 > 0$, since 
$U^{(f)}/U_0$ is negative, see panel (a)) and $S' = B_3/B_1$ as functions of $Da$ 
for the model Janus particle with first-order kinetics reaction. In both (a) and (b) 
the phoretic mobility is position-independent and negative ($g(\theta) 
= -1$).
}
\end{figure}

For a given value of $Da$, once the coefficients $f_n^{(kin)}$ are calculated 
and known, the solute number density follows from Eq. (\ref{eq:nB_series}). 
Since this series expansion has exactly the same form as the one in 
Eq. (\ref{eq:c_series}), the whole machinery, developed in 
Sec. \ref{squirm} for establishing the mapping onto an effective squirmer, 
can be employed directly. In view of this context, we shall discuss 
only a couple of aspects pertaining to such models. For a particle with uniform phoretic 
mobility, we thus know that the amplitudes of the squirming modes are given by 
$B_n^{(kin)}(Da) = n f_n^{(kin)}(Da)/2$, where we explicitly indicated the 
dependence on $Da$. The velocity of the squirmer as a 
function of $Da$ immediately follows from $f_1^{(kin)}(Da)$ and 
Eqs. (\ref{U_free}) and (\ref{Bn_unif_b}). This result is shown in 
Fig. \ref{fig8}(a) in units of $U_0$, the latter being defined via 
Eq. (\ref{eq:def_U0}) with the replacement 
$C_0 \to {\bar C}_0$. (Accordingly, here $U_0$ is a function of 
$Da \, \times \, C_0^{(\infty)}$. Therefore the limits $Da \ll 1$ and 
$Da \gg 1$ should be taken under the constraint of a finite, non-vanishing $U_0$, 
i.e., $0 < Da\,\times\,C_0^{(\infty)} < \infty$.)

Figure \ref{fig8}(a) shows also the results obtained by 
employing the Boundary Element Method (BEM) in order to numerically 
solve the corresponding Laplace and Stokes equations directly. These 
results render a successful cross-check of the calculation outlined above. 
Additionally, we show the analytical approximation $- 1/(4+ Da)$ (dotted line), 
which is obtained by setting to zero the off-diagonal terms in 
Eq. (\ref{eq:sys_coef_nB}), i.e., $H_{n,k} = H_{n,n} \delta_{n,k}$. This 
amounts to a ``zeroth-order'' approximation suggested 
in Ref. \cite{Traytak1995a}). This approximation works 
well qualitatively, although it is quantitatively inaccurate as far as 
the location is concerned of the cross-over between the two plateaux values 
at low and large $Da$, respectively. Finally, the dashed line shows the heuristic 
``fit'' $- 1/(4+ 3 Da)$; such an accurate description provided by a simple 
functional form with integer coefficients strongly hints towards the existence 
of an exact analytical solution of Eq. (\ref{eq:coef_exp_nB}); however, 
so far our efforts to derive such a solution have been unsuccessful.

\begin{figure}[!htb]
\centering
\includegraphics[width=.77\columnwidth]{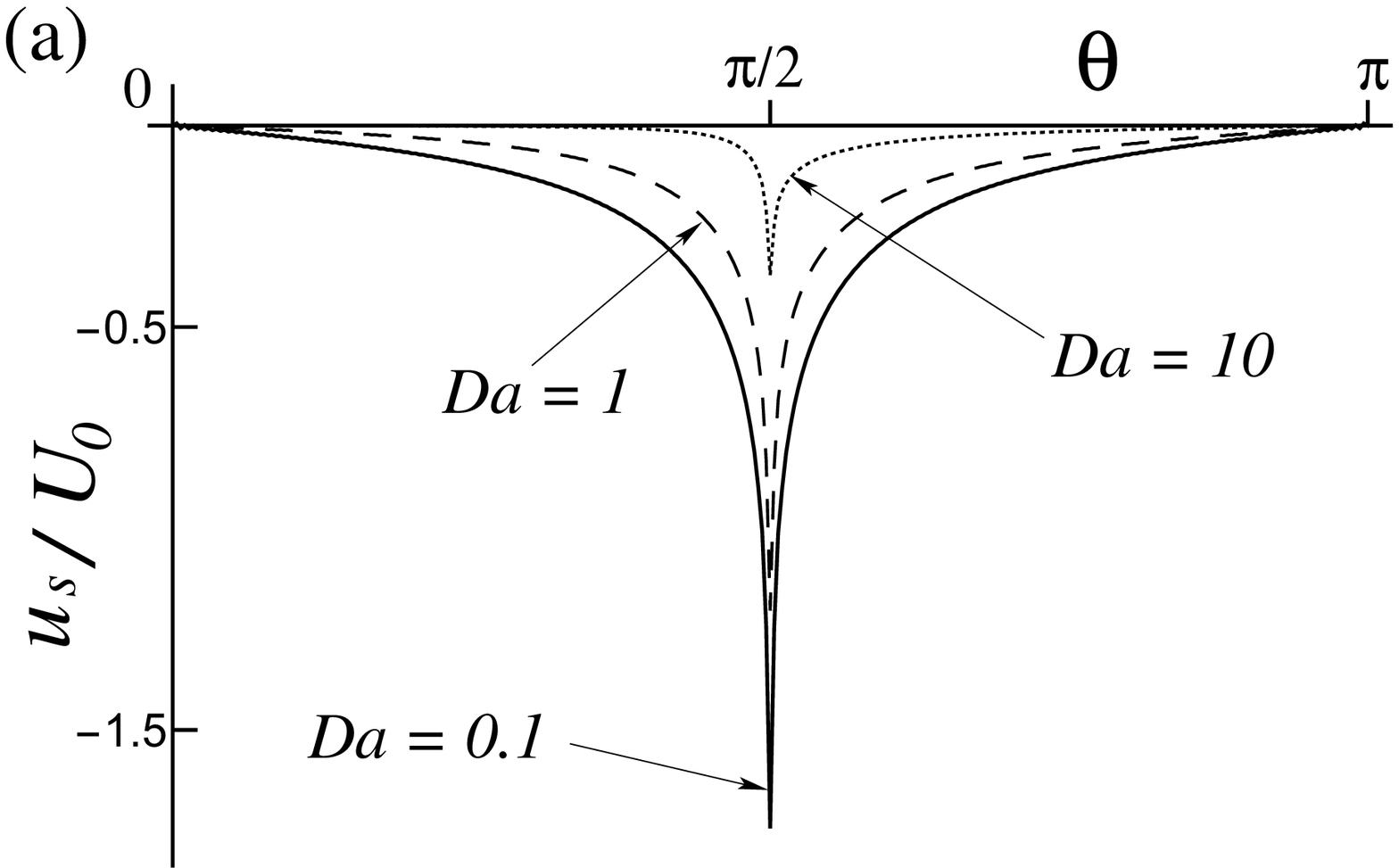}\hspace*{1.cm}
\vspace*{0.15cm}\hfill\\
\includegraphics[width=.61\columnwidth]{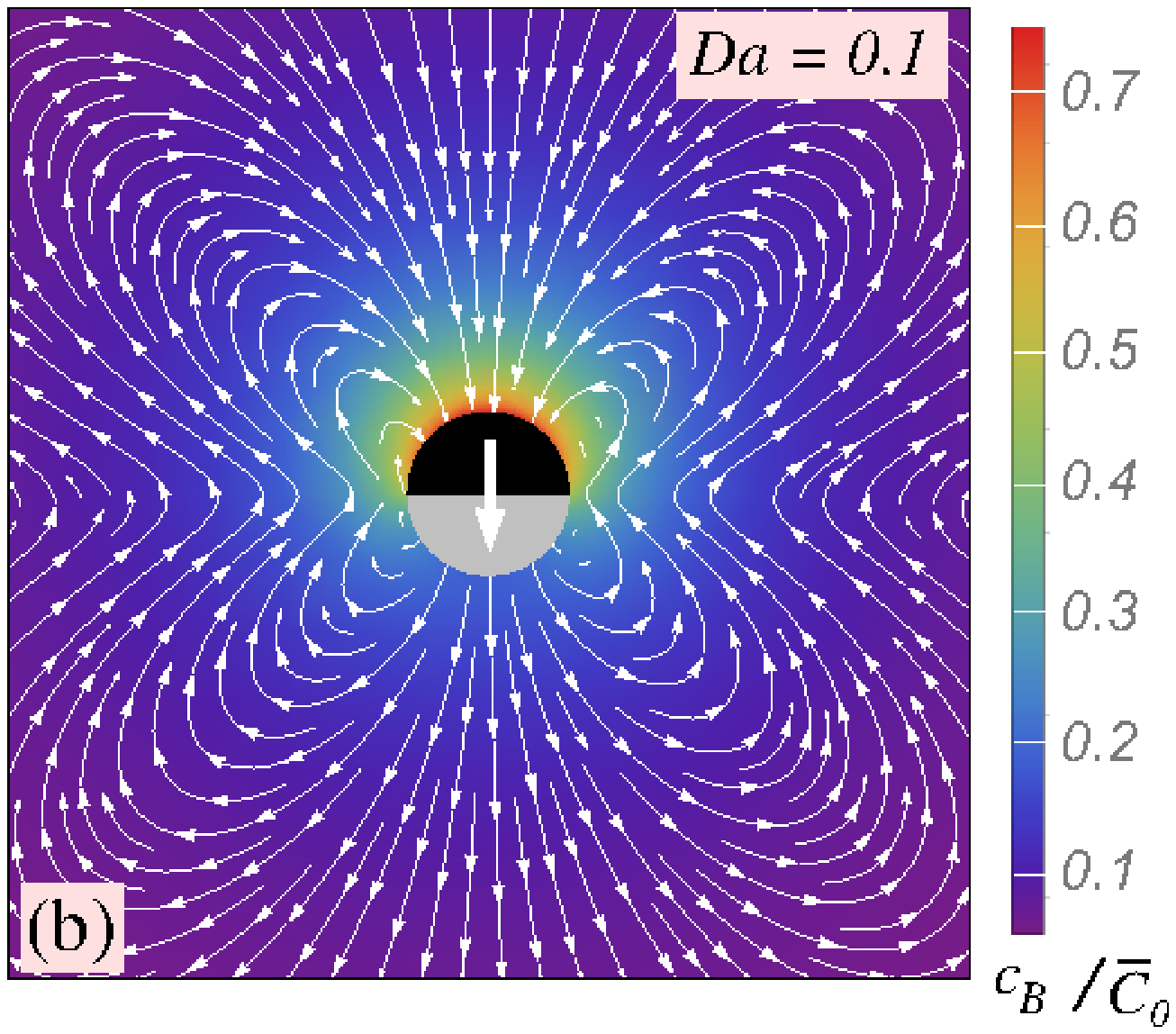}
\vspace*{0.2cm}\hfill\\
\includegraphics[width=.61\columnwidth]{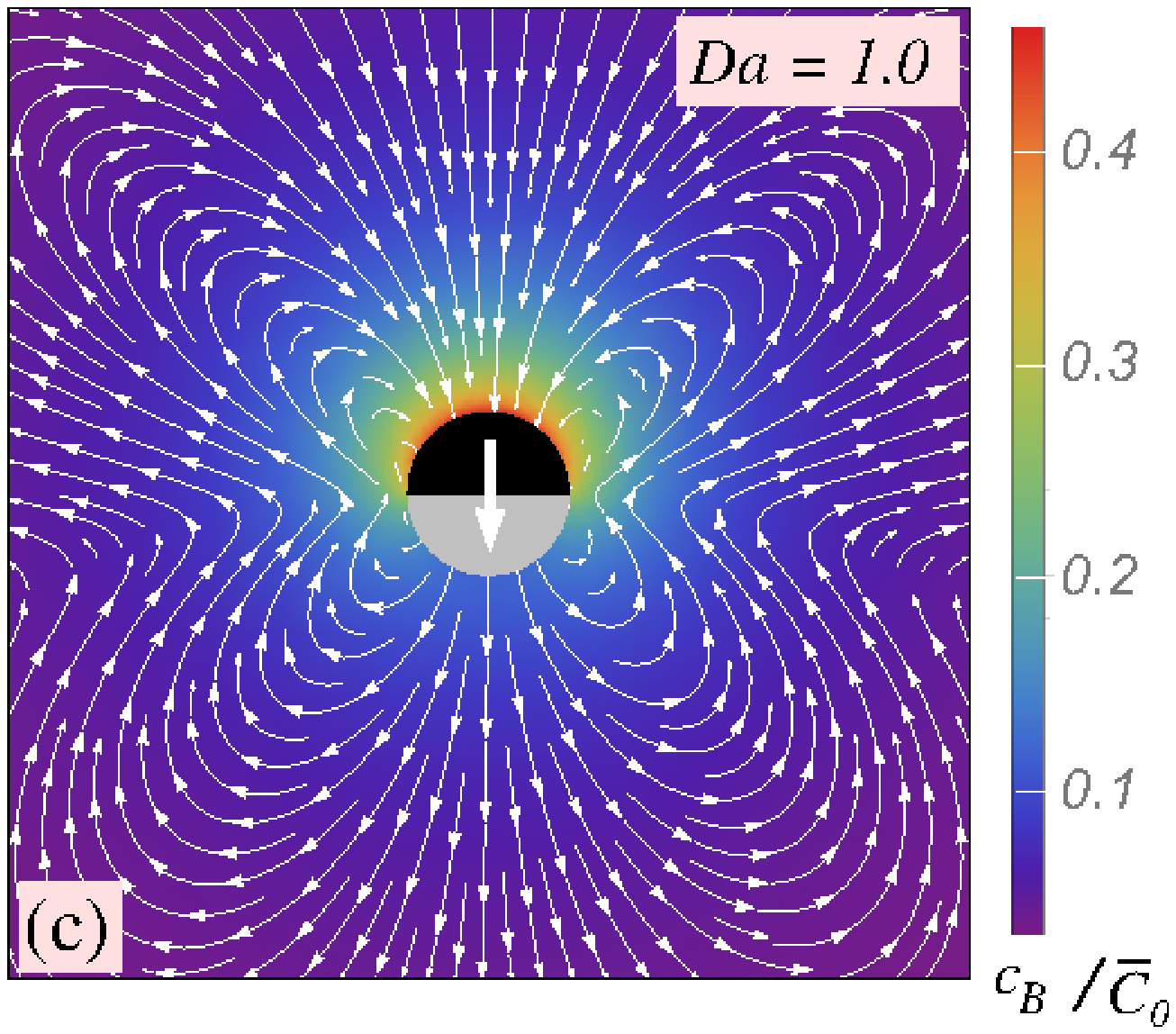}
\vspace*{0.2cm}\hfill\\
\includegraphics[width=.61\columnwidth]{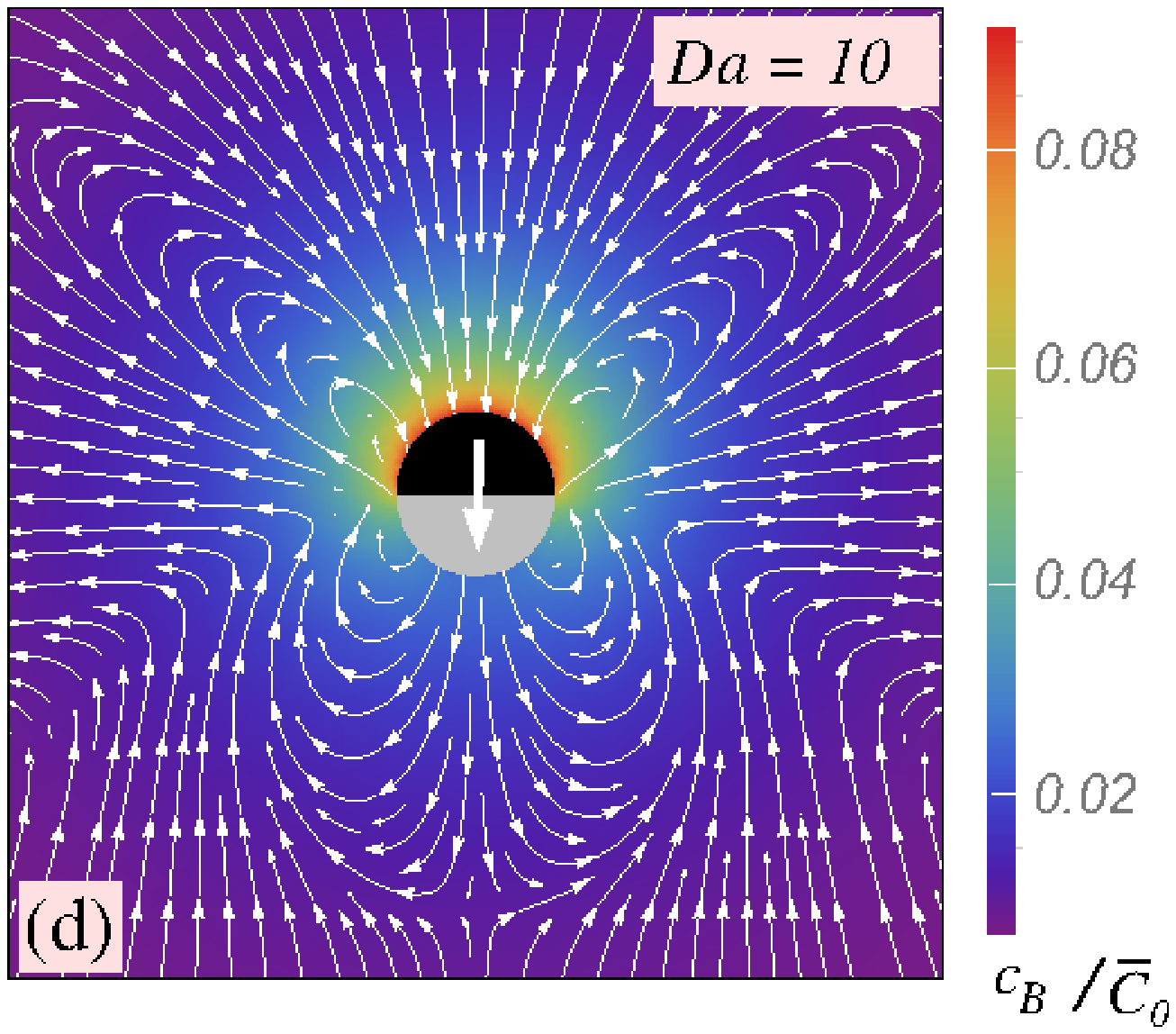}
\caption{
\label{fig9}
(a) The phoretic slip as a function of angular position $\theta$ 
[Eq. (\ref{eq:phor_slip_squirm}), truncated at the first 300 terms] 
for an active particle with chemical reaction of first order kinetics and 
$Da = 0.1,~1,~\mathrm{and}~10$, respectively. 
(b)-(d) The force-free flow in the laboratory system (lines, Eqs. 
(\ref{eq:flow_lab_syst})) and the distribution of solute (color coded, 
Eq. (\ref{eq:nB_series})), for an active particle as in (a). For both the 
density and the flow the series are truncated after the first 50 terms. The 
thick white arrows at the particle show its direction of 
motion. In (a)-(d) the phoretic mobility is position independent and negative 
($g(\theta) = -1$). 
}
\end{figure}

The corresponding squirmer parameters $S$ and $S'$ are shown in Fig. \ref{fig8}(b). 
While the parameter $S'(Da) < 0$ does not exhibit any noticeable dependence on the 
Damk{\"o}hler number $Da$, this is different for $S(Da)$, which 
exhibits an interesting transition as a function of $Da$. As expected, if 
$Da \ll 1$, i.e., the limit in which -- as noted above -- the system maps back 
onto the model $(pi)$ analyzed in Sec. \ref{C_and_flow}, one has 
$S \simeq 0$ so that the half-covered Janus particle behaves hydrodynamically 
like a neutral squirmer. However, upon increasing the Damk{\"o}hler number to 
$Da \gtrsim {\cal O}(1)$, the parameter $S$ increases and 
becomes significantly positive. Therefore, the far-field flow of this model particle 
with $Da \geq 1$ exhibits a ``puller'' character. To conclude, the far-field 
hydrodynamics exhibited by this type of model particles depends significantly on 
the details of the first-order reaction via the magnitude of the 
Damk{\"o}hler number. These aspects are illustrated in Fig. \ref{fig9}, where we show 
how around the Janus particle both the distribution of the phoretic slip 
along the surface of the particle (Fig. \ref{fig9}(a)) and the hydrodynamic 
flow, as well as the number density of species $B$ change upon increasing $Da$ 
from $10^{-1}$ (for which $S \simeq 0.03$) to $10$ (for which 
$S \simeq 0.96$). 

Figure \ref{fig9}(a) shows that the increase in the Damk{\"o}hler number is 
correlated with an increasing asymmetry of the distribution of the phoretic slip 
along the surface: the larger the value of $Da$, the more sharply the phoretic 
slip is concentrated on the inert part. There are also visible changes in 
the magnitude of the phoretic slip. However, as emphasized above, the 
velocity scale $U_0$ itself depends on $Da$. The number density of product 
molecules (color coded background in Fig. \ref{fig9}(b)-(d)) mainly exhibits 
an overall change in magnitude upon increasing $Da$. (Note that also the 
density scale ${\bar C}_0$ itself depends on $Da$.) This translates into a 
reduced range of variations in $c$, consistent with the features in 
Fig. \ref{fig9}(a). On the other hand, the flow changes qualitatively. As 
expected, at $Da = 10^{-1}$ the flow field is basically the 
one of a neutral squirmer (see the above discussion, Fig. \ref{fig8}(b), 
and Sec. \ref{C_and_flow}.1). Upon increasing the Damk{\"o}hler number to 
$Da = 1$ a front-back asymmetry of the flow pattern emerges, while further 
increasing the Damk{\"o}hler number to $Da = 10$ produces a clear puller-type 
flow, with a well defined stagnation point in front of the particle.

We conclude this section by remarking that the case of spatially varying phoretic 
mobilities can be straightforwardly addressed along the lines used for 
the models discussed in Sec. \ref{C_and_flow}.2. All necessary relations 
are provided in Sec. \ref{squirm}. As in the other cases discussed in 
Sec. \ref{squirm}, upon varying the sign and the magnitude of the mobilities 
on the active and inert parts, respectively, far-field hydrodynamic behaviors 
corresponding to a ``pusher'', ``puller'', ``neutral'', or ``shaker'' particle 
can be observed.

\section{\label{spatial_conf} Effective interactions between active 
particles and between corresponding effective squirmers}

The analyses in the previous sections illustrate the exact mapping of the 
hydrodynamic flow exhibited by various active colloids -- suspended in 
an unbounded solution -- onto effective hydrodynamic squirmers. Inter alia, for 
classical hydrodynamic \textit{squirmers} an attractive or repulsive interaction 
between a squirmer and a wall can be used to infer a pusher, puller, 
or neutral character of the squirmer, respectively \cite{Spagnolie2012,Ignacio2010}. 
This naturally raises the question of whether the interaction of 
\textit{chemically active} particles with confining  boundaries (such as solid 
walls, fluid interfaces, inert particles, or other active colloids) can be 
captured -- at least qualitatively -- from the effective interaction with 
the same boundary of the corresponding effective squirmer (i.e., the one obtained 
via the mapping, as described in the previous sections, of the active particle 
in unbounded fluid). Moreover, one can further specify the above question as 
follows. Suppose that, e.g., an active particle exhibits, in the far field, an 
effective attractive or repulsive interaction  with a wall. Can one then robustly 
infer from this observation that the effective squirmer corresponding to that 
particle (see above) has a pusher, puller, or neutral 
character \cite{Brown2016b,Howse2015}?

The complex behavior of the flow and of the particle motion depends sensitively 
on the details of both the chemical activity and the phoretic mobility. The analyses 
of this behavior in the previous sections point towards a negative answer to the 
above questions. The results in Sec. \ref{const_b} show that -- up to an overall 
constant scale factor -- identical flow patterns may emerge from seemingly 
different physical mechanisms of the chemical activity (e.g., the distinct models 
$(pi)$ and $(pa)$ are characterized by the same set of squirmer modes $B_n/B_1$). On 
the other hand, one can anticipate that the distinct mechanisms of activity will 
introduce distinct interactions with the boundary, obscuring the ``effective 
squirmer'' character of the mechanism of activity. In Sec. \ref{first_order_kin} we 
have seen that changes in the importance of 
transport by diffusion relative to that of the reaction rate, i.e., the variation 
of the Damk{\"o}hler number $Da$, may change the associated hydrodynamic flow from 
exhibiting a neutral squirmer character to a puller-like one. Finally, 
in Sec. \ref{pos_dep_mob} we have seen that for a given chemical activity the 
character of the hydrodynamic flow induced by the active particle may vary 
across the whole spectrum of ``pusher'', ``puller'', and ``neutral'' behaviors 
via varying the pattern of phoretic mobility at the surface of the 
particle.

\begin{figure}[!htb]
    \centering
   \includegraphics[width=.95\columnwidth]{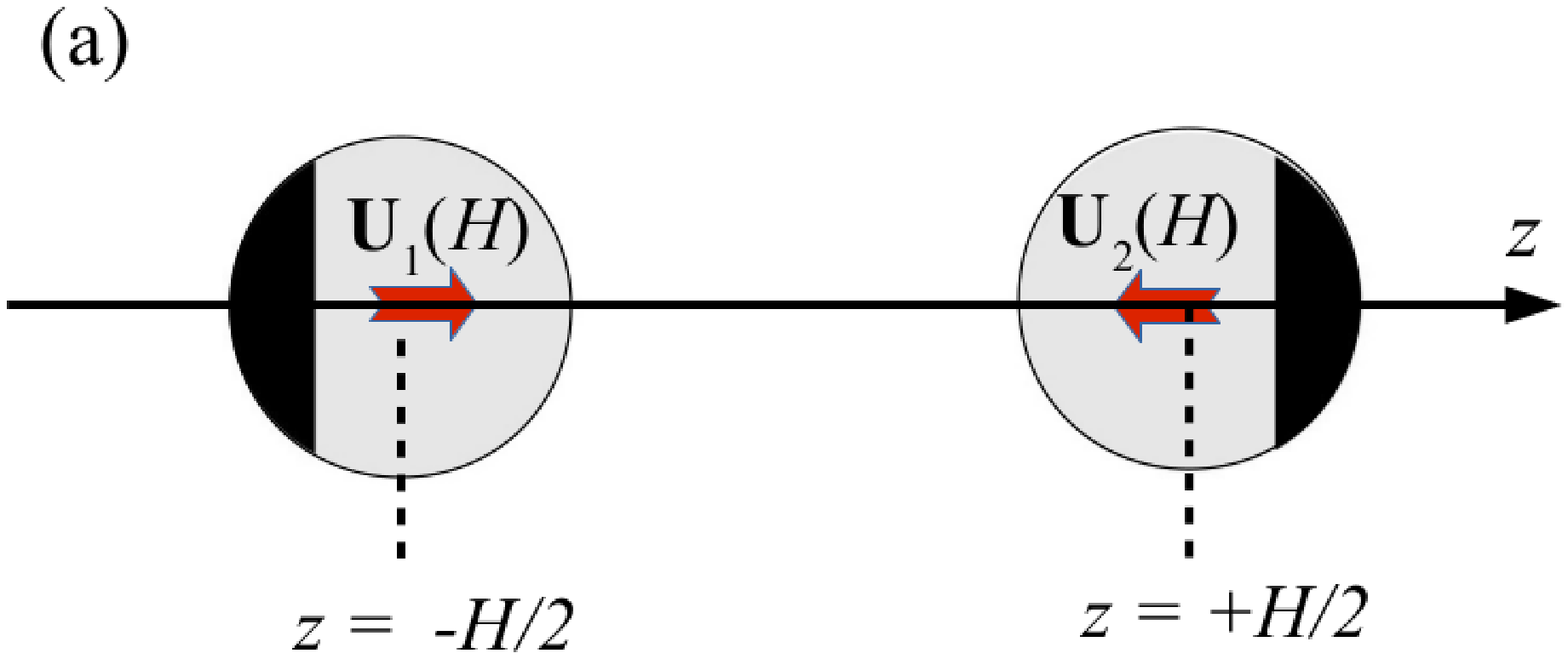}
   \vspace*{1.cm}\hfill\\
    \includegraphics[width=.95\columnwidth]{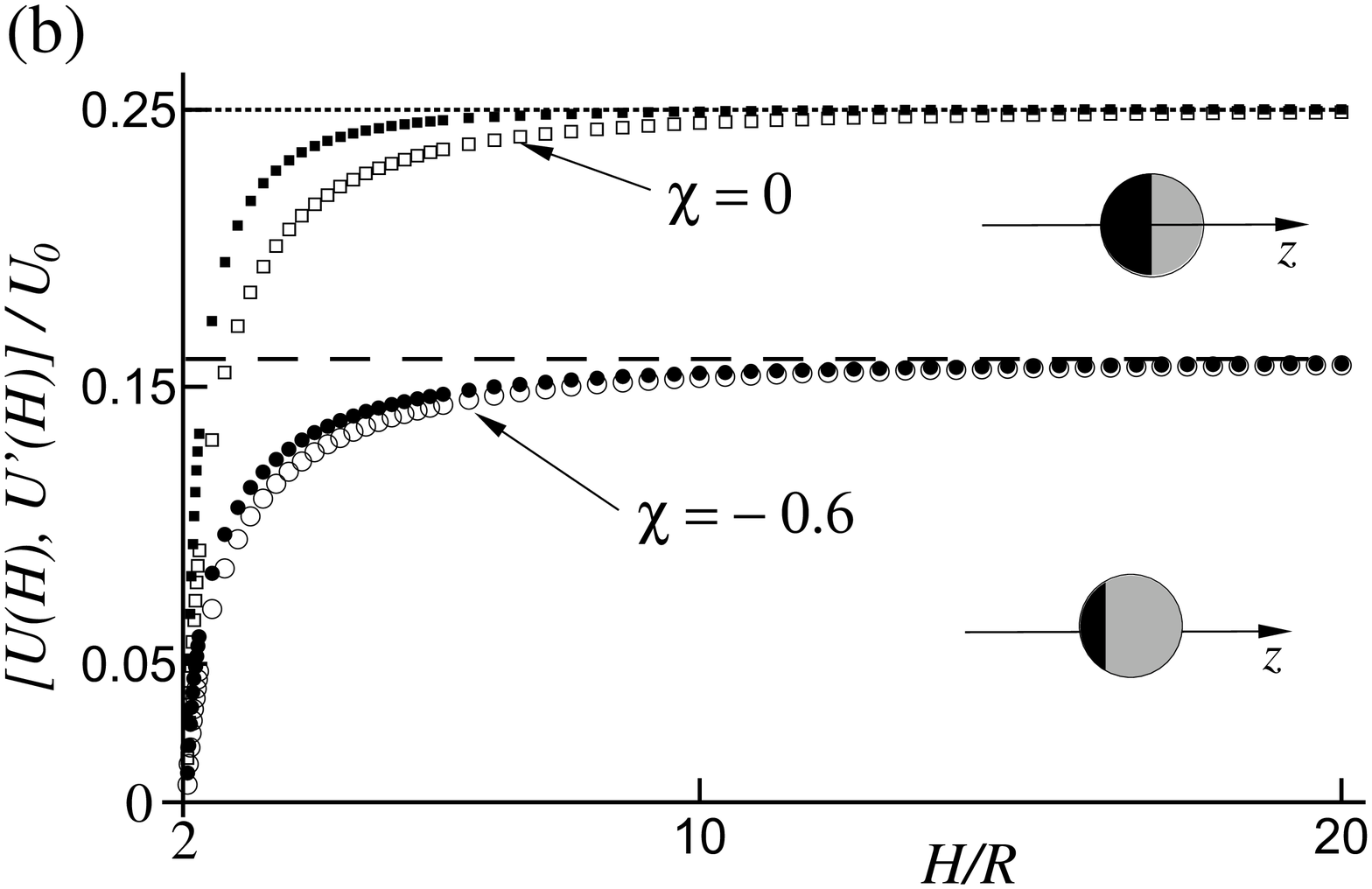}   
    \vspace*{1.cm}\hfill\\
    \includegraphics[width=.95\columnwidth]{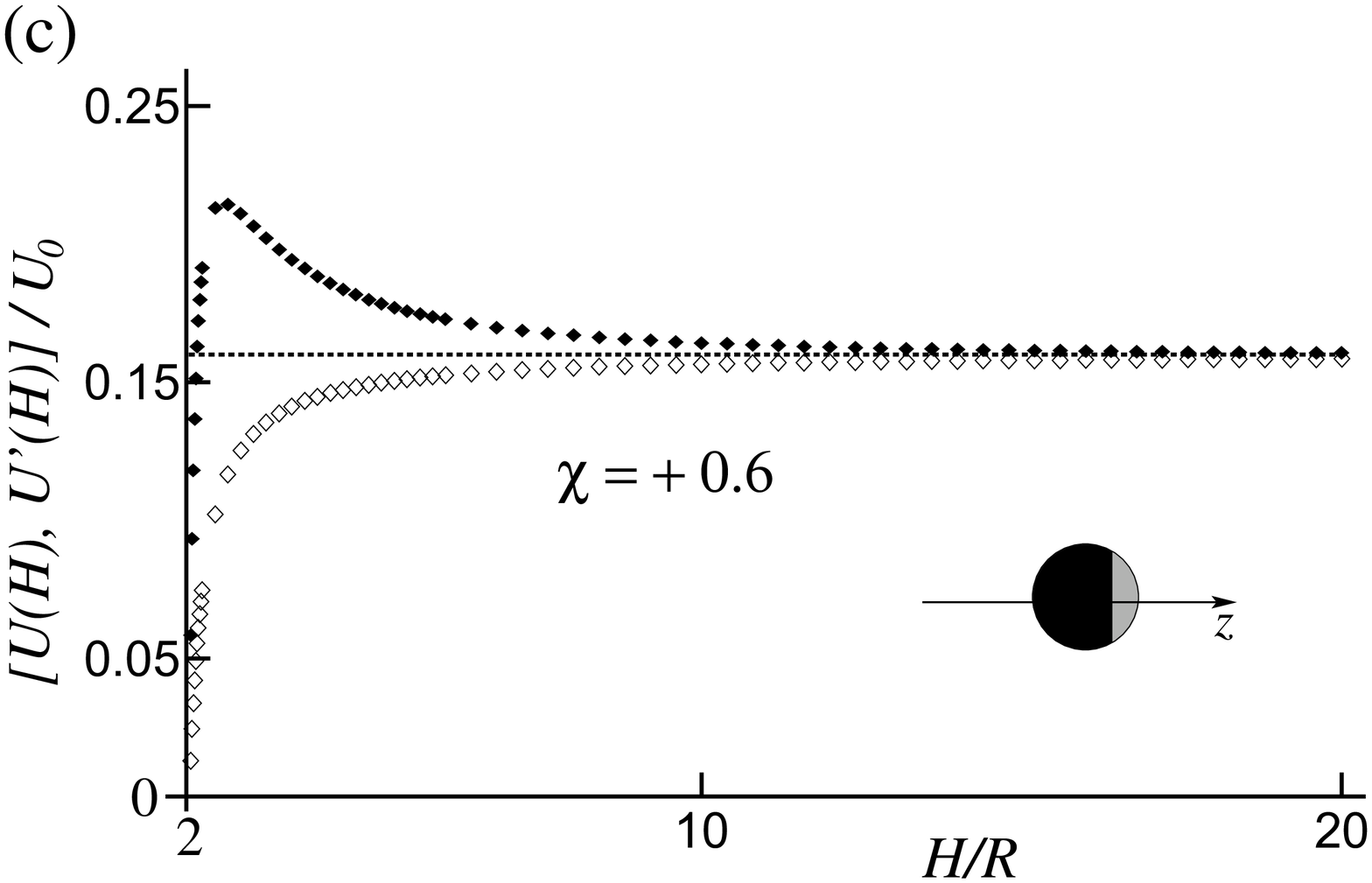}   
     \caption{
\label{fig10}
(a) Schematic diagram illustrating a central collision between two active 
particles of model $(pi)$ which are force and torque free. (b), (c) 
Velocity $U(H)$ 
(open symbols) of the active particle 1 and $U'(H)$ (filled symbols) of the 
effective squirmer 1 (see the main text), respectively, in units of 
$U_0$. 
The results 
correspond to a position independent, negative phoretic mobility and to 
coverages (Eq. (\ref{eq:chi_cov_def})) (b): $\chi = -0.6$ (effective ``pusher'') 
and $0$ (effective ``neutral''), and (c): $\chi = 0.6$ (effective ``puller''). 
The dashed line in (b) and the dotted lines in (b) and (c) indicate the value 
of the corresponding velocity $U^{(f)}(\chi)$ (Eq. (\ref{Uf_pi})) of the particle 
in free space.
}
\end{figure}
As a simple but insightful example, in this section we consider the set-up of the 
central collision between two identical active particles (i) belonging to 
model $(pi)$ with a position-independent phoretic mobility (see Fig. \ref{fig10}(a)), 
and their corresponding effective squirmer \textit{ersatz} particles (ii). The 
results discussed below illustrate that there can be significant qualitative 
differences between the behaviors exhibited in the above two cases (i) and (ii). 
This implies that in the presence of geometrical confinements significant 
qualitative differences can exist between the behaviors exhibited by an active 
particle and by its corresponding ``effective squirmer'', respectively. 
Therefore, the ``inverse problem'', i.e., inferring a pusher, puller, or neutral 
character of the \textit{effective squirmer} from the observation of a far-field 
attraction or repulsion of the \textit{active particle} with a boundary, is in 
general ill-posed.

Turning to the example depicted schematically in Fig. \ref{fig10}(a), the 
two spherical, active particles are described by model $(pi)$ with a 
position-independent surface mobility (see the analysis in Sec. \ref{const_b}); 
furthermore, we focus on the case that the particles are 
force and torque free. The number density $c(\mathbf{r})$ of solute molecules 
obeys the Laplace equation subject to the BC of solute-production at the catalytic 
caps (black areas in Fig. \ref{fig10}) on each of the two particles; 
it vanishes far from the particles. The hydrodynamic flow $\mathbf{u}(\mathbf{r})$ 
is the solution of the Stokes equations with the phoretic slip BC 
(Eq. (\ref{eq:phor_slip_connect})) which holds on each of the two particles 
and is determined by $c(\mathbf{r})$. The BC of a quiescent 
fluid applies far from the particles. Both these fields depend parametrically on 
$H = h/R$, where $h \geq 2 R$ denotes the distance between the centers of the two 
particles of radius $R$. Due to symmetry, the two particles move along 
the $z$-axis, which connects their centers, with velocities $\mathbf{U}_{1,2}$ 
which are equal in magnitude but opposite in sign. $\mathbf{U}_{1,2}$ depend 
on $H$; for $H \to \infty$, the velocities take their corresponding values 
$\mathbf{U}^{(f)}$ for a single particle in an unbounded fluid 
(see Sec. \ref{const_b}). The difference with the reference system of 
colliding ``effective squirmers'' is that for the latter the hydrodynamic slip 
at the surface is prescribed to be independent of $H$. It is set to 
that distribution which corresponds to the active particle as if it 
would be immersed in an unbounded fluid (see 
Eqs. (\ref{eq:phor_slip_squirm}), (\ref{Bn_unif_b}), and (\ref{fn_pi})). 
In order to discriminate between the active particles and the effective squirmer 
cases, we shall denote the corresponding velocities of the squirmers, which also 
depend on $H$, by a prime, i.e., $\mathbf{U}'_{1,2}$. 

Both boundary value problems (i.e., for the number density of the solute 
and for the hydrodynamic flow) can be solved exactly in terms of series 
representations in bi-polar coordinates \cite{Popescu2011EPL,Alexander2017}; here 
we simply adapt and employ the solutions available in 
Refs. \cite{Alvaro2016,Paolo2018}.\footnote{The symmetry of the diffusion problem 
shows that the current of $c(\mathbf{r})$ through the midplane normal to the 
$z$-axis vanishes. Thus, the diffusion problem is equivalent to 
that in the set-up of a Janus particle facing a wall. Similarly, for 
the hydrodynamics the same plane is a surface of zero normal flow and zero 
tangential stress (see also Ref. \cite{HaBr73}); therefore the Stokes flow is the 
same as that for the set-up of a particle moving towards a free (liquid-vapor) 
interface. Thus both situations of interest follow from the general 
solutions given in Refs. \cite{Alvaro2016,Paolo2018} for the problem of an 
active particle facing a liquid 1 - liquid 2 interface as limiting 
cases (i.e., by taking the appropriate limits for the parameters of liquid 2.} 
We shall use 
the velocities of particle 1, $U_1(H):= U(H) > 0$ and $U'_1(H):=U'(H)$, in order 
to characterize the effective interactions: an increase (decrease) relative to 
the free-space value $U^{(f)} = U(H \to \infty) \equiv U'(H \to \infty)$ 
defines an effective attraction (repulsion) induced by the presence of particle 
$2$.

As discussed in Sec. \ref{const_b}, the size of the catalytic cap is 
determined by the parameter
\begin{equation}
\label{eq:chi_cov_def}
 \chi = -\cos\theta_0\,,
\end{equation}
which in the followings will be referred to as the coverage. For 
model $(pi)$ with a position independent phoretic mobility, this 
parameter allows one to tune the hydrodynamic flow of the active particle. 
For negative phoretic mobility the corresponding effective squirmer shows 
the far-field characteristics of a pusher for $\chi < 0$, of a neutral squirmer 
for $\chi = 0$, and of a puller for $\chi > 0$ (see also Fig. \ref{fig2}(b)). 
We therefore choose suitable values for $\chi$ such that the model system 
under consideration exhibits each one of these desired behaviors. 

The results $U(H)$ (open symbols) and $U'(H)$ (filled symbols), in units of 
$U_0$ and corresponding to $\chi = -0.6$ (``pusher'')  and $0$ (``neutral'') are 
shown in Fig. \ref{fig10}(b), while those corresponding to $\chi = 0.6$ (``puller'') 
are shown in Fig. \ref{fig10}(c). While in the first two cases the somewhat 
na\"ive replacement of the active colloid with an effective squirmer results in 
a rather accurate approximation, in the last case this procedure fails. For 
$\chi = 0.6$, the behavior is not only qualitatively different -- 
the effective squirmers exhibit an attractive interaction, while the active 
colloids exhibit repulsion -- but this qualitatively different behavior persists 
all the way to large separations $H$ (i.e., to the far field). However, 
this is 
not too surprising: it has been already pointed out in the literature (see, e.g., 
Refs. \cite{Golestanian2007,Uspal2015a,Simmchen2016}) that the distortions of 
the chemical field and those of the hydrodynamic flow induce, in general, changes in 
the velocity of the particle of the same order of decay as a function of the 
distance from the cause of distortions (boundary). As we show below, this is the 
case here, too. Accounting for both these distortions (rather than just for the 
hydrodynamic one, as it is the case for the effective squirmers) fully 
explains the effective repulsion exhibited in all cases by the active particle. 
Before we proceed, we re-emphasize that attributing the repulsion observed 
in Fig. \ref{fig10}(c) entirely to hydrodynamics and attempting to assign to 
this a type of squirmer leads to the erroneous inference of a ``pusher'' 
character. As a consequence, a follow-up analysis would attempt to find 
a surface activity such that it could be mapped onto an effective 
pusher, instead of searching for the correct ``puller''-like mechanism (which 
is known to be the case for the active particle considered in Fig. \ref{fig10}(c)).

Focusing now solely on the results for the active particles, we look at providing 
a simple far-field approximation which allows one to understand their 
interaction. With this aim, we resort to a point particle approximation according 
to which we simply replace the second particle by the chemical, 
$c^{(2)} (\mathbf{r})$, and hydrodynamic flow field, 
$\mathbf{u}^{(2)} (\mathbf{r})$, which it produces as if it was single, 
and calculate the effects, in leading order, of these fields on particle 1. 
We start with the effects of the chemical field. 

Colloid 1, located at 
$\mathbf{R}_p = (0,0, z = -H/2)$ in the external field $c^{(2)} (\mathbf{r})$ 
(see Eqs. (\ref{eq:c_series}) and (\ref{fn_pi}) with the origin shifted to the 
point $(0,0,z = +H/2)$) acquires a phoretic velocity \cite{Anderson1989}
\begin{eqnarray}
 \label{eq:cor_C}
 && \delta \mathbf{U}_c = (- b_0) (\nabla c^{(2)})|_{\mathbf{r} = \mathbf{R}_p} 
 \Rightarrow \\
 && \frac{\delta \mathbf{U}_c}{U_0} \simeq -\left\lbrace 
 f_0 \left(\frac{R}{H}\right)^2 - 
 f_1 \left(\frac{R}{H}\right)^3 + {\cal O} \left[\left(\frac{R}{H}\right)^4 \right] 
 \right \rbrace \mathbf{e}_z \,.\nonumber
\end{eqnarray}
For the case of interest here, the particle is a net source of solute ($f_0 \neq 0$) and 
the first term provides the leading order. However, there can be cases, such as the model 
activity $(pa)$ discussed in Sec. \ref{C_and_flow}, for which there is no net production 
(or annihilation), i.e., $f_0 = 0$. In such cases, the leading order term is the second 
one in the brackets. This latter term is proportional to $f_1$, which in turn determines 
the motility of the particle (see Eqs. (\ref{U_free}) and (\ref{Bn_unif_b})) and is nonzero 
for an active particle exhibiting motility in an unbounded solution.

We now turn to the effects of the flow $\mathbf{u}^{(2)} (\mathbf{r})$ on the motion 
of the particle 1. Fax{\'e}n's laws \cite{HaBr73} imply that, in the geometry shown 
in Fig. \ref{fig10}(a), particle 1 immersed in the flow 
$\mathbf{u}^{(2)}(\mathbf{r})$ (see Eq. (\ref{eq:flow_lab_syst}) with $F = 0$ 
and the origin shifted to the point $(0,0,z = +H/2)$) acquires the 
velocity \cite{Ignacio2010,Ignacio2010b,Pedley2006}
\begin{eqnarray}
 \label{eq:cor_U}
 && \delta \mathbf{U}_h =  \mathbf{u}^{(2)}|_{\mathbf{r} = \mathbf{R}_p} 
 \Rightarrow \\
 && \frac{\delta \mathbf{U}_h}{U_0} 
 \simeq -\left[ 
 f_2 \left(\frac{R}{H}\right)^2 + \left(\frac{1}{3} f_1 - \frac{3}{2} f_3 \right)  
 \left(\frac{R}{H}\right)^3 \right] \mathbf{e}_z\nonumber\\
 && \hspace*{1.2 cm} + \,{\cal O} \left[\left(\frac{R}{H}\right)^4 \right] 
 \mathbf{e}_z \,.
\end{eqnarray}
If the effective squirmer model, corresponding to the particle with 
uniform mobility, is either a pusher or a puller, one has $f_2 \neq 0$ and the 
first term provides the leading order behavior (note that it has opposite signs for 
pusher and puller, respectively). On the other hand, if the effective squirmer 
mapping leads to a neutral squirmer, i.e., $f_2 = 0$, the first term vanishes 
and the leading order behavior is given by the second term, the amplitude 
of which involves both the modes $f_1$ and $f_3$ of the activity function.

Combining the two results above, and noting that $f_0 > 0$ for model $(pi)$ of 
interest here, one arrives at the following expression for the 
deviation $\delta \mathbf{U}$, at leading order,  of the velocity of particle 1 
from the ``single-particle'' value $U^{(f)} \mathbf{e}_z$:
\begin{eqnarray}
 \label{eq:tot_corr}
 \frac{\delta \mathbf{U} \cdot \mathbf{e}_z}{U_0} &\simeq& - 
 \left( f_0 + f_2 \right) 
 \left(\frac{R}{H}\right)^2 \\
 &\overset{(pi)}{=}& -\left( \frac{1+\chi}{2} - 
 \frac{5}{4} \chi \sqrt{1-\chi^2}\right) 
 \left(\frac{R}{H}\right)^2 \nonumber\,,
\end{eqnarray}
where in the last line we have specialized the more general result to the 
specific choice of the coefficients $f_n$ of model $(pi)$ (Eq. (\ref{fn_pi})).
\begin{figure}[!htb]
    \centering
    \includegraphics[width=.95\columnwidth]{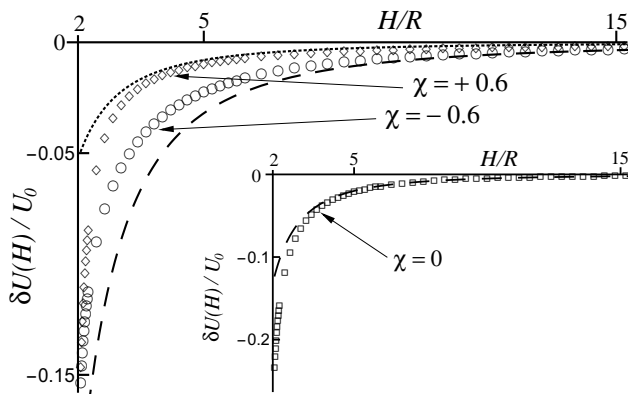}   
     \caption{
\label{fig11}
The deviation of the velocity of particle 1 with radius $R$ from the 
``single particle'' value  $U^{(f)}$, in  a central collision as depicted in 
Fig. \ref{fig10}(a), as function of the center-to-center separation 
$H/R$. The data correspond to a position independent, 
negative phoretic mobility and to coverages $\chi = \mp \,0.6$ (main panel), 
corresponding to an effective ``pusher'' or ``puller'', respectively, and 
$\chi = 0$ (inset), i.e., effectively ``neutral''. The symbols show 
exact results obtained from series in terms of bi-polar coordinates; the 
data are taken from Figs. \ref{fig10}(b) and (c). The dashed and dotted 
lines show the corresponding predictions of the far-field approximation 
(Eq. (\ref{eq:tot_corr})).
}
\end{figure}

The first term in parentheses in the second line of Eq. (\ref{eq:tot_corr}), 
which accounts for the effective interaction induced by the chemical distortions, 
is positive and dominates the second one (which emerges from hydrodynamic 
interactions and changes sign at $\chi = 0$) for all values of $-1 < \chi < 1$. 
Thus, the leading order far-field analysis above indeed correctly captures the 
main qualitative feature of the collision between these model active particles, 
in that they exhibit an effective repulsion irrespective of the coverage (see 
Figs. \ref{fig10}(b), \ref{fig10}(c), and \ref{fig11}). Furthermore, as can be seen 
in Fig. \ref{fig11}, in all three cases the predicted deviations from the 
single-particle velocity (Eq. (\ref{eq:tot_corr})), i.e., the 
``effective interaction'' between the two active particles,  quantitatively capture 
the exact results at large values $H/R$, and, for $\chi \geq 0$, 
quasi-quantitatively down to separations as small as $H/R \gtrsim 4$ (see, also, 
similar reports on the accuracy of far-field approximations by, e.g., 
Ref. \cite{Spagnolie2012,Uspal2016,Uspal2018a}).

\section{\label{sum} Summary and conclusions}

We have analyzed in detail the connection between simple models of chemically 
active, spherical, axisymmetric particles moving by self-diffusiophoresis, as 
defined in Sects. \ref{model_def} and \ref{math_form}, and the classic hydrodynamic 
squirmer model \cite{Lighthill1952,Blake1971}. 

For the motion in an unbounded fluid, this connection takes the form of an 
exact mapping from the model chemical activity and phoretic mobility of the particle 
to a set of squirming modes $B_n$, as reported previously 
\cite{Majee2013,Michelin2014,Liverpool2016}. For completeness, and because 
these results are somewhat isolated in the literature, the steps involved in 
constructing this mapping have been succinctly summarized in Sect. \ref{squirm}. 
Such mappings provide the hydrodynamic flow around the active particle, i.e., 
the solution of the Stokes problem, in explicit form. Along the construction 
of this mapping, one finds as a by-product a simple derivation of 
the relation between the external force needed to immobilize the active particle 
(so-called ``stall'' force) and the velocity with which the particle would move in 
the absence of external forces (Eq. (\ref{stallF_Uf})).

In Sect. \ref{C_and_flow} we have illustrated such mappings for 
various, commonly used models of chemical activity and for particles with 
either position-independent or position-dependent phoretic mobility. For each 
case considered, the mapping was derived explicitly in terms of the functions 
describing the chemical activity and the phoretic mobility. The results have been 
illustrated as functions of geometrical parameters (e.g., the coverage) of the 
system and for various ratios of phoretic mobilities over the two parts of the 
particle surface: plots of (i) the number density of the 
solute, together with the associated hydrodynamic flow field, (ii) the 
distribution of the phoretic slip around the particle, and (iii) the sign 
and magnitude of the first two squirmer modes, based on which the particles 
can be classified as ``pushers'', ``pullers'', ``neutral'', and, 
eventually, ``shakers''. Two important conclusions have emerged from the 
examples discussed in Sect. \ref{C_and_flow}. The first is that the models of 
chemical activity involving arguably very distinct physical mechanisms, may 
map onto the same (up to a change of the velocity scale) effective 
squirmer (Sects. \ref{const_b}.1. and \ref{const_b}.2.). The second is that, 
for a given model of chemical activity, solely by varying the value of the ratio 
of phoretic mobilities over the two parts of the particle surface one arrives at 
effective squirmers which can exhibit any of the ``pusher'', ``puller'', 
or ``neutral'' characters (Sect. \ref{pos_dep_mob}). This highlights that even if 
the hydrodynamic flow around a squirmer is perfectly known, the pattern of 
chemical activity cannot be reliably and robustly inferred from it without knowing 
the phoretic mobility function and de-convoluting its contributions.

In Sect. \ref{first_order_kin} we have shown how the mapping procedure can be 
straightforwardly extended to include the more complex case of a chemical reaction 
with a first order chemical kinetics. For the particular case of a Janus 
particle, this has been illustrated, in terms of number densities of the 
chemical species, the hydrodynamic flow field, and the slip 
distribution around the particle. By studying these at various values of the 
Damk\"ohler number $Da$, which characterizes the relative importance of the 
transport by diffusion with respect to the reaction rate, we found changes in 
the resulting ``effective'' squirmer type from ``neutral'' at $Da \ll 1$ 
(reaction-limited kinetics) to ``puller'' at $Da \gg 1$ 
(diffusion-limited kinetics).

Finally, in order to build counter-examples, in Sect. \ref{spatial_conf} we 
have illustrated, by using the simple case of a central collision between two  
identical model active particles, that in the presence of geometric confinement 
it is not justified to replace the active particle by its effective squirmer, 
even if the interest is solely in the far-field behavior. Moreover, the 
examples shown in Figs. \ref{fig10}(b) and (c) highlight that attributing an 
observed effective interaction exhibited by the active particles 
entirely to hydrodynamics, and attempting to associate a type of squirmer based on 
such a hypothesis, may easily lead to flawed conclusions about the 
characteristics of a surface activity 
function compatible with that type of squirmer. While the direct use of the 
corresponding effective squirmer -- if the particle operates under 
geometrical  confinements -- is thus unwise, the mapping onto an 
effective squirmer remains a very useful tool. By resorting to it, we have 
derived a far-field approximation (Eq. (\ref{eq:tot_corr})) which 
rationalizes and captures
(quasi)-quantitatively the effective interaction between the two active particles 
in the particular configuration shown in Fig. \ref{fig11}.

To conclude, we have highlighted both the advantages of the mapping of model 
chemically active colloids to effective hydrodynamic squirmers, as well as 
the drawbacks of unwarranted use of such approaches. As we have shown in 
Sect. \ref{spatial_conf}, carrying over the descriptions in terms of 
squirmers to studies of, e.g., the collective behavior in suspensions of 
active particles, is not well grounded even as an approximation of a 
``dilute limit'' focused on far-field interactions. However, the success of the 
procedure in terms of effective interactions, which leads to 
Eq. (\ref{eq:tot_corr}), raises the intriguing question of 
whether similar results may hold in more complex, general situations. Such results 
would pave the way for finding simplified, yet robust and reliable models, to be 
employed, e.g., in large scale computer simulations of suspensions of active 
particles \cite{Ripoll2017,Sano2016}. We thus consider it as rewarding to study 
this issue in the context of more complex geometries, such as the motion of an 
active 
particle parallel or normal to an interface, or non-central binary collisions. 
For hydrodynamic squirmers in such geometries a wealth of results is 
available (see, e.g., 
Refs. \cite{Spagnolie2012,Ignacio2010,Ignacio2010b,Pedley2006,Ishimoto2013}), 
which 
can facilitate a study as suggested above. Finally, we note that squirmer models 
are available for other axisymmetric shapes, such as prolate and oblate spheroids 
(see, e.g., Ref. \cite{Leshansky2007,Winkler2016}), and thus similar mappings 
from chemically active particles to squirmers should also be possible for particles 
with elongated shapes (e.g., rods) as encountered in actual experimental studies.

\acknowledgments 

This research has benefited from the scientific interactions facilitated by the COST 
Action MP1305 ``Flowing Matter'', supported by COST (European Cooperation in Science 
and Technology). M.T. acknowledges financial support from the Portuguese
 Foundation for Science and Technology (FCT) under Contract no. IF/00322/2015.
 
\appendix

\section{\label{formulas} Useful relations}

At various steps, the following relations \cite{abram} involving the Legendre 
polynomials $P_n$ and the associated Legendre functions $P_n^m$ of degree $n$ and order 
$m = 1$  have been used:
\begin{equation}
 \label{Pn1_vs_Pn}
 P_n^1(\cos\theta):= - \sin\theta \,\frac{d P_n(\cos\theta)}{d(\cos\theta)} = 
\frac{d P_n(\cos\theta)}{d \theta}\,,
\end{equation}
\begin{equation}
 \label{Pn_norm}
 \int\limits_0^\pi d\theta \sin \theta P_n(\cos\theta) P_k(\cos\theta) = 
 \frac{2}{(2 n + 1)}\, \delta_{n,k}\,,
\end{equation}
\begin{equation}
 \label{Pn1_norm}
 \int\limits_0^\pi d\theta \sin \theta P_n^1(\cos\theta) P_k^1(\cos\theta) = 
 \frac{2 n (n+1)}{(2 n + 1)}\, \delta_{n,k}\,,
\end{equation}
and
\begin{eqnarray}
 \label{subint_integr_Pn}
&&{\cal H}_n (\omega_0:= \cos\theta_0):= \int\limits_{\omega_0}^1 du P_n(u) \\
&& \hspace*{1.cm}= 
 \begin{cases}
 1- \omega_0\,,  &n = 0\,,\\
 ~~ \\
 - \frac{(1-\omega_0^2)^{1/2}}{n (n + 1)} P_n^1(\omega_0)\,, &n \geq 1\,.
 \nonumber
 \end{cases}
\end{eqnarray}
\newline\newline
\noindent For $m \neq n$, one has
\begin{eqnarray}
 \label{subint_integr_Pn_Pm}
 && H_{n,m}(\omega_0:= \cos\theta_0) := \int\limits_{\omega_0}^1 du P_m(u) P_n(u) \\
 && \hspace*{1.cm} =-(1-\omega_0^2)^{1/2} \,\,
 \frac{P_n(\omega_0) P_m^1(\omega_0) - P_m(\omega_0) P_n^1(\omega_0)}{m(m+1) - 
 n(n + 1)}\,.
 \nonumber
\end{eqnarray}
\newpage

\bibliography{active_Janus_particle_as_squirmer}

\begin{thebibliography}{118}%
\makeatletter
\providecommand \@ifxundefined [1]{%
 \@ifx{#1\undefined}
}%
\providecommand \@ifnum [1]{%
 \ifnum #1\expandafter \@firstoftwo
 \else \expandafter \@secondoftwo
 \fi
}%
\providecommand \@ifx [1]{%
 \ifx #1\expandafter \@firstoftwo
 \else \expandafter \@secondoftwo
 \fi
}%
\providecommand \natexlab [1]{#1}%
\providecommand \enquote  [1]{``#1''}%
\providecommand \bibnamefont  [1]{#1}%
\providecommand \bibfnamefont [1]{#1}%
\providecommand \citenamefont [1]{#1}%
\providecommand \href@noop [0]{\@secondoftwo}%
\providecommand \href [0]{\begingroup \@sanitize@url \@href}%
\providecommand \@href[1]{\@@startlink{#1}\@@href}%
\providecommand \@@href[1]{\endgroup#1\@@endlink}%
\providecommand \@sanitize@url [0]{\catcode `\\12\catcode `\$12\catcode
  `\&12\catcode `\#12\catcode `\^12\catcode `\_12\catcode `\%12\relax}%
\providecommand \@@startlink[1]{}%
\providecommand \@@endlink[0]{}%
\providecommand \url  [0]{\begingroup\@sanitize@url \@url }%
\providecommand \@url [1]{\endgroup\@href {#1}{\urlprefix }}%
\providecommand \urlprefix  [0]{URL }%
\providecommand \Eprint [0]{\href }%
\providecommand \doibase [0]{http://dx.doi.org/}%
\providecommand \selectlanguage [0]{\@gobble}%
\providecommand \bibinfo  [0]{\@secondoftwo}%
\providecommand \bibfield  [0]{\@secondoftwo}%
\providecommand \translation [1]{[#1]}%
\providecommand \BibitemOpen [0]{}%
\providecommand \bibitemStop [0]{}%
\providecommand \bibitemNoStop [0]{.\EOS\space}%
\providecommand \EOS [0]{\spacefactor3000\relax}%
\providecommand \BibitemShut  [1]{\csname bibitem#1\endcsname}%
\let\auto@bib@innerbib\@empty
\bibitem [{\citenamefont {Ismagilov}\ \emph {et~al.}(2002)\citenamefont
  {Ismagilov}, \citenamefont {Schwartz}, \citenamefont {Bowden},\ and\
  \citenamefont {Whitesides}}]{Ismagilov2002}%
  \BibitemOpen
  \bibfield  {author} {\bibinfo {author} {\bibfnamefont {R.F.}\ \bibnamefont
  {Ismagilov}}, \bibinfo {author} {\bibfnamefont {A.}~\bibnamefont {Schwartz}},
  \bibinfo {author} {\bibfnamefont {N.}~\bibnamefont {Bowden}}, \ and\ \bibinfo
  {author} {\bibfnamefont {G.M.}\ \bibnamefont {Whitesides}},\ }\bibfield
  {title} {\enquote {\bibinfo {title} {Autonomous movement and
  self-assembly},}\ }\href@noop {} {\bibfield  {journal} {\bibinfo  {journal}
  {Angew. Chem. Int. Ed.}\ }\textbf {\bibinfo {volume} {41}},\ \bibinfo {pages}
  {652--654} (\bibinfo {year} {2002})}\BibitemShut {NoStop}%
\bibitem [{\citenamefont {Paxton}\ \emph {et~al.}(2004)\citenamefont {Paxton},
  \citenamefont {Kistler}, \citenamefont {Olmeda}, \citenamefont {Sen},
  \citenamefont {St.~Angelo}, \citenamefont {Cao}, \citenamefont {Mallouk},
  \citenamefont {Lammert},\ and\ \citenamefont {Crespi}}]{Paxton2004}%
  \BibitemOpen
  \bibfield  {author} {\bibinfo {author} {\bibfnamefont {W.F.}\ \bibnamefont
  {Paxton}}, \bibinfo {author} {\bibfnamefont {K.C.}\ \bibnamefont {Kistler}},
  \bibinfo {author} {\bibfnamefont {C.C.}\ \bibnamefont {Olmeda}}, \bibinfo
  {author} {\bibfnamefont {A.}~\bibnamefont {Sen}}, \bibinfo {author}
  {\bibfnamefont {S.K.}\ \bibnamefont {St.~Angelo}}, \bibinfo {author}
  {\bibfnamefont {Y.Y.}\ \bibnamefont {Cao}}, \bibinfo {author} {\bibfnamefont
  {T.E.}\ \bibnamefont {Mallouk}}, \bibinfo {author} {\bibfnamefont {P.E.}\
  \bibnamefont {Lammert}}, \ and\ \bibinfo {author} {\bibfnamefont {V.H.}\
  \bibnamefont {Crespi}},\ }\bibfield  {title} {\enquote {\bibinfo {title}
  {Catalytic nanomotors: Autonomous movement of striped nanorods},}\
  }\href@noop {} {\bibfield  {journal} {\bibinfo  {journal} {J. Am. Chem.
  Soc.}\ }\textbf {\bibinfo {volume} {126}},\ \bibinfo {pages} {13424--13431}
  (\bibinfo {year} {2004})}\BibitemShut {NoStop}%
\bibitem [{\citenamefont {Ozin}\ \emph {et~al.}(2005)\citenamefont {Ozin},
  \citenamefont {Manners}, \citenamefont {Fournier-Bidoz},\ and\ \citenamefont
  {Arsenault}}]{ozin2005}%
  \BibitemOpen
  \bibfield  {author} {\bibinfo {author} {\bibfnamefont {G.A.}\ \bibnamefont
  {Ozin}}, \bibinfo {author} {\bibfnamefont {I.}~\bibnamefont {Manners}},
  \bibinfo {author} {\bibfnamefont {S.}~\bibnamefont {Fournier-Bidoz}}, \ and\
  \bibinfo {author} {\bibfnamefont {A.}~\bibnamefont {Arsenault}},\ }\bibfield
  {title} {\enquote {\bibinfo {title} {Dream nanomachines},}\ }\href@noop {}
  {\bibfield  {journal} {\bibinfo  {journal} {Adv. Mater.}\ }\textbf {\bibinfo
  {volume} {17}},\ \bibinfo {pages} {3011--3018} (\bibinfo {year}
  {2005})}\BibitemShut {NoStop}%
\bibitem [{\citenamefont {Paxton}\ \emph
  {et~al.}(2006{\natexlab{a}})\citenamefont {Paxton}, \citenamefont
  {Sundararajan}, \citenamefont {Mallouk},\ and\ \citenamefont
  {Sen}}]{Paxton2006}%
  \BibitemOpen
  \bibfield  {author} {\bibinfo {author} {\bibfnamefont {W.F.}\ \bibnamefont
  {Paxton}}, \bibinfo {author} {\bibfnamefont {S.}~\bibnamefont
  {Sundararajan}}, \bibinfo {author} {\bibfnamefont {T.E.}\ \bibnamefont
  {Mallouk}}, \ and\ \bibinfo {author} {\bibfnamefont {A.}~\bibnamefont
  {Sen}},\ }\bibfield  {title} {\enquote {\bibinfo {title} {Chemical
  locomotion},}\ }\href@noop {} {\bibfield  {journal} {\bibinfo  {journal}
  {Angew. Chem. Int. Ed.}\ }\textbf {\bibinfo {volume} {45}},\ \bibinfo {pages}
  {5420--5429} (\bibinfo {year} {2006}{\natexlab{a}})}\BibitemShut {NoStop}%
\bibitem [{\citenamefont {Paxton}\ \emph
  {et~al.}(2006{\natexlab{b}})\citenamefont {Paxton}, \citenamefont {Baker},
  \citenamefont {Kline}, \citenamefont {Wang}, \citenamefont {Mallouk},\ and\
  \citenamefont {Sen}}]{Paxton2006a}%
  \BibitemOpen
  \bibfield  {author} {\bibinfo {author} {\bibfnamefont {W.F.}\ \bibnamefont
  {Paxton}}, \bibinfo {author} {\bibfnamefont {P.T.}\ \bibnamefont {Baker}},
  \bibinfo {author} {\bibfnamefont {T.R.}\ \bibnamefont {Kline}}, \bibinfo
  {author} {\bibfnamefont {Y.}~\bibnamefont {Wang}}, \bibinfo {author}
  {\bibfnamefont {T.E.}\ \bibnamefont {Mallouk}}, \ and\ \bibinfo {author}
  {\bibfnamefont {A.}~\bibnamefont {Sen}},\ }\bibfield  {title} {\enquote
  {\bibinfo {title} {Catalytically induced electrokinetics for motors and
  micropumps},}\ }\href@noop {} {\bibfield  {journal} {\bibinfo  {journal} {J.
  Am. Chem. Soc.}\ }\textbf {\bibinfo {volume} {128}},\ \bibinfo {pages}
  {14881--14888} (\bibinfo {year} {2006}{\natexlab{b}})}\BibitemShut {NoStop}%
\bibitem [{\citenamefont {Solovev}\ \emph {et~al.}(2009)\citenamefont
  {Solovev}, \citenamefont {Mei}, \citenamefont {Urena}, \citenamefont
  {Huang},\ and\ \citenamefont {Schmidt}}]{solovev2009}%
  \BibitemOpen
  \bibfield  {author} {\bibinfo {author} {\bibfnamefont {A.A.}\ \bibnamefont
  {Solovev}}, \bibinfo {author} {\bibfnamefont {Y.F.}\ \bibnamefont {Mei}},
  \bibinfo {author} {\bibfnamefont {E.B.}\ \bibnamefont {Urena}}, \bibinfo
  {author} {\bibfnamefont {G.S.}\ \bibnamefont {Huang}}, \ and\ \bibinfo
  {author} {\bibfnamefont {O.G.}\ \bibnamefont {Schmidt}},\ }\bibfield  {title}
  {\enquote {\bibinfo {title} {Catalytic microtubular jet engines
  self-propelled by accumulated gas bubbles},}\ }\href@noop {} {\bibfield
  {journal} {\bibinfo  {journal} {Small}\ }\textbf {\bibinfo {volume} {5}},\
  \bibinfo {pages} {1688--1692} (\bibinfo {year} {2009})}\BibitemShut {NoStop}%
\bibitem [{\citenamefont {Mirkovic}\ \emph {et~al.}(2010)\citenamefont
  {Mirkovic}, \citenamefont {Zacharia}, \citenamefont {Scholes},\ and\
  \citenamefont {Ozin}}]{mirkovic2010}%
  \BibitemOpen
  \bibfield  {author} {\bibinfo {author} {\bibfnamefont {T.}~\bibnamefont
  {Mirkovic}}, \bibinfo {author} {\bibfnamefont {N.S.}\ \bibnamefont
  {Zacharia}}, \bibinfo {author} {\bibfnamefont {G.D.}\ \bibnamefont
  {Scholes}}, \ and\ \bibinfo {author} {\bibfnamefont {G.A.}\ \bibnamefont
  {Ozin}},\ }\bibfield  {title} {\enquote {\bibinfo {title} {Nanolocomotion -
  catalytic nanomotors and nanorotors},}\ }\href@noop {} {\bibfield  {journal}
  {\bibinfo  {journal} {Small}\ }\textbf {\bibinfo {volume} {6}},\ \bibinfo
  {pages} {159--167} (\bibinfo {year} {2010})}\BibitemShut {NoStop}%
\bibitem [{\citenamefont {Fournier-Bidoz}\ \emph {et~al.}(2005)\citenamefont
  {Fournier-Bidoz}, \citenamefont {Arsenault}, \citenamefont {Manners},\ and\
  \citenamefont {Ozin}}]{Fournier-Bidoz2005}%
  \BibitemOpen
  \bibfield  {author} {\bibinfo {author} {\bibfnamefont {S.}~\bibnamefont
  {Fournier-Bidoz}}, \bibinfo {author} {\bibfnamefont {A.C.}\ \bibnamefont
  {Arsenault}}, \bibinfo {author} {\bibfnamefont {I.}~\bibnamefont {Manners}},
  \ and\ \bibinfo {author} {\bibfnamefont {G.A.}\ \bibnamefont {Ozin}},\
  }\bibfield  {title} {\enquote {\bibinfo {title} {Synthetic self-propelled
  nanorotors},}\ }\href@noop {} {\bibfield  {journal} {\bibinfo  {journal}
  {Chem. Commun.}\ }\textbf {\bibinfo {volume} {{}}},\ \bibinfo {pages}
  {441--443} (\bibinfo {year} {2005})}\BibitemShut {NoStop}%
\bibitem [{\citenamefont {Howse}\ \emph {et~al.}(2007)\citenamefont {Howse},
  \citenamefont {Jones}, \citenamefont {Ryan}, \citenamefont {Gough},
  \citenamefont {Vafabakhsh},\ and\ \citenamefont {Golestanian}}]{Howse2007}%
  \BibitemOpen
  \bibfield  {author} {\bibinfo {author} {\bibfnamefont {J.R.}\ \bibnamefont
  {Howse}}, \bibinfo {author} {\bibfnamefont {R.A.L.}\ \bibnamefont {Jones}},
  \bibinfo {author} {\bibfnamefont {A.J.}\ \bibnamefont {Ryan}}, \bibinfo
  {author} {\bibfnamefont {T.}~\bibnamefont {Gough}}, \bibinfo {author}
  {\bibfnamefont {R.}~\bibnamefont {Vafabakhsh}}, \ and\ \bibinfo {author}
  {\bibfnamefont {R.}~\bibnamefont {Golestanian}},\ }\bibfield  {title}
  {\enquote {\bibinfo {title} {Self-motile colloidal particles: From directed
  propulsion to random walk},}\ }\href@noop {} {\bibfield  {journal} {\bibinfo
  {journal} {Phys. Rev. Lett.}\ }\textbf {\bibinfo {volume} {99}},\ \bibinfo
  {pages} {048102:1--4} (\bibinfo {year} {2007})}\BibitemShut {NoStop}%
\bibitem [{\citenamefont {Volpe}\ \emph {et~al.}(2011)\citenamefont {Volpe},
  \citenamefont {Buttinoni}, \citenamefont {Vogt}, \citenamefont
  {K{\"u}mmerer},\ and\ \citenamefont {Bechinger}}]{volpe11}%
  \BibitemOpen
  \bibfield  {author} {\bibinfo {author} {\bibfnamefont {G.}~\bibnamefont
  {Volpe}}, \bibinfo {author} {\bibfnamefont {I.}~\bibnamefont {Buttinoni}},
  \bibinfo {author} {\bibfnamefont {D.}~\bibnamefont {Vogt}}, \bibinfo {author}
  {\bibfnamefont {H.-J.}\ \bibnamefont {K{\"u}mmerer}}, \ and\ \bibinfo
  {author} {\bibfnamefont {C.}~\bibnamefont {Bechinger}},\ }\bibfield  {title}
  {\enquote {\bibinfo {title} {Microswimmers in patterned environments},}\
  }\href@noop {} {\bibfield  {journal} {\bibinfo  {journal} {Soft Matter}\
  }\textbf {\bibinfo {volume} {7}},\ \bibinfo {pages} {8810--8815} (\bibinfo
  {year} {2011})}\BibitemShut {NoStop}%
\bibitem [{\citenamefont {Ebbens}\ \emph {et~al.}(2012)\citenamefont {Ebbens},
  \citenamefont {Tu}, \citenamefont {Howse},\ and\ \citenamefont
  {Golestanian}}]{Golestanian2012}%
  \BibitemOpen
  \bibfield  {author} {\bibinfo {author} {\bibfnamefont {S.}~\bibnamefont
  {Ebbens}}, \bibinfo {author} {\bibfnamefont {M.-H.}\ \bibnamefont {Tu}},
  \bibinfo {author} {\bibfnamefont {J.R.}\ \bibnamefont {Howse}}, \ and\
  \bibinfo {author} {\bibfnamefont {R.}~\bibnamefont {Golestanian}},\
  }\bibfield  {title} {\enquote {\bibinfo {title} {Size dependence of the
  propulsion velocity for catalytic {Janus}-sphere swimmers},}\ }\href@noop {}
  {\bibfield  {journal} {\bibinfo  {journal} {Phys. Rev. E}\ }\textbf {\bibinfo
  {volume} {85}},\ \bibinfo {pages} {020401(R):1--4} (\bibinfo {year}
  {2012})}\BibitemShut {NoStop}%
\bibitem [{\citenamefont {K\"ummel}\ \emph {et~al.}(2013)\citenamefont
  {K\"ummel}, \citenamefont {ten Hagen}, \citenamefont {Wittkowski},
  \citenamefont {Buttinoni}, \citenamefont {Eichhorn}, \citenamefont {Volpe},
  \citenamefont {L\"owen},\ and\ \citenamefont {Bechinger}}]{Bechinger2013a}%
  \BibitemOpen
  \bibfield  {author} {\bibinfo {author} {\bibfnamefont {F.}~\bibnamefont
  {K\"ummel}}, \bibinfo {author} {\bibfnamefont {B.}~\bibnamefont {ten Hagen}},
  \bibinfo {author} {\bibfnamefont {R.}~\bibnamefont {Wittkowski}}, \bibinfo
  {author} {\bibfnamefont {I.}~\bibnamefont {Buttinoni}}, \bibinfo {author}
  {\bibfnamefont {R.}~\bibnamefont {Eichhorn}}, \bibinfo {author}
  {\bibfnamefont {G.}~\bibnamefont {Volpe}}, \bibinfo {author} {\bibfnamefont
  {H.}~\bibnamefont {L\"owen}}, \ and\ \bibinfo {author} {\bibfnamefont
  {C.}~\bibnamefont {Bechinger}},\ }\bibfield  {title} {\enquote {\bibinfo
  {title} {Circular motion of asymmetric self-propelling particles},}\
  }\href@noop {} {\bibfield  {journal} {\bibinfo  {journal} {Phys. Rev. Lett.}\
  }\textbf {\bibinfo {volume} {110}},\ \bibinfo {pages} {198302:1--5} (\bibinfo
  {year} {2013})}\BibitemShut {NoStop}%
\bibitem [{\citenamefont {Buttinoni}\ \emph {et~al.}(2013)\citenamefont
  {Buttinoni}, \citenamefont {Bialk\'e}, \citenamefont {K\"ummel},
  \citenamefont {L\"owen}, \citenamefont {Bechinger},\ and\ \citenamefont
  {Speck}}]{Bechinger2013b}%
  \BibitemOpen
  \bibfield  {author} {\bibinfo {author} {\bibfnamefont {I.}~\bibnamefont
  {Buttinoni}}, \bibinfo {author} {\bibfnamefont {J.}~\bibnamefont {Bialk\'e}},
  \bibinfo {author} {\bibfnamefont {F.}~\bibnamefont {K\"ummel}}, \bibinfo
  {author} {\bibfnamefont {H.}~\bibnamefont {L\"owen}}, \bibinfo {author}
  {\bibfnamefont {C.}~\bibnamefont {Bechinger}}, \ and\ \bibinfo {author}
  {\bibfnamefont {T.}~\bibnamefont {Speck}},\ }\bibfield  {title} {\enquote
  {\bibinfo {title} {Dynamical clustering and phase separation in suspensions
  of self-propelled colloidal particles},}\ }\href@noop {} {\bibfield
  {journal} {\bibinfo  {journal} {Phys. Rev. Lett.}\ }\textbf {\bibinfo
  {volume} {110}},\ \bibinfo {pages} {238301:1--5} (\bibinfo {year}
  {2013})}\BibitemShut {NoStop}%
\bibitem [{\citenamefont {Lee}\ \emph {et~al.}(2014)\citenamefont {Lee},
  \citenamefont {Alarc{\'o}n-Correa}, \citenamefont {Miksch}, \citenamefont
  {Hahn}, \citenamefont {Gibbs},\ and\ \citenamefont {Fischer}}]{Fisher2014}%
  \BibitemOpen
  \bibfield  {author} {\bibinfo {author} {\bibfnamefont {T.-C.}\ \bibnamefont
  {Lee}}, \bibinfo {author} {\bibfnamefont {M.}~\bibnamefont
  {Alarc{\'o}n-Correa}}, \bibinfo {author} {\bibfnamefont {C.}~\bibnamefont
  {Miksch}}, \bibinfo {author} {\bibfnamefont {K.}~\bibnamefont {Hahn}},
  \bibinfo {author} {\bibfnamefont {J.G.}\ \bibnamefont {Gibbs}}, \ and\
  \bibinfo {author} {\bibfnamefont {P.}~\bibnamefont {Fischer}},\ }\bibfield
  {title} {\enquote {\bibinfo {title} {Self-propelling nanomotors in the
  presence of strong {Brownian} forces},}\ }\href@noop {} {\bibfield  {journal}
  {\bibinfo  {journal} {Nano Lett.}\ }\textbf {\bibinfo {volume} {14}},\
  \bibinfo {pages} {2407--2412} (\bibinfo {year} {2014})}\BibitemShut {NoStop}%
\bibitem [{\citenamefont {Ebbens}\ \emph {et~al.}(2014)\citenamefont {Ebbens},
  \citenamefont {Gregory}, \citenamefont {Dunderdale}, \citenamefont {Howse},
  \citenamefont {Ibrahim}, \citenamefont {Liverpool},\ and\ \citenamefont
  {Golestanian}}]{Golestanian2014}%
  \BibitemOpen
  \bibfield  {author} {\bibinfo {author} {\bibfnamefont {S.}~\bibnamefont
  {Ebbens}}, \bibinfo {author} {\bibfnamefont {D.A.}\ \bibnamefont {Gregory}},
  \bibinfo {author} {\bibfnamefont {G.}~\bibnamefont {Dunderdale}}, \bibinfo
  {author} {\bibfnamefont {J.R.}\ \bibnamefont {Howse}}, \bibinfo {author}
  {\bibfnamefont {Y.}~\bibnamefont {Ibrahim}}, \bibinfo {author} {\bibfnamefont
  {T.B.}\ \bibnamefont {Liverpool}}, \ and\ \bibinfo {author} {\bibfnamefont
  {R.}~\bibnamefont {Golestanian}},\ }\bibfield  {title} {\enquote {\bibinfo
  {title} {Electrokinetic effects in catalytic {Pt}-insulator {Janus}
  swimmers},}\ }\href@noop {} {\bibfield  {journal} {\bibinfo  {journal} {EPL}\
  }\textbf {\bibinfo {volume} {106}},\ \bibinfo {pages} {58003:1--6} (\bibinfo
  {year} {2014})}\BibitemShut {NoStop}%
\bibitem [{\citenamefont {ten Hagen}\ \emph {et~al.}(2014)\citenamefont {ten
  Hagen}, \citenamefont {K{\"u}mmel}, \citenamefont {Wittkowski}, \citenamefont
  {Takagi}, \citenamefont {L\"owen},\ and\ \citenamefont
  {Bechinger}}]{Bechinger2014}%
  \BibitemOpen
  \bibfield  {author} {\bibinfo {author} {\bibfnamefont {B.}~\bibnamefont {ten
  Hagen}}, \bibinfo {author} {\bibfnamefont {F.}~\bibnamefont {K{\"u}mmel}},
  \bibinfo {author} {\bibfnamefont {R.}~\bibnamefont {Wittkowski}}, \bibinfo
  {author} {\bibfnamefont {D.}~\bibnamefont {Takagi}}, \bibinfo {author}
  {\bibfnamefont {H.}~\bibnamefont {L\"owen}}, \ and\ \bibinfo {author}
  {\bibfnamefont {C.}~\bibnamefont {Bechinger}},\ }\bibfield  {title} {\enquote
  {\bibinfo {title} {Gravitaxis of asymmetric self-propelled colloidal
  particles},}\ }\href@noop {} {\bibfield  {journal} {\bibinfo  {journal}
  {Nature Comm.}\ }\textbf {\bibinfo {volume} {5}},\ \bibinfo {pages}
  {4829:1--7} (\bibinfo {year} {2014})}\BibitemShut {NoStop}%
\bibitem [{\citenamefont {Ma}\ \emph {et~al.}(2016)\citenamefont {Ma},
  \citenamefont {Jang}, \citenamefont {Popescu}, \citenamefont {Uspal},
  \citenamefont {Miguel-L{\'o}pez}, \citenamefont {Hahn}, \citenamefont {Kim},\
  and\ \citenamefont {S{\'a}nchez}}]{Ma2016}%
  \BibitemOpen
  \bibfield  {author} {\bibinfo {author} {\bibfnamefont {X.}~\bibnamefont
  {Ma}}, \bibinfo {author} {\bibfnamefont {S.}~\bibnamefont {Jang}}, \bibinfo
  {author} {\bibfnamefont {M.N.}\ \bibnamefont {Popescu}}, \bibinfo {author}
  {\bibfnamefont {W.E.}\ \bibnamefont {Uspal}}, \bibinfo {author}
  {\bibfnamefont {A.}~\bibnamefont {Miguel-L{\'o}pez}}, \bibinfo {author}
  {\bibfnamefont {K.}~\bibnamefont {Hahn}}, \bibinfo {author} {\bibfnamefont
  {D.-P.}\ \bibnamefont {Kim}}, \ and\ \bibinfo {author} {\bibfnamefont
  {S.}~\bibnamefont {S{\'a}nchez}},\ }\bibfield  {title} {\enquote {\bibinfo
  {title} {Reversed {Janus} micro/nanomotors with internal chemical engine},}\
  }\href@noop {} {\bibfield  {journal} {\bibinfo  {journal} {ACS Nano}\
  }\textbf {\bibinfo {volume} {10}},\ \bibinfo {pages} {8751--8759} (\bibinfo
  {year} {2016})}\BibitemShut {NoStop}%
\bibitem [{\citenamefont {Herminghaus}\ \emph {et~al.}(2014)\citenamefont
  {Herminghaus}, \citenamefont {Maas}, \citenamefont {Kr\"uger}, \citenamefont
  {Thutupalli}, \citenamefont {Goehring},\ and\ \citenamefont
  {Bahr}}]{Herminghaus2014}%
  \BibitemOpen
  \bibfield  {author} {\bibinfo {author} {\bibfnamefont {S.}~\bibnamefont
  {Herminghaus}}, \bibinfo {author} {\bibfnamefont {C.C.}\ \bibnamefont
  {Maas}}, \bibinfo {author} {\bibfnamefont {C.}~\bibnamefont {Kr\"uger}},
  \bibinfo {author} {\bibfnamefont {S.}~\bibnamefont {Thutupalli}}, \bibinfo
  {author} {\bibfnamefont {L.}~\bibnamefont {Goehring}}, \ and\ \bibinfo
  {author} {\bibfnamefont {C.}~\bibnamefont {Bahr}},\ }\bibfield  {title}
  {\enquote {\bibinfo {title} {Interfacial mechanisms in active emulsions},}\
  }\href@noop {} {\bibfield  {journal} {\bibinfo  {journal} {Soft Matter}\
  }\textbf {\bibinfo {volume} {10}},\ \bibinfo {pages} {7008--7022} (\bibinfo
  {year} {2014})}\BibitemShut {NoStop}%
\bibitem [{\citenamefont {Seemann}\ \emph {et~al.}(2016)\citenamefont
  {Seemann}, \citenamefont {Fleury},\ and\ \citenamefont {Maas}}]{Seemann2016}%
  \BibitemOpen
  \bibfield  {author} {\bibinfo {author} {\bibfnamefont {R.}~\bibnamefont
  {Seemann}}, \bibinfo {author} {\bibfnamefont {J.-B.}\ \bibnamefont {Fleury}},
  \ and\ \bibinfo {author} {\bibfnamefont {C.C.}\ \bibnamefont {Maas}},\
  }\bibfield  {title} {\enquote {\bibinfo {title} {Self-propelled droplets},}\
  }\href@noop {} {\bibfield  {journal} {\bibinfo  {journal} {Eur. Phys. J.
  Special Topics}\ }\textbf {\bibinfo {volume} {225}},\ \bibinfo {pages}
  {2227--2240} (\bibinfo {year} {2016})}\BibitemShut {NoStop}%
\bibitem [{\citenamefont {Kroy}\ \emph {et~al.}(2016)\citenamefont {Kroy},
  \citenamefont {Chakraborty},\ and\ \citenamefont {Cichos}}]{Kroy2016}%
  \BibitemOpen
  \bibfield  {author} {\bibinfo {author} {\bibfnamefont {K.}~\bibnamefont
  {Kroy}}, \bibinfo {author} {\bibfnamefont {D.}~\bibnamefont {Chakraborty}}, \
  and\ \bibinfo {author} {\bibfnamefont {F.}~\bibnamefont {Cichos}},\
  }\bibfield  {title} {\enquote {\bibinfo {title} {Hot microswimmers},}\
  }\href@noop {} {\bibfield  {journal} {\bibinfo  {journal} {Eur. Phys. J.
  Special Topics}\ }\textbf {\bibinfo {volume} {225}},\ \bibinfo {pages}
  {2207--2226} (\bibinfo {year} {2016})}\BibitemShut {NoStop}%
\bibitem [{\citenamefont {Lozano}\ \emph {et~al.}(2016)\citenamefont {Lozano},
  \citenamefont {ten Hagen}, \citenamefont {L{\"o}wen},\ and\ \citenamefont
  {Bechinger}}]{Bechinger2016}%
  \BibitemOpen
  \bibfield  {author} {\bibinfo {author} {\bibfnamefont {C.}~\bibnamefont
  {Lozano}}, \bibinfo {author} {\bibfnamefont {B.}~\bibnamefont {ten Hagen}},
  \bibinfo {author} {\bibfnamefont {H.}~\bibnamefont {L{\"o}wen}}, \ and\
  \bibinfo {author} {\bibfnamefont {C.}~\bibnamefont {Bechinger}},\ }\bibfield
  {title} {\enquote {\bibinfo {title} {Phototaxis of synthetic microswimmers in
  optical landscapes},}\ }\href@noop {} {\bibfield  {journal} {\bibinfo
  {journal} {Nature Comm.}\ }\textbf {\bibinfo {volume} {7}},\ \bibinfo {pages}
  {12828:1--10} (\bibinfo {year} {2016})}\BibitemShut {NoStop}%
\bibitem [{\citenamefont {Golestanian}\ \emph {et~al.}(2005)\citenamefont
  {Golestanian}, \citenamefont {Liverpool},\ and\ \citenamefont
  {Ajdari}}]{Golestanian2005}%
  \BibitemOpen
  \bibfield  {author} {\bibinfo {author} {\bibfnamefont {R.}~\bibnamefont
  {Golestanian}}, \bibinfo {author} {\bibfnamefont {T.B.}\ \bibnamefont
  {Liverpool}}, \ and\ \bibinfo {author} {\bibfnamefont {A.}~\bibnamefont
  {Ajdari}},\ }\bibfield  {title} {\enquote {\bibinfo {title} {Propulsion of a
  molecular machine by asymmetric distribution of reaction products},}\
  }\href@noop {} {\bibfield  {journal} {\bibinfo  {journal} {Phys. Rev. Lett.}\
  }\textbf {\bibinfo {volume} {94}},\ \bibinfo {pages} {220801:1--4} (\bibinfo
  {year} {2005})}\BibitemShut {NoStop}%
\bibitem [{\citenamefont {Golestanian}\ \emph {et~al.}(2007)\citenamefont
  {Golestanian}, \citenamefont {Liverpool},\ and\ \citenamefont
  {Ajdari}}]{Golestanian2007}%
  \BibitemOpen
  \bibfield  {author} {\bibinfo {author} {\bibfnamefont {R.}~\bibnamefont
  {Golestanian}}, \bibinfo {author} {\bibfnamefont {T.B.}\ \bibnamefont
  {Liverpool}}, \ and\ \bibinfo {author} {\bibfnamefont {A.}~\bibnamefont
  {Ajdari}},\ }\bibfield  {title} {\enquote {\bibinfo {title} {Designing
  phoretic micro- and nano-swimmers},}\ }\href@noop {} {\bibfield  {journal}
  {\bibinfo  {journal} {New J. Phys.}\ }\textbf {\bibinfo {volume} {9}},\
  \bibinfo {pages} {126:1--8} (\bibinfo {year} {2007})}\BibitemShut {NoStop}%
\bibitem [{\citenamefont {R{\"u}ckner}\ and\ \citenamefont
  {Kapral}(2007)}]{Kapral2007}%
  \BibitemOpen
  \bibfield  {author} {\bibinfo {author} {\bibfnamefont {G.R.}\ \bibnamefont
  {R{\"u}ckner}}\ and\ \bibinfo {author} {\bibfnamefont {R.}~\bibnamefont
  {Kapral}},\ }\bibfield  {title} {\enquote {\bibinfo {title} {Chemically
  powered nanodimers},}\ }\href@noop {} {\bibfield  {journal} {\bibinfo
  {journal} {Phys. Rev. Lett.}\ }\textbf {\bibinfo {volume} {98}},\ \bibinfo
  {pages} {150603:1--4} (\bibinfo {year} {2007})}\BibitemShut {NoStop}%
\bibitem [{\citenamefont {J{\"u}licher}\ and\ \citenamefont
  {Prost}(2009)}]{Julicher2009}%
  \BibitemOpen
  \bibfield  {author} {\bibinfo {author} {\bibfnamefont {F.}~\bibnamefont
  {J{\"u}licher}}\ and\ \bibinfo {author} {\bibfnamefont {J.}~\bibnamefont
  {Prost}},\ }\bibfield  {title} {\enquote {\bibinfo {title} {Generic theory of
  colloidal transport},}\ }\href@noop {} {\bibfield  {journal} {\bibinfo
  {journal} {Eur. Phys. J. E}\ }\textbf {\bibinfo {volume} {29}},\ \bibinfo
  {pages} {27--36} (\bibinfo {year} {2009})}\BibitemShut {NoStop}%
\bibitem [{\citenamefont {Popescu}\ \emph {et~al.}(2011)\citenamefont
  {Popescu}, \citenamefont {Tasinkevych},\ and\ \citenamefont
  {Dietrich}}]{Popescu2011EPL}%
  \BibitemOpen
  \bibfield  {author} {\bibinfo {author} {\bibfnamefont {M.N.}\ \bibnamefont
  {Popescu}}, \bibinfo {author} {\bibfnamefont {M.}~\bibnamefont
  {Tasinkevych}}, \ and\ \bibinfo {author} {\bibfnamefont {S.}~\bibnamefont
  {Dietrich}},\ }\bibfield  {title} {\enquote {\bibinfo {title} {Pulling and
  pushing a cargo with a catalytically active carrier},}\ }\href@noop {}
  {\bibfield  {journal} {\bibinfo  {journal} {EPL}\ }\textbf {\bibinfo {volume}
  {95}},\ \bibinfo {pages} {28004:1--6} (\bibinfo {year} {2011})}\BibitemShut
  {NoStop}%
\bibitem [{\citenamefont {Sabass}\ and\ \citenamefont
  {Seifert}(2012{\natexlab{a}})}]{Seifert2012a}%
  \BibitemOpen
  \bibfield  {author} {\bibinfo {author} {\bibfnamefont {B.}~\bibnamefont
  {Sabass}}\ and\ \bibinfo {author} {\bibfnamefont {U.}~\bibnamefont
  {Seifert}},\ }\bibfield  {title} {\enquote {\bibinfo {title} {Dynamics and
  efficiency of a self-propelled, diffusiophoretic swimmer},}\ }\href@noop {}
  {\bibfield  {journal} {\bibinfo  {journal} {J. Chem. Phys.}\ }\textbf
  {\bibinfo {volume} {136}},\ \bibinfo {pages} {064508:1--15} (\bibinfo {year}
  {2012}{\natexlab{a}})}\BibitemShut {NoStop}%
\bibitem [{\citenamefont {Sabass}\ and\ \citenamefont
  {Seifert}(2012{\natexlab{b}})}]{Seifert2012b}%
  \BibitemOpen
  \bibfield  {author} {\bibinfo {author} {\bibfnamefont {B.}~\bibnamefont
  {Sabass}}\ and\ \bibinfo {author} {\bibfnamefont {U.}~\bibnamefont
  {Seifert}},\ }\bibfield  {title} {\enquote {\bibinfo {title} {Nonlinear,
  electrocatalytic swimming in the presence of salt},}\ }\href@noop {}
  {\bibfield  {journal} {\bibinfo  {journal} {J. Chem. Phys.}\ }\textbf
  {\bibinfo {volume} {136}},\ \bibinfo {pages} {214507:1--13} (\bibinfo {year}
  {2012}{\natexlab{b}})}\BibitemShut {NoStop}%
\bibitem [{\citenamefont {Kapral}(2013)}]{Kapral2013}%
  \BibitemOpen
  \bibfield  {author} {\bibinfo {author} {\bibfnamefont {R.}~\bibnamefont
  {Kapral}},\ }\bibfield  {title} {\enquote {\bibinfo {title} {Nanomotors
  without moving parts that propel themselves in solution},}\ }\href@noop {}
  {\bibfield  {journal} {\bibinfo  {journal} {J. Chem. Phys.}\ }\textbf
  {\bibinfo {volume} {138}},\ \bibinfo {pages} {202901:1--10} (\bibinfo {year}
  {2013})}\BibitemShut {NoStop}%
\bibitem [{\citenamefont {Sharifi-Mood}\ \emph {et~al.}(2013)\citenamefont
  {Sharifi-Mood}, \citenamefont {Koplik},\ and\ \citenamefont
  {Maldarelli}}]{Koplik2013}%
  \BibitemOpen
  \bibfield  {author} {\bibinfo {author} {\bibfnamefont {N.}~\bibnamefont
  {Sharifi-Mood}}, \bibinfo {author} {\bibfnamefont {J.}~\bibnamefont
  {Koplik}}, \ and\ \bibinfo {author} {\bibfnamefont {C.}~\bibnamefont
  {Maldarelli}},\ }\bibfield  {title} {\enquote {\bibinfo {title}
  {Diffusiophoretic self-propulsion of colloids driven by a surface reaction:
  The sub-micron particle regime for exponential and van der {Waals}
  interactions},}\ }\href@noop {} {\bibfield  {journal} {\bibinfo  {journal}
  {Phys. Fluids}\ }\textbf {\bibinfo {volume} {25}},\ \bibinfo {pages}
  {012001:1--34} (\bibinfo {year} {2013})}\BibitemShut {NoStop}%
\bibitem [{\citenamefont {ten Hagen}\ \emph {et~al.}(2011)\citenamefont {ten
  Hagen}, \citenamefont {van Teeffelen},\ and\ \citenamefont
  {L{\"o}wen}}]{Lowen2011}%
  \BibitemOpen
  \bibfield  {author} {\bibinfo {author} {\bibfnamefont {B.}~\bibnamefont {ten
  Hagen}}, \bibinfo {author} {\bibfnamefont {S.}~\bibnamefont {van Teeffelen}},
  \ and\ \bibinfo {author} {\bibfnamefont {H.}~\bibnamefont {L{\"o}wen}},\
  }\bibfield  {title} {\enquote {\bibinfo {title} {Brownian motion of a
  self-propelled particle},}\ }\href@noop {} {\bibfield  {journal} {\bibinfo
  {journal} {J. Phys.: Condens. Matter}\ }\textbf {\bibinfo {volume} {23}},\
  \bibinfo {pages} {194119:1--12} (\bibinfo {year} {2011})}\BibitemShut
  {NoStop}%
\bibitem [{\citenamefont {Michelin}\ and\ \citenamefont
  {Lauga}(2015)}]{Michelin2015}%
  \BibitemOpen
  \bibfield  {author} {\bibinfo {author} {\bibfnamefont {S.}~\bibnamefont
  {Michelin}}\ and\ \bibinfo {author} {\bibfnamefont {E.}~\bibnamefont
  {Lauga}},\ }\bibfield  {title} {\enquote {\bibinfo {title} {Autophoretic
  locomotion from geometric asymmetry},}\ }\href@noop {} {\bibfield  {journal}
  {\bibinfo  {journal} {Eur. Phys. J. E}\ }\textbf {\bibinfo {volume} {38}},\
  \bibinfo {pages} {7:1--16} (\bibinfo {year} {2015})}\BibitemShut {NoStop}%
\bibitem [{\citenamefont {Hu}\ \emph {et~al.}(2015)\citenamefont {Hu},
  \citenamefont {Wysocki}, \citenamefont {Winkler},\ and\ \citenamefont
  {Gompper}}]{Gommper2015}%
  \BibitemOpen
  \bibfield  {author} {\bibinfo {author} {\bibfnamefont {J.}~\bibnamefont
  {Hu}}, \bibinfo {author} {\bibfnamefont {A.}~\bibnamefont {Wysocki}},
  \bibinfo {author} {\bibfnamefont {R.G.}\ \bibnamefont {Winkler}}, \ and\
  \bibinfo {author} {\bibfnamefont {G.}~\bibnamefont {Gompper}},\ }\bibfield
  {title} {\enquote {\bibinfo {title} {Physical sensing of surface properties
  by microswimmers - directing bacterial motion via wall slip},}\ }\href@noop
  {} {\bibfield  {journal} {\bibinfo  {journal} {Sci. Rep.}\ }\textbf {\bibinfo
  {volume} {5}},\ \bibinfo {pages} {9586:1--7} (\bibinfo {year}
  {2015})}\BibitemShut {NoStop}%
\bibitem [{\citenamefont {Popescu}\ \emph {et~al.}(2016)\citenamefont
  {Popescu}, \citenamefont {Uspal},\ and\ \citenamefont
  {Dietrich}}]{Popescu2016}%
  \BibitemOpen
  \bibfield  {author} {\bibinfo {author} {\bibfnamefont {M.N.}\ \bibnamefont
  {Popescu}}, \bibinfo {author} {\bibfnamefont {W.E.}\ \bibnamefont {Uspal}}, \
  and\ \bibinfo {author} {\bibfnamefont {S.}~\bibnamefont {Dietrich}},\
  }\bibfield  {title} {\enquote {\bibinfo {title} {Self-diffusiophoresis of
  chemically active coloids},}\ }\href@noop {} {\bibfield  {journal} {\bibinfo
  {journal} {Eur. Phys. J. Special Topics}\ }\textbf {\bibinfo {volume}
  {225}},\ \bibinfo {pages} {2189--2206} (\bibinfo {year} {2016})}\BibitemShut
  {NoStop}%
\bibitem [{\citenamefont {Z{\"o}ttl}\ and\ \citenamefont
  {Stark}(2016)}]{Stark2016}%
  \BibitemOpen
  \bibfield  {author} {\bibinfo {author} {\bibfnamefont {A.}~\bibnamefont
  {Z{\"o}ttl}}\ and\ \bibinfo {author} {\bibfnamefont {H.}~\bibnamefont
  {Stark}},\ }\bibfield  {title} {\enquote {\bibinfo {title} {Emergent behavior
  in active colloids},}\ }\href@noop {} {\bibfield  {journal} {\bibinfo
  {journal} {J. Phys.: Condens. Matter}\ }\textbf {\bibinfo {volume} {28}},\
  \bibinfo {pages} {253001:1--28} (\bibinfo {year} {2016})}\BibitemShut
  {NoStop}%
\bibitem [{\citenamefont {de~Graaf}\ \emph {et~al.}(2015)\citenamefont
  {de~Graaf}, \citenamefont {Rempfer},\ and\ \citenamefont
  {Holm}}]{deGraaf2015}%
  \BibitemOpen
  \bibfield  {author} {\bibinfo {author} {\bibfnamefont {J.}~\bibnamefont
  {de~Graaf}}, \bibinfo {author} {\bibfnamefont {G.}~\bibnamefont {Rempfer}}, \
  and\ \bibinfo {author} {\bibfnamefont {C.}~\bibnamefont {Holm}},\ }\bibfield
  {title} {\enquote {\bibinfo {title} {Diffusiophoretic self-propulsion for
  partially catalytic spherical colloids},}\ }\href@noop {} {\bibfield
  {journal} {\bibinfo  {journal} {IEEE Trans. NanoBiosci.}\ }\textbf {\bibinfo
  {volume} {14}},\ \bibinfo {pages} {272--288} (\bibinfo {year}
  {2015})}\BibitemShut {NoStop}%
\bibitem [{\citenamefont {Oshanin}\ \emph {et~al.}(2017)\citenamefont
  {Oshanin}, \citenamefont {Popescu},\ and\ \citenamefont
  {Dietrich}}]{Gleb2017}%
  \BibitemOpen
  \bibfield  {author} {\bibinfo {author} {\bibfnamefont {G.}~\bibnamefont
  {Oshanin}}, \bibinfo {author} {\bibfnamefont {M.N.}\ \bibnamefont {Popescu}},
  \ and\ \bibinfo {author} {\bibfnamefont {S.}~\bibnamefont {Dietrich}},\
  }\bibfield  {title} {\enquote {\bibinfo {title} {Active colloids in the
  context of chemical kinetics},}\ }\href@noop {} {\bibfield  {journal}
  {\bibinfo  {journal} {J. Phys. A}\ }\textbf {\bibinfo {volume} {50}},\
  \bibinfo {pages} {134001:1--50} (\bibinfo {year} {2017})}\BibitemShut
  {NoStop}%
\bibitem [{\citenamefont {Lammert}\ \emph {et~al.}(2016)\citenamefont
  {Lammert}, \citenamefont {Crespi},\ and\ \citenamefont
  {Nourhani}}]{Lammert2016}%
  \BibitemOpen
  \bibfield  {author} {\bibinfo {author} {\bibfnamefont {P.E.}\ \bibnamefont
  {Lammert}}, \bibinfo {author} {\bibfnamefont {V.H.}\ \bibnamefont {Crespi}},
  \ and\ \bibinfo {author} {\bibfnamefont {A.}~\bibnamefont {Nourhani}},\
  }\bibfield  {title} {\enquote {\bibinfo {title} {Bypassing slip velocity:
  rotational and translational velocities of autophoretic colloids in terms of
  surface flux},}\ }\href@noop {} {\bibfield  {journal} {\bibinfo  {journal}
  {J. Fluid Mech.}\ }\textbf {\bibinfo {volume} {802}},\ \bibinfo {pages}
  {294--304} (\bibinfo {year} {2016})}\BibitemShut {NoStop}%
\bibitem [{\citenamefont {Brown}\ \emph {et~al.}(2017)\citenamefont {Brown},
  \citenamefont {Poon}, \citenamefont {Holm},\ and\ \citenamefont
  {de~Graaf}}]{Brown2017}%
  \BibitemOpen
  \bibfield  {author} {\bibinfo {author} {\bibfnamefont {A.T.}\ \bibnamefont
  {Brown}}, \bibinfo {author} {\bibfnamefont {W.C.K.}\ \bibnamefont {Poon}},
  \bibinfo {author} {\bibfnamefont {C.}~\bibnamefont {Holm}}, \ and\ \bibinfo
  {author} {\bibfnamefont {J.}~\bibnamefont {de~Graaf}},\ }\bibfield  {title}
  {\enquote {\bibinfo {title} {Ionic screening and dissociation are crucial for
  understanding chemical self-propulsion in polar solvents},}\ }\href@noop {}
  {\bibfield  {journal} {\bibinfo  {journal} {Soft Matter}\ }\textbf {\bibinfo
  {volume} {13}},\ \bibinfo {pages} {1200--1222} (\bibinfo {year}
  {2017})}\BibitemShut {NoStop}%
\bibitem [{\citenamefont {Lauga}\ and\ \citenamefont
  {Powers}(2009)}]{Lauga2009}%
  \BibitemOpen
  \bibfield  {author} {\bibinfo {author} {\bibfnamefont {E.}~\bibnamefont
  {Lauga}}\ and\ \bibinfo {author} {\bibfnamefont {T.R.}\ \bibnamefont
  {Powers}},\ }\bibfield  {title} {\enquote {\bibinfo {title} {The
  hydrodynamics of swimming microorganisms},}\ }\href@noop {} {\bibfield
  {journal} {\bibinfo  {journal} {Rep. Prog. Phys.}\ }\textbf {\bibinfo
  {volume} {72}},\ \bibinfo {pages} {096601:1--36} (\bibinfo {year}
  {2009})}\BibitemShut {NoStop}%
\bibitem [{\citenamefont {Ebbens}\ and\ \citenamefont
  {Howse}(2010)}]{Ebbens2010}%
  \BibitemOpen
  \bibfield  {author} {\bibinfo {author} {\bibfnamefont {S.J.}\ \bibnamefont
  {Ebbens}}\ and\ \bibinfo {author} {\bibfnamefont {J.R.}\ \bibnamefont
  {Howse}},\ }\bibfield  {title} {\enquote {\bibinfo {title} {In pursuit of
  propulsion at the nanoscale},}\ }\href@noop {} {\bibfield  {journal}
  {\bibinfo  {journal} {Soft Matter}\ }\textbf {\bibinfo {volume} {6}},\
  \bibinfo {pages} {726--738} (\bibinfo {year} {2010})}\BibitemShut {NoStop}%
\bibitem [{\citenamefont {Hong}\ \emph {et~al.}(2010)\citenamefont {Hong},
  \citenamefont {Velegol}, \citenamefont {Chaturvedi},\ and\ \citenamefont
  {Sen}}]{SenRev}%
  \BibitemOpen
  \bibfield  {author} {\bibinfo {author} {\bibfnamefont {Y.}~\bibnamefont
  {Hong}}, \bibinfo {author} {\bibfnamefont {D.}~\bibnamefont {Velegol}},
  \bibinfo {author} {\bibfnamefont {N.}~\bibnamefont {Chaturvedi}}, \ and\
  \bibinfo {author} {\bibfnamefont {A.}~\bibnamefont {Sen}},\ }\bibfield
  {title} {\enquote {\bibinfo {title} {Biomimetic behavior of synthetic
  particles: from microscopic randomness to macroscopic control},}\ }\href@noop
  {} {\bibfield  {journal} {\bibinfo  {journal} {Phys. Chem. Chem. Phys.}\
  }\textbf {\bibinfo {volume} {12}},\ \bibinfo {pages} {1423--1435} (\bibinfo
  {year} {2010})}\BibitemShut {NoStop}%
\bibitem [{\citenamefont {Elgeti}\ \emph {et~al.}(2015)\citenamefont {Elgeti},
  \citenamefont {Winkler},\ and\ \citenamefont {Gompper}}]{Gommper2015_rev}%
  \BibitemOpen
  \bibfield  {author} {\bibinfo {author} {\bibfnamefont {J.}~\bibnamefont
  {Elgeti}}, \bibinfo {author} {\bibfnamefont {R.G.}\ \bibnamefont {Winkler}},
  \ and\ \bibinfo {author} {\bibfnamefont {G.}~\bibnamefont {Gompper}},\
  }\bibfield  {title} {\enquote {\bibinfo {title} {Physics of microswimmers --
  single particle motion and collective behavior: a review},}\ }\href@noop {}
  {\bibfield  {journal} {\bibinfo  {journal} {Rep. Prog. Phys.}\ }\textbf
  {\bibinfo {volume} {78}},\ \bibinfo {pages} {056601:1--50} (\bibinfo {year}
  {2015})}\BibitemShut {NoStop}%
\bibitem [{\citenamefont {Bechinger}\ \emph {et~al.}(2016)\citenamefont
  {Bechinger}, \citenamefont {Di~Leonardo}, \citenamefont {L\"owen},
  \citenamefont {Reichhardt}, \citenamefont {Volpe},\ and\ \citenamefont
  {Volpe}}]{Bechinger2016_rev}%
  \BibitemOpen
  \bibfield  {author} {\bibinfo {author} {\bibfnamefont {C.}~\bibnamefont
  {Bechinger}}, \bibinfo {author} {\bibfnamefont {R.}~\bibnamefont
  {Di~Leonardo}}, \bibinfo {author} {\bibfnamefont {H.}~\bibnamefont
  {L\"owen}}, \bibinfo {author} {\bibfnamefont {C.}~\bibnamefont {Reichhardt}},
  \bibinfo {author} {\bibfnamefont {G.}~\bibnamefont {Volpe}}, \ and\ \bibinfo
  {author} {\bibfnamefont {G.}~\bibnamefont {Volpe}},\ }\bibfield  {title}
  {\enquote {\bibinfo {title} {Active particles in complex and crowded
  environments},}\ }\href@noop {} {\bibfield  {journal} {\bibinfo  {journal}
  {Rev. Mod. Phys.}\ }\textbf {\bibinfo {volume} {88}},\ \bibinfo {pages}
  {045006:1--50} (\bibinfo {year} {2016})}\BibitemShut {NoStop}%
\bibitem [{\citenamefont {Moran}\ and\ \citenamefont
  {Posner}(2016)}]{Posner2017}%
  \BibitemOpen
  \bibfield  {author} {\bibinfo {author} {\bibfnamefont {J.L.}\ \bibnamefont
  {Moran}}\ and\ \bibinfo {author} {\bibfnamefont {J.D.}\ \bibnamefont
  {Posner}},\ }\bibfield  {title} {\enquote {\bibinfo {title} {Phoretic
  self-propulsion},}\ }\href@noop {} {\bibfield  {journal} {\bibinfo  {journal}
  {Ann. Rev. Fluid Mech.}\ }\textbf {\bibinfo {volume} {49}},\ \bibinfo {pages}
  {511--540} (\bibinfo {year} {2016})}\BibitemShut {NoStop}%
\bibitem [{\citenamefont {Derjaguin}\ \emph {et~al.}(1966)\citenamefont
  {Derjaguin}, \citenamefont {Yalamov},\ and\ \citenamefont
  {Storozhilova}}]{Derjaguin1966}%
  \BibitemOpen
  \bibfield  {author} {\bibinfo {author} {\bibfnamefont {B.V.}\ \bibnamefont
  {Derjaguin}}, \bibinfo {author} {\bibfnamefont {Yu.I.}\ \bibnamefont
  {Yalamov}}, \ and\ \bibinfo {author} {\bibfnamefont {A.I.}\ \bibnamefont
  {Storozhilova}},\ }\bibfield  {title} {\enquote {\bibinfo {title}
  {Diffusiophoresis of large aerosol particles},}\ }\href@noop {} {\bibfield
  {journal} {\bibinfo  {journal} {J. Colloid Interface Sci.}\ }\textbf
  {\bibinfo {volume} {22}},\ \bibinfo {pages} {117--125} (\bibinfo {year}
  {1966})}\BibitemShut {NoStop}%
\bibitem [{\citenamefont {Anderson}(1989)}]{Anderson1989}%
  \BibitemOpen
  \bibfield  {author} {\bibinfo {author} {\bibfnamefont {J.L.}\ \bibnamefont
  {Anderson}},\ }\bibfield  {title} {\enquote {\bibinfo {title} {Colloid
  transport by interfacial forces},}\ }\href@noop {} {\bibfield  {journal}
  {\bibinfo  {journal} {Ann. Rev. Fluid Mech.}\ }\textbf {\bibinfo {volume}
  {21}},\ \bibinfo {pages} {61--99} (\bibinfo {year} {1989})}\BibitemShut
  {NoStop}%
\bibitem [{\citenamefont {Pozrikidis}(2002)}]{pozrikidis02}%
  \BibitemOpen
  \bibfield  {author} {\bibinfo {author} {\bibfnamefont {C.}~\bibnamefont
  {Pozrikidis}},\ }\href@noop {} {\emph {\bibinfo {title} {A Practical Guide to
  {Boundary Element Methods} with the Software Library BEMLIB}}}\ (\bibinfo
  {publisher} {CRC Press},\ \bibinfo {address} {Boca Raton},\ \bibinfo {year}
  {2002})\BibitemShut {NoStop}%
\bibitem [{\citenamefont {Happel}\ and\ \citenamefont
  {Brenner}(1973)}]{HaBr73}%
  \BibitemOpen
  \bibfield  {author} {\bibinfo {author} {\bibfnamefont {J.}~\bibnamefont
  {Happel}}\ and\ \bibinfo {author} {\bibfnamefont {H.}~\bibnamefont
  {Brenner}},\ }\href@noop {} {\emph {\bibinfo {title} {Low {Reynolds} number
  hydrodynamics}}}\ (\bibinfo  {publisher} {Noordhoff Int. Pub.},\ \bibinfo
  {address} {Leyden, The Netherlands},\ \bibinfo {year} {1973})\BibitemShut
  {NoStop}%
\bibitem [{\citenamefont {Lighthill}(1952)}]{Lighthill1952}%
  \BibitemOpen
  \bibfield  {author} {\bibinfo {author} {\bibfnamefont {M.J.}\ \bibnamefont
  {Lighthill}},\ }\bibfield  {title} {\enquote {\bibinfo {title} {On the
  squirming motion of nearly spherical deformable bodies through liquids at
  very small {Reynolds} numbers},}\ }\href@noop {} {\bibfield  {journal}
  {\bibinfo  {journal} {Commun. Pure App. Math.}\ }\textbf {\bibinfo {volume}
  {5}},\ \bibinfo {pages} {109--118} (\bibinfo {year} {1952})}\BibitemShut
  {NoStop}%
\bibitem [{\citenamefont {Blake}(1971)}]{Blake1971}%
  \BibitemOpen
  \bibfield  {author} {\bibinfo {author} {\bibfnamefont {J.R.}\ \bibnamefont
  {Blake}},\ }\bibfield  {title} {\enquote {\bibinfo {title} {A spherical
  envelope approach to ciliary propulsion},}\ }\href@noop {} {\bibfield
  {journal} {\bibinfo  {journal} {J. Fluid Mech.}\ }\textbf {\bibinfo {volume}
  {46}},\ \bibinfo {pages} {199--208} (\bibinfo {year} {1971})}\BibitemShut
  {NoStop}%
\bibitem [{\citenamefont {Pak}\ and\ \citenamefont {Lauga}(2014)}]{Lauga2014}%
  \BibitemOpen
  \bibfield  {author} {\bibinfo {author} {\bibfnamefont {O.S.}\ \bibnamefont
  {Pak}}\ and\ \bibinfo {author} {\bibfnamefont {E.}~\bibnamefont {Lauga}},\
  }\bibfield  {title} {\enquote {\bibinfo {title} {Generalized squirming motion
  of a sphere},}\ }\href@noop {} {\bibfield  {journal} {\bibinfo  {journal} {J.
  Eng. Math.}\ }\textbf {\bibinfo {volume} {88}},\ \bibinfo {pages} {1--28}
  (\bibinfo {year} {2014})}\BibitemShut {NoStop}%
\bibitem [{\citenamefont {Lauga}\ \emph {et~al.}(2006)\citenamefont {Lauga},
  \citenamefont {DiLuzio}, \citenamefont {Whitesides},\ and\ \citenamefont
  {Stone}}]{Lauga2006}%
  \BibitemOpen
  \bibfield  {author} {\bibinfo {author} {\bibfnamefont {E.}~\bibnamefont
  {Lauga}}, \bibinfo {author} {\bibfnamefont {W.R.}\ \bibnamefont {DiLuzio}},
  \bibinfo {author} {\bibfnamefont {G.M.}\ \bibnamefont {Whitesides}}, \ and\
  \bibinfo {author} {\bibfnamefont {H.A.}\ \bibnamefont {Stone}},\ }\bibfield
  {title} {\enquote {\bibinfo {title} {Swimming in circles: Motion of bacteria
  near solid boundaries},}\ }\href@noop {} {\bibfield  {journal} {\bibinfo
  {journal} {Biophys. J.}\ }\textbf {\bibinfo {volume} {90}},\ \bibinfo {pages}
  {400--412} (\bibinfo {year} {2006})}\BibitemShut {NoStop}%
\bibitem [{\citenamefont {Berke}\ \emph {et~al.}(2008)\citenamefont {Berke},
  \citenamefont {Turner}, \citenamefont {Berg},\ and\ \citenamefont
  {Lauga}}]{Lauga2008}%
  \BibitemOpen
  \bibfield  {author} {\bibinfo {author} {\bibfnamefont {A.P.}\ \bibnamefont
  {Berke}}, \bibinfo {author} {\bibfnamefont {L.}~\bibnamefont {Turner}},
  \bibinfo {author} {\bibfnamefont {H.C.}\ \bibnamefont {Berg}}, \ and\
  \bibinfo {author} {\bibfnamefont {E.}~\bibnamefont {Lauga}},\ }\bibfield
  {title} {\enquote {\bibinfo {title} {Hydrodynamic attraction of swimming
  microorganisms by surfaces},}\ }\href@noop {} {\bibfield  {journal} {\bibinfo
   {journal} {Phys. Rev. Lett.}\ }\textbf {\bibinfo {volume} {101}},\ \bibinfo
  {pages} {038102:1--4} (\bibinfo {year} {2008})}\BibitemShut {NoStop}%
\bibitem [{\citenamefont {Drescher}\ \emph {et~al.}(2010)\citenamefont
  {Drescher}, \citenamefont {Goldstein}, \citenamefont {Michel}, \citenamefont
  {Polin},\ and\ \citenamefont {Tuval}}]{Goldstein2010}%
  \BibitemOpen
  \bibfield  {author} {\bibinfo {author} {\bibfnamefont {K.}~\bibnamefont
  {Drescher}}, \bibinfo {author} {\bibfnamefont {R.E.}\ \bibnamefont
  {Goldstein}}, \bibinfo {author} {\bibfnamefont {N.}~\bibnamefont {Michel}},
  \bibinfo {author} {\bibfnamefont {M.}~\bibnamefont {Polin}}, \ and\ \bibinfo
  {author} {\bibfnamefont {I.}~\bibnamefont {Tuval}},\ }\bibfield  {title}
  {\enquote {\bibinfo {title} {Direct measurement of the flow field around
  swimming microorganisms},}\ }\href@noop {} {\bibfield  {journal} {\bibinfo
  {journal} {Phys. Rev. Lett.}\ }\textbf {\bibinfo {volume} {105}},\ \bibinfo
  {pages} {168101:1--4} (\bibinfo {year} {2010})}\BibitemShut {NoStop}%
\bibitem [{\citenamefont {Lopez}\ and\ \citenamefont
  {Lauga}(2014)}]{Lauga2014b}%
  \BibitemOpen
  \bibfield  {author} {\bibinfo {author} {\bibfnamefont {D.}~\bibnamefont
  {Lopez}}\ and\ \bibinfo {author} {\bibfnamefont {E.}~\bibnamefont {Lauga}},\
  }\bibfield  {title} {\enquote {\bibinfo {title} {Dynamics of swimming
  bacteria at complex interfaces},}\ }\href@noop {} {\bibfield  {journal}
  {\bibinfo  {journal} {Phys. Fluids}\ }\textbf {\bibinfo {volume} {26}},\
  \bibinfo {pages} {071902:1--23} (\bibinfo {year} {2014})}\BibitemShut
  {NoStop}%
\bibitem [{\citenamefont {Mathijssen}\ \emph {et~al.}(2016)\citenamefont
  {Mathijssen}, \citenamefont {Doostmohammadi}, \citenamefont {Yeomans},\ and\
  \citenamefont {Shendruk}}]{Yeomans2016}%
  \BibitemOpen
  \bibfield  {author} {\bibinfo {author} {\bibfnamefont {A.J.T.M.}\
  \bibnamefont {Mathijssen}}, \bibinfo {author} {\bibfnamefont
  {A.}~\bibnamefont {Doostmohammadi}}, \bibinfo {author} {\bibfnamefont {J.M.}\
  \bibnamefont {Yeomans}}, \ and\ \bibinfo {author} {\bibfnamefont {T.N.}\
  \bibnamefont {Shendruk}},\ }\bibfield  {title} {\enquote {\bibinfo {title}
  {Hydrodynamics of micro-swimmers in films},}\ }\href@noop {} {\bibfield
  {journal} {\bibinfo  {journal} {J. Fluid Mech.}\ }\textbf {\bibinfo {volume}
  {806}},\ \bibinfo {pages} {35--70} (\bibinfo {year} {2016})}\BibitemShut
  {NoStop}%
\bibitem [{\citenamefont {Matas~Navarro}\ and\ \citenamefont
  {Pagonabarraga}(2010{\natexlab{a}})}]{Ignacio2010}%
  \BibitemOpen
  \bibfield  {author} {\bibinfo {author} {\bibfnamefont {R.}~\bibnamefont
  {Matas~Navarro}}\ and\ \bibinfo {author} {\bibfnamefont {I.}~\bibnamefont
  {Pagonabarraga}},\ }\bibfield  {title} {\enquote {\bibinfo {title}
  {Hydrodynamic interactions in squirmer motion: swimming with a neighbour and
  close to a wall},}\ }\href@noop {} {\bibfield  {journal} {\bibinfo  {journal}
  {J. Non-Newtonian Fluid Mech.}\ }\textbf {\bibinfo {volume} {165}},\ \bibinfo
  {pages} {946--952} (\bibinfo {year} {2010}{\natexlab{a}})}\BibitemShut
  {NoStop}%
\bibitem [{\citenamefont {Ishimoto}\ and\ \citenamefont
  {Gaffney}(2013)}]{Ishimoto2013}%
  \BibitemOpen
  \bibfield  {author} {\bibinfo {author} {\bibfnamefont {K.}~\bibnamefont
  {Ishimoto}}\ and\ \bibinfo {author} {\bibfnamefont {E.A.}\ \bibnamefont
  {Gaffney}},\ }\bibfield  {title} {\enquote {\bibinfo {title} {Squirmer
  dynamics near a boundary},}\ }\href@noop {} {\bibfield  {journal} {\bibinfo
  {journal} {Phys. Rev. E}\ }\textbf {\bibinfo {volume} {88}},\ \bibinfo
  {pages} {062702:1--12} (\bibinfo {year} {2013})}\BibitemShut {NoStop}%
\bibitem [{\citenamefont {de~Graaf}\ \emph {et~al.}(2016)\citenamefont
  {de~Graaf}, \citenamefont {Mathijssen}, \citenamefont {Fabritius},
  \citenamefont {Menke}, \citenamefont {Holm},\ and\ \citenamefont
  {Shendruk}}]{Holm2016}%
  \BibitemOpen
  \bibfield  {author} {\bibinfo {author} {\bibfnamefont {J.}~\bibnamefont
  {de~Graaf}}, \bibinfo {author} {\bibfnamefont {A.J.T.M.}\ \bibnamefont
  {Mathijssen}}, \bibinfo {author} {\bibfnamefont {M.}~\bibnamefont
  {Fabritius}}, \bibinfo {author} {\bibfnamefont {H.}~\bibnamefont {Menke}},
  \bibinfo {author} {\bibfnamefont {C.}~\bibnamefont {Holm}}, \ and\ \bibinfo
  {author} {\bibfnamefont {T.N.}\ \bibnamefont {Shendruk}},\ }\bibfield
  {title} {\enquote {\bibinfo {title} {Understanding the onset of oscillatory
  swimming in microchannels},}\ }\href@noop {} {\bibfield  {journal} {\bibinfo
  {journal} {Soft Matter}\ }\textbf {\bibinfo {volume} {12}},\ \bibinfo {pages}
  {4704--4708} (\bibinfo {year} {2016})}\BibitemShut {NoStop}%
\bibitem [{\citenamefont {Lintuvuori}\ \emph {et~al.}(2016)\citenamefont
  {Lintuvuori}, \citenamefont {Brown}, \citenamefont {Stratford},\ and\
  \citenamefont {Marenduzzo}}]{Brown2016}%
  \BibitemOpen
  \bibfield  {author} {\bibinfo {author} {\bibfnamefont {J.S.}\ \bibnamefont
  {Lintuvuori}}, \bibinfo {author} {\bibfnamefont {A.T.}\ \bibnamefont
  {Brown}}, \bibinfo {author} {\bibfnamefont {K.}~\bibnamefont {Stratford}}, \
  and\ \bibinfo {author} {\bibfnamefont {D.}~\bibnamefont {Marenduzzo}},\
  }\bibfield  {title} {\enquote {\bibinfo {title} {Hydrodynamic oscillations
  and variable swimming speed in squirmers close to repulsive walls},}\
  }\href@noop {} {\bibfield  {journal} {\bibinfo  {journal} {Soft Matter}\
  }\textbf {\bibinfo {volume} {12}},\ \bibinfo {pages} {7959--7968} (\bibinfo
  {year} {2016})}\BibitemShut {NoStop}%
\bibitem [{\citenamefont {Spagnolie}\ and\ \citenamefont
  {Lauga}(2012)}]{Spagnolie2012}%
  \BibitemOpen
  \bibfield  {author} {\bibinfo {author} {\bibfnamefont {S.}~\bibnamefont
  {Spagnolie}}\ and\ \bibinfo {author} {\bibfnamefont {E.}~\bibnamefont
  {Lauga}},\ }\bibfield  {title} {\enquote {\bibinfo {title} {Hydrodynamics of
  self-propulsion near a boundary: predictions and accuracy of far-field
  approximations},}\ }\href@noop {} {\bibfield  {journal} {\bibinfo  {journal}
  {J. Fluid Mech.}\ }\textbf {\bibinfo {volume} {700}},\ \bibinfo {pages}
  {105--147} (\bibinfo {year} {2012})}\BibitemShut {NoStop}%
\bibitem [{\citenamefont {Spagnolie}\ \emph {et~al.}(2015)\citenamefont
  {Spagnolie}, \citenamefont {Moreno-Flores}, \citenamefont {Bartolo},\ and\
  \citenamefont {Lauga}}]{Spagnolie2015}%
  \BibitemOpen
  \bibfield  {author} {\bibinfo {author} {\bibfnamefont {S.E.}\ \bibnamefont
  {Spagnolie}}, \bibinfo {author} {\bibfnamefont {G.R.}\ \bibnamefont
  {Moreno-Flores}}, \bibinfo {author} {\bibfnamefont {D.}~\bibnamefont
  {Bartolo}}, \ and\ \bibinfo {author} {\bibfnamefont {E.}~\bibnamefont
  {Lauga}},\ }\bibfield  {title} {\enquote {\bibinfo {title} {Geometric capture
  and escape of a microswimmer colliding with an obstacle},}\ }\href@noop {}
  {\bibfield  {journal} {\bibinfo  {journal} {Soft Matter}\ }\textbf {\bibinfo
  {volume} {11}},\ \bibinfo {pages} {3396--3411} (\bibinfo {year}
  {2015})}\BibitemShut {NoStop}%
\bibitem [{\citenamefont {Takagi}\ \emph {et~al.}(2014)\citenamefont {Takagi},
  \citenamefont {Palacci}, \citenamefont {Braunschweig}, \citenamefont
  {Shelley},\ and\ \citenamefont {Zhang}}]{Shelley2014}%
  \BibitemOpen
  \bibfield  {author} {\bibinfo {author} {\bibfnamefont {D.}~\bibnamefont
  {Takagi}}, \bibinfo {author} {\bibfnamefont {J.}~\bibnamefont {Palacci}},
  \bibinfo {author} {\bibfnamefont {A.B.}\ \bibnamefont {Braunschweig}},
  \bibinfo {author} {\bibfnamefont {M.J.}\ \bibnamefont {Shelley}}, \ and\
  \bibinfo {author} {\bibfnamefont {J.}~\bibnamefont {Zhang}},\ }\bibfield
  {title} {\enquote {\bibinfo {title} {Hydrodynamic capture of microswimmers
  into sphere-bound orbits},}\ }\href@noop {} {\bibfield  {journal} {\bibinfo
  {journal} {Soft Matter}\ }\textbf {\bibinfo {volume} {10}},\ \bibinfo {pages}
  {1784--1789} (\bibinfo {year} {2014})}\BibitemShut {NoStop}%
\bibitem [{\citenamefont {Ishikawa}\ \emph {et~al.}(2006)\citenamefont
  {Ishikawa}, \citenamefont {Simmonds},\ and\ \citenamefont
  {Pedley}}]{Pedley2006}%
  \BibitemOpen
  \bibfield  {author} {\bibinfo {author} {\bibfnamefont {T.}~\bibnamefont
  {Ishikawa}}, \bibinfo {author} {\bibfnamefont {M.P.}\ \bibnamefont
  {Simmonds}}, \ and\ \bibinfo {author} {\bibfnamefont {T.J.}\ \bibnamefont
  {Pedley}},\ }\bibfield  {title} {\enquote {\bibinfo {title} {Hydrodynamic
  interaction of two swimming model micro-organisms},}\ }\href@noop {}
  {\bibfield  {journal} {\bibinfo  {journal} {J. Fluid Mech.}\ }\textbf
  {\bibinfo {volume} {568}},\ \bibinfo {pages} {119--160} (\bibinfo {year}
  {2006})}\BibitemShut {NoStop}%
\bibitem [{\citenamefont {Papavassiliou}\ and\ \citenamefont
  {Alexander}(2017)}]{Alexander2017}%
  \BibitemOpen
  \bibfield  {author} {\bibinfo {author} {\bibfnamefont {D.}~\bibnamefont
  {Papavassiliou}}\ and\ \bibinfo {author} {\bibfnamefont {G.P.}\ \bibnamefont
  {Alexander}},\ }\bibfield  {title} {\enquote {\bibinfo {title} {Exact
  solutions for hydrodynamic interactions of two squirming spheres},}\
  }\href@noop {} {\bibfield  {journal} {\bibinfo  {journal} {J. Fluid Mech.}\
  }\textbf {\bibinfo {volume} {813}},\ \bibinfo {pages} {618--646} (\bibinfo
  {year} {2017})}\BibitemShut {NoStop}%
\bibitem [{\citenamefont {Matas~Navarro}\ and\ \citenamefont
  {Pagonabarraga}(2010{\natexlab{b}})}]{Ignacio2010b}%
  \BibitemOpen
  \bibfield  {author} {\bibinfo {author} {\bibfnamefont {R.}~\bibnamefont
  {Matas~Navarro}}\ and\ \bibinfo {author} {\bibfnamefont {I.}~\bibnamefont
  {Pagonabarraga}},\ }\bibfield  {title} {\enquote {\bibinfo {title}
  {Hydrodynamic interaction between two trapped swimming model
  micro-organisms},}\ }\href@noop {} {\bibfield  {journal} {\bibinfo  {journal}
  {Eur. Phys. J. E}\ }\textbf {\bibinfo {volume} {33}},\ \bibinfo {pages}
  {27--39} (\bibinfo {year} {2010}{\natexlab{b}})}\BibitemShut {NoStop}%
\bibitem [{\citenamefont {Alarc{\'o}n}\ and\ \citenamefont
  {Pagonabarraga}(2013)}]{Ignacio2013}%
  \BibitemOpen
  \bibfield  {author} {\bibinfo {author} {\bibfnamefont {F.}~\bibnamefont
  {Alarc{\'o}n}}\ and\ \bibinfo {author} {\bibfnamefont {I.}~\bibnamefont
  {Pagonabarraga}},\ }\bibfield  {title} {\enquote {\bibinfo {title}
  {Spontaneous aggregation and global polar ordering in squirmer
  suspensions},}\ }\href@noop {} {\bibfield  {journal} {\bibinfo  {journal} {J.
  Molecular Liquids}\ }\textbf {\bibinfo {volume} {185}},\ \bibinfo {pages}
  {56--61} (\bibinfo {year} {2013})}\BibitemShut {NoStop}%
\bibitem [{\citenamefont {Delfau}\ \emph {et~al.}(2016)\citenamefont {Delfau},
  \citenamefont {Molina},\ and\ \citenamefont {Sano}}]{Sano2016}%
  \BibitemOpen
  \bibfield  {author} {\bibinfo {author} {\bibfnamefont {J.-B.}\ \bibnamefont
  {Delfau}}, \bibinfo {author} {\bibfnamefont {J.}~\bibnamefont {Molina}}, \
  and\ \bibinfo {author} {\bibfnamefont {M.}~\bibnamefont {Sano}},\ }\bibfield
  {title} {\enquote {\bibinfo {title} {Collective behavior of strongly confined
  suspensions of squirmers},}\ }\href@noop {} {\bibfield  {journal} {\bibinfo
  {journal} {EPL}\ }\textbf {\bibinfo {volume} {114}},\ \bibinfo {pages}
  {24001:1--6} (\bibinfo {year} {2016})}\BibitemShut {NoStop}%
\bibitem [{\citenamefont {Saintillan}\ and\ \citenamefont
  {Shelley}(2008)}]{Shelley2008}%
  \BibitemOpen
  \bibfield  {author} {\bibinfo {author} {\bibfnamefont {D.}~\bibnamefont
  {Saintillan}}\ and\ \bibinfo {author} {\bibfnamefont {M.J.}\ \bibnamefont
  {Shelley}},\ }\bibfield  {title} {\enquote {\bibinfo {title} {Instabilities,
  pattern formation, and mixing in active suspensions},}\ }\href@noop {}
  {\bibfield  {journal} {\bibinfo  {journal} {Phys. Fluids}\ }\textbf {\bibinfo
  {volume} {20}},\ \bibinfo {pages} {123304:1--16} (\bibinfo {year}
  {2008})}\BibitemShut {NoStop}%
\bibitem [{\citenamefont {Lauga}\ and\ \citenamefont
  {Nadal}(2016)}]{Lauga2016}%
  \BibitemOpen
  \bibfield  {author} {\bibinfo {author} {\bibfnamefont {E.}~\bibnamefont
  {Lauga}}\ and\ \bibinfo {author} {\bibfnamefont {F.}~\bibnamefont {Nadal}},\
  }\bibfield  {title} {\enquote {\bibinfo {title} {Clustering instability of
  focused swimmers},}\ }\href@noop {} {\bibfield  {journal} {\bibinfo
  {journal} {EPL}\ }\textbf {\bibinfo {volume} {116}},\ \bibinfo {pages}
  {64004:1--6} (\bibinfo {year} {2016})}\BibitemShut {NoStop}%
\bibitem [{\citenamefont {Bickel}\ \emph {et~al.}(2013)\citenamefont {Bickel},
  \citenamefont {Majee},\ and\ \citenamefont {W{\"u}rger}}]{Majee2013}%
  \BibitemOpen
  \bibfield  {author} {\bibinfo {author} {\bibfnamefont {T.}~\bibnamefont
  {Bickel}}, \bibinfo {author} {\bibfnamefont {A.}~\bibnamefont {Majee}}, \
  and\ \bibinfo {author} {\bibfnamefont {A.}~\bibnamefont {W{\"u}rger}},\
  }\bibfield  {title} {\enquote {\bibinfo {title} {Flow pattern in the vicinity
  of self-propelling hot {Janus} particles},}\ }\href@noop {} {\bibfield
  {journal} {\bibinfo  {journal} {Phys. Rev. E}\ }\textbf {\bibinfo {volume}
  {88}},\ \bibinfo {pages} {012301:1--6} (\bibinfo {year} {2013})}\BibitemShut
  {NoStop}%
\bibitem [{\citenamefont {Michelin}\ and\ \citenamefont
  {Lauga}(2014)}]{Michelin2014}%
  \BibitemOpen
  \bibfield  {author} {\bibinfo {author} {\bibfnamefont {S.}~\bibnamefont
  {Michelin}}\ and\ \bibinfo {author} {\bibfnamefont {E.}~\bibnamefont
  {Lauga}},\ }\bibfield  {title} {\enquote {\bibinfo {title} {Phoretic
  self-propulsion at finite {Pecl\'et} numbers},}\ }\href@noop {} {\bibfield
  {journal} {\bibinfo  {journal} {J. Fluid Mech.}\ }\textbf {\bibinfo {volume}
  {747}},\ \bibinfo {pages} {572--604} (\bibinfo {year} {2014})}\BibitemShut
  {NoStop}%
\bibitem [{\citenamefont {Ibrahim}\ and\ \citenamefont
  {Liverpool}(2016)}]{Liverpool2016}%
  \BibitemOpen
  \bibfield  {author} {\bibinfo {author} {\bibfnamefont {Y.}~\bibnamefont
  {Ibrahim}}\ and\ \bibinfo {author} {\bibfnamefont {T.B.}\ \bibnamefont
  {Liverpool}},\ }\bibfield  {title} {\enquote {\bibinfo {title} {How walls
  affect the dynamics of self-phoretic microswimmers},}\ }\href@noop {}
  {\bibfield  {journal} {\bibinfo  {journal} {Eur. Phys. J. Special Topics}\
  }\textbf {\bibinfo {volume} {225}},\ \bibinfo {pages} {1843--1874} (\bibinfo
  {year} {2016})}\BibitemShut {NoStop}%
\bibitem [{\citenamefont {Baraban}\ \emph {et~al.}(2012)\citenamefont
  {Baraban}, \citenamefont {Tasinkevych}, \citenamefont {Popescu},
  \citenamefont {S{\'a}nchez}, \citenamefont {Dietrich},\ and\ \citenamefont
  {Schmidt}}]{Baraban2012}%
  \BibitemOpen
  \bibfield  {author} {\bibinfo {author} {\bibfnamefont {L.}~\bibnamefont
  {Baraban}}, \bibinfo {author} {\bibfnamefont {M.}~\bibnamefont
  {Tasinkevych}}, \bibinfo {author} {\bibfnamefont {M.N.}\ \bibnamefont
  {Popescu}}, \bibinfo {author} {\bibfnamefont {S.}~\bibnamefont
  {S{\'a}nchez}}, \bibinfo {author} {\bibfnamefont {S.}~\bibnamefont
  {Dietrich}}, \ and\ \bibinfo {author} {\bibfnamefont {O.G.}\ \bibnamefont
  {Schmidt}},\ }\bibfield  {title} {\enquote {\bibinfo {title} {Transport of
  cargo by catalytic {Janus} micro-motors},}\ }\href@noop {} {\bibfield
  {journal} {\bibinfo  {journal} {Soft Matter}\ }\textbf {\bibinfo {volume}
  {8}},\ \bibinfo {pages} {48--52} (\bibinfo {year} {2012})}\BibitemShut
  {NoStop}%
\bibitem [{\citenamefont {Palacci}\ \emph {et~al.}(2013)\citenamefont
  {Palacci}, \citenamefont {Sacanna}, \citenamefont {Steinberg}, \citenamefont
  {Pine},\ and\ \citenamefont {Chaikin}}]{Pine2013}%
  \BibitemOpen
  \bibfield  {author} {\bibinfo {author} {\bibfnamefont {J.}~\bibnamefont
  {Palacci}}, \bibinfo {author} {\bibfnamefont {S.}~\bibnamefont {Sacanna}},
  \bibinfo {author} {\bibfnamefont {A.S.}\ \bibnamefont {Steinberg}}, \bibinfo
  {author} {\bibfnamefont {D.J.}\ \bibnamefont {Pine}}, \ and\ \bibinfo
  {author} {\bibfnamefont {P.M.}\ \bibnamefont {Chaikin}},\ }\bibfield  {title}
  {\enquote {\bibinfo {title} {Living crystals of light-activated colloidal
  surfers},}\ }\href@noop {} {\bibfield  {journal} {\bibinfo  {journal}
  {Science}\ }\textbf {\bibinfo {volume} {339}},\ \bibinfo {pages} {936--940}
  (\bibinfo {year} {2013})}\BibitemShut {NoStop}%
\bibitem [{\citenamefont {Uspal}\ \emph
  {et~al.}(2015{\natexlab{a}})\citenamefont {Uspal}, \citenamefont {Popescu},
  \citenamefont {Dietrich},\ and\ \citenamefont {Tasinkevych}}]{Uspal2015a}%
  \BibitemOpen
  \bibfield  {author} {\bibinfo {author} {\bibfnamefont {W.E.}\ \bibnamefont
  {Uspal}}, \bibinfo {author} {\bibfnamefont {M.N.}\ \bibnamefont {Popescu}},
  \bibinfo {author} {\bibfnamefont {S.}~\bibnamefont {Dietrich}}, \ and\
  \bibinfo {author} {\bibfnamefont {M.}~\bibnamefont {Tasinkevych}},\
  }\bibfield  {title} {\enquote {\bibinfo {title} {Self-propulsion of a
  catalytically active particle near a planar wall: from reflection to sliding
  and hovering},}\ }\href@noop {} {\bibfield  {journal} {\bibinfo  {journal}
  {Soft Matter}\ }\textbf {\bibinfo {volume} {11}},\ \bibinfo {pages}
  {434--438} (\bibinfo {year} {2015}{\natexlab{a}})}\BibitemShut {NoStop}%
\bibitem [{\citenamefont {Mozaffari}\ \emph {et~al.}(2016)\citenamefont
  {Mozaffari}, \citenamefont {Sharifi-Mood}, \citenamefont {Koplik},\ and\
  \citenamefont {Maldarelli}}]{Koplik2016}%
  \BibitemOpen
  \bibfield  {author} {\bibinfo {author} {\bibfnamefont {A.}~\bibnamefont
  {Mozaffari}}, \bibinfo {author} {\bibfnamefont {N.}~\bibnamefont
  {Sharifi-Mood}}, \bibinfo {author} {\bibfnamefont {J.}~\bibnamefont
  {Koplik}}, \ and\ \bibinfo {author} {\bibfnamefont {C.}~\bibnamefont
  {Maldarelli}},\ }\bibfield  {title} {\enquote {\bibinfo {title}
  {Self-diffusiophoretic colloidal propulsion near a solid boundary},}\
  }\href@noop {} {\bibfield  {journal} {\bibinfo  {journal} {Phys. Fluids}\
  }\textbf {\bibinfo {volume} {28}},\ \bibinfo {pages} {053107:1--35} (\bibinfo
  {year} {2016})}\BibitemShut {NoStop}%
\bibitem [{\citenamefont {Brown}\ \emph {et~al.}(2016)\citenamefont {Brown},
  \citenamefont {Vladescu}, \citenamefont {Dawson}, \citenamefont {Vissers},
  \citenamefont {Schwarz-Linek}, \citenamefont {Lintuvuori},\ and\
  \citenamefont {Poon}}]{Brown2016b}%
  \BibitemOpen
  \bibfield  {author} {\bibinfo {author} {\bibfnamefont {A.T.}\ \bibnamefont
  {Brown}}, \bibinfo {author} {\bibfnamefont {I.D.}\ \bibnamefont {Vladescu}},
  \bibinfo {author} {\bibfnamefont {A.}~\bibnamefont {Dawson}}, \bibinfo
  {author} {\bibfnamefont {T.}~\bibnamefont {Vissers}}, \bibinfo {author}
  {\bibfnamefont {J.}~\bibnamefont {Schwarz-Linek}}, \bibinfo {author}
  {\bibfnamefont {J.S.}\ \bibnamefont {Lintuvuori}}, \ and\ \bibinfo {author}
  {\bibfnamefont {W.C.K.}\ \bibnamefont {Poon}},\ }\bibfield  {title} {\enquote
  {\bibinfo {title} {Swimming in a crystal},}\ }\href@noop {} {\bibfield
  {journal} {\bibinfo  {journal} {Soft Matter}\ }\textbf {\bibinfo {volume}
  {12}},\ \bibinfo {pages} {131--140} (\bibinfo {year} {2016})}\BibitemShut
  {NoStop}%
\bibitem [{\citenamefont {Leshansky}\ \emph {et~al.}(1997)\citenamefont
  {Leshansky}, \citenamefont {Golovin},\ and\ \citenamefont
  {Nir}}]{Leshansky1997}%
  \BibitemOpen
  \bibfield  {author} {\bibinfo {author} {\bibfnamefont {A.M.}\ \bibnamefont
  {Leshansky}}, \bibinfo {author} {\bibfnamefont {A.A.}\ \bibnamefont
  {Golovin}}, \ and\ \bibinfo {author} {\bibfnamefont {A.}~\bibnamefont
  {Nir}},\ }\bibfield  {title} {\enquote {\bibinfo {title} {Thermocapillary
  interaction between a solid particle and a liquid-gas interface},}\
  }\href@noop {} {\bibfield  {journal} {\bibinfo  {journal} {Phys. Fluids}\
  }\textbf {\bibinfo {volume} {9}},\ \bibinfo {pages} {2818--2827} (\bibinfo
  {year} {1997})}\BibitemShut {NoStop}%
\bibitem [{\citenamefont {Dom{\'i}nguez}\ \emph {et~al.}(2016)\citenamefont
  {Dom{\'i}nguez}, \citenamefont {Malgaretti}, \citenamefont {Popescu},\ and\
  \citenamefont {Dietrich}}]{Alvaro2016}%
  \BibitemOpen
  \bibfield  {author} {\bibinfo {author} {\bibfnamefont {A.}~\bibnamefont
  {Dom{\'i}nguez}}, \bibinfo {author} {\bibfnamefont {P.}~\bibnamefont
  {Malgaretti}}, \bibinfo {author} {\bibfnamefont {M.N.}\ \bibnamefont
  {Popescu}}, \ and\ \bibinfo {author} {\bibfnamefont {S.}~\bibnamefont
  {Dietrich}},\ }\bibfield  {title} {\enquote {\bibinfo {title} {Effective
  interaction between active colloids and fluid interfaces induced by
  {Marangoni} flows},}\ }\href@noop {} {\bibfield  {journal} {\bibinfo
  {journal} {Phys. Rev. Lett.}\ }\textbf {\bibinfo {volume} {116}},\ \bibinfo
  {pages} {078301:1--5} (\bibinfo {year} {2016})}\BibitemShut {NoStop}%
\bibitem [{\citenamefont {Das}\ \emph {et~al.}(2015)\citenamefont {Das},
  \citenamefont {Garg}, \citenamefont {Campbell}, \citenamefont {Howse},
  \citenamefont {Sen}, \citenamefont {Velegol}, \citenamefont {Golestanian},\
  and\ \citenamefont {Ebbens}}]{Howse2015}%
  \BibitemOpen
  \bibfield  {author} {\bibinfo {author} {\bibfnamefont {S.}~\bibnamefont
  {Das}}, \bibinfo {author} {\bibfnamefont {A.}~\bibnamefont {Garg}}, \bibinfo
  {author} {\bibfnamefont {A.I.}\ \bibnamefont {Campbell}}, \bibinfo {author}
  {\bibfnamefont {J.R.}\ \bibnamefont {Howse}}, \bibinfo {author}
  {\bibfnamefont {A.}~\bibnamefont {Sen}}, \bibinfo {author} {\bibfnamefont
  {D.}~\bibnamefont {Velegol}}, \bibinfo {author} {\bibfnamefont
  {R.}~\bibnamefont {Golestanian}}, \ and\ \bibinfo {author} {\bibfnamefont
  {S.J.}\ \bibnamefont {Ebbens}},\ }\bibfield  {title} {\enquote {\bibinfo
  {title} {Boundaries can steer active {Janus} spheres},}\ }\href@noop {}
  {\bibfield  {journal} {\bibinfo  {journal} {Nature Comm.}\ }\textbf {\bibinfo
  {volume} {6}},\ \bibinfo {pages} {8999:1--10} (\bibinfo {year}
  {2015})}\BibitemShut {NoStop}%
\bibitem [{\citenamefont {Simmchen}\ \emph {et~al.}(2016)\citenamefont
  {Simmchen}, \citenamefont {Katuri}, \citenamefont {Uspal}, \citenamefont
  {Popescu}, \citenamefont {Tasinkevych},\ and\ \citenamefont
  {S{\'a}nchez}}]{Simmchen2016}%
  \BibitemOpen
  \bibfield  {author} {\bibinfo {author} {\bibfnamefont {J.}~\bibnamefont
  {Simmchen}}, \bibinfo {author} {\bibfnamefont {J.}~\bibnamefont {Katuri}},
  \bibinfo {author} {\bibfnamefont {W.E.}\ \bibnamefont {Uspal}}, \bibinfo
  {author} {\bibfnamefont {M.N.}\ \bibnamefont {Popescu}}, \bibinfo {author}
  {\bibfnamefont {M.}~\bibnamefont {Tasinkevych}}, \ and\ \bibinfo {author}
  {\bibfnamefont {S.}~\bibnamefont {S{\'a}nchez}},\ }\bibfield  {title}
  {\enquote {\bibinfo {title} {Topographical pathways guide chemical
  microswimmers},}\ }\href@noop {} {\bibfield  {journal} {\bibinfo  {journal}
  {Nature Comm.}\ }\textbf {\bibinfo {volume} {7}},\ \bibinfo {pages}
  {10598:1--9} (\bibinfo {year} {2016})}\BibitemShut {NoStop}%
\bibitem [{\citenamefont {Uspal}\ \emph {et~al.}(2016)\citenamefont {Uspal},
  \citenamefont {Popescu}, \citenamefont {Dietrich},\ and\ \citenamefont
  {Tasinkevych}}]{Uspal2016}%
  \BibitemOpen
  \bibfield  {author} {\bibinfo {author} {\bibfnamefont {W.E.}\ \bibnamefont
  {Uspal}}, \bibinfo {author} {\bibfnamefont {M.N.}\ \bibnamefont {Popescu}},
  \bibinfo {author} {\bibfnamefont {S.}~\bibnamefont {Dietrich}}, \ and\
  \bibinfo {author} {\bibfnamefont {M.}~\bibnamefont {Tasinkevych}},\
  }\bibfield  {title} {\enquote {\bibinfo {title} {Guiding catalytically active
  particles with chemically patterned surfaces},}\ }\href@noop {} {\bibfield
  {journal} {\bibinfo  {journal} {Phys. Rev. Lett.}\ }\textbf {\bibinfo
  {volume} {117}},\ \bibinfo {pages} {048002:1--5} (\bibinfo {year}
  {2016})}\BibitemShut {NoStop}%
\bibitem [{\citenamefont {Popescu}\ \emph
  {et~al.}(2017{\natexlab{a}})\citenamefont {Popescu}, \citenamefont {Uspal},\
  and\ \citenamefont {Dietrich}}]{Popescu2017a}%
  \BibitemOpen
  \bibfield  {author} {\bibinfo {author} {\bibfnamefont {M.N.}\ \bibnamefont
  {Popescu}}, \bibinfo {author} {\bibfnamefont {W.E.}\ \bibnamefont {Uspal}}, \
  and\ \bibinfo {author} {\bibfnamefont {S.}~\bibnamefont {Dietrich}},\
  }\bibfield  {title} {\enquote {\bibinfo {title} {Chemically active colloids
  near osmotic-responsive walls with surface-chemistry gradients},}\
  }\href@noop {} {\bibfield  {journal} {\bibinfo  {journal} {J. Phys.: Condens.
  Matter}\ }\textbf {\bibinfo {volume} {29}},\ \bibinfo {pages} {134001:1--13}
  (\bibinfo {year} {2017}{\natexlab{a}})}\BibitemShut {NoStop}%
\bibitem [{\citenamefont {Uspal}\ \emph {et~al.}(2018)\citenamefont {Uspal},
  \citenamefont {Popescu}, \citenamefont {Tasinkevych},\ and\ \citenamefont
  {Dietrich}}]{Uspal2018a}%
  \BibitemOpen
  \bibfield  {author} {\bibinfo {author} {\bibfnamefont {W.E.}\ \bibnamefont
  {Uspal}}, \bibinfo {author} {\bibfnamefont {M.N.}\ \bibnamefont {Popescu}},
  \bibinfo {author} {\bibfnamefont {M.}~\bibnamefont {Tasinkevych}}, \ and\
  \bibinfo {author} {\bibfnamefont {S.}~\bibnamefont {Dietrich}},\ }\bibfield
  {title} {\enquote {\bibinfo {title} {Shape-dependent guidance of active
  {Janus} particles by chemically patterned surfaces},}\ }\href@noop {}
  {\bibfield  {journal} {\bibinfo  {journal} {New J. Phys.}\ }\textbf {\bibinfo
  {volume} {20}},\ \bibinfo {pages} {015013} (\bibinfo {year}
  {2018})}\BibitemShut {NoStop}%
\bibitem [{\citenamefont {Liu}\ \emph {et~al.}(2016)\citenamefont {Liu},
  \citenamefont {Zhou}, \citenamefont {Wang},\ and\ \citenamefont
  {Zhang}}]{Wei2016}%
  \BibitemOpen
  \bibfield  {author} {\bibinfo {author} {\bibfnamefont {C.}~\bibnamefont
  {Liu}}, \bibinfo {author} {\bibfnamefont {C.}~\bibnamefont {Zhou}}, \bibinfo
  {author} {\bibfnamefont {W.}~\bibnamefont {Wang}}, \ and\ \bibinfo {author}
  {\bibfnamefont {H.P.}\ \bibnamefont {Zhang}},\ }\bibfield  {title} {\enquote
  {\bibinfo {title} {Bimetallic microswimmers speed up in confining
  channels},}\ }\href@noop {} {\bibfield  {journal} {\bibinfo  {journal} {Phys.
  Rev. Lett.}\ }\textbf {\bibinfo {volume} {117}},\ \bibinfo {pages}
  {198001:1--6} (\bibinfo {year} {2016})}\BibitemShut {NoStop}%
\bibitem [{\citenamefont {Popescu}\ \emph {et~al.}(2009)\citenamefont
  {Popescu}, \citenamefont {Dietrich},\ and\ \citenamefont
  {Oshanin}}]{Popescu2009}%
  \BibitemOpen
  \bibfield  {author} {\bibinfo {author} {\bibfnamefont {M.N.}\ \bibnamefont
  {Popescu}}, \bibinfo {author} {\bibfnamefont {S.}~\bibnamefont {Dietrich}}, \
  and\ \bibinfo {author} {\bibfnamefont {G.}~\bibnamefont {Oshanin}},\
  }\bibfield  {title} {\enquote {\bibinfo {title} {Confinement effects on
  diffusiophoretic self-propellers},}\ }\href@noop {} {\bibfield  {journal}
  {\bibinfo  {journal} {J. Chem. Phys.}\ }\textbf {\bibinfo {volume} {130}},\
  \bibinfo {pages} {194702:1--16} (\bibinfo {year} {2009})}\BibitemShut
  {NoStop}%
\bibitem [{\citenamefont {Uspal}\ \emph
  {et~al.}(2015{\natexlab{b}})\citenamefont {Uspal}, \citenamefont {Popescu},
  \citenamefont {Dietrich},\ and\ \citenamefont {Tasinkevych}}]{Uspal2015b}%
  \BibitemOpen
  \bibfield  {author} {\bibinfo {author} {\bibfnamefont {W.E.}\ \bibnamefont
  {Uspal}}, \bibinfo {author} {\bibfnamefont {M.N.}\ \bibnamefont {Popescu}},
  \bibinfo {author} {\bibfnamefont {S.}~\bibnamefont {Dietrich}}, \ and\
  \bibinfo {author} {\bibfnamefont {M.}~\bibnamefont {Tasinkevych}},\
  }\bibfield  {title} {\enquote {\bibinfo {title} {Rheotaxis of spherical
  active particles near a planar wall},}\ }\href@noop {} {\bibfield  {journal}
  {\bibinfo  {journal} {Soft Matter}\ }\textbf {\bibinfo {volume} {11}},\
  \bibinfo {pages} {6613--6632} (\bibinfo {year}
  {2015}{\natexlab{b}})}\BibitemShut {NoStop}%
\bibitem [{\citenamefont {Katuri}\ \emph {et~al.}(2018)\citenamefont {Katuri},
  \citenamefont {Uspal}, \citenamefont {Simmchen}, \citenamefont
  {Miguel~L{\'o}pez},\ and\ \citenamefont {S{\'a}nchez}}]{Uspal2018b}%
  \BibitemOpen
  \bibfield  {author} {\bibinfo {author} {\bibfnamefont {J.}~\bibnamefont
  {Katuri}}, \bibinfo {author} {\bibfnamefont {W.E.}\ \bibnamefont {Uspal}},
  \bibinfo {author} {\bibfnamefont {J.}~\bibnamefont {Simmchen}}, \bibinfo
  {author} {\bibfnamefont {A.}~\bibnamefont {Miguel~L{\'o}pez}}, \ and\
  \bibinfo {author} {\bibfnamefont {S.}~\bibnamefont {S{\'a}nchez}},\
  }\bibfield  {title} {\enquote {\bibinfo {title} {Cross-stream migration of
  active particles},}\ }\href@noop {} {\bibfield  {journal} {\bibinfo
  {journal} {Sci. Adv.}\ }\textbf {\bibinfo {volume} {4}},\ \bibinfo {pages}
  {eaao1755} (\bibinfo {year} {2018})}\BibitemShut {NoStop}%
\bibitem [{\citenamefont {Campbell}\ and\ \citenamefont
  {Ebbens}(2013)}]{Ebbens2013}%
  \BibitemOpen
  \bibfield  {author} {\bibinfo {author} {\bibfnamefont {A.I.}\ \bibnamefont
  {Campbell}}\ and\ \bibinfo {author} {\bibfnamefont {S.J.}\ \bibnamefont
  {Ebbens}},\ }\bibfield  {title} {\enquote {\bibinfo {title} {Gravitaxis in
  spherical {Janus} swimming devices},}\ }\href@noop {} {\bibfield  {journal}
  {\bibinfo  {journal} {Langmuir}\ }\textbf {\bibinfo {volume} {29}},\ \bibinfo
  {pages} {14066--14073} (\bibinfo {year} {2013})}\BibitemShut {NoStop}%
\bibitem [{\citenamefont {Enculescu}\ and\ \citenamefont
  {Stark}(2011)}]{Stark2011}%
  \BibitemOpen
  \bibfield  {author} {\bibinfo {author} {\bibfnamefont {M.}~\bibnamefont
  {Enculescu}}\ and\ \bibinfo {author} {\bibfnamefont {H.}~\bibnamefont
  {Stark}},\ }\bibfield  {title} {\enquote {\bibinfo {title} {Active colloidal
  suspensions exhibit polar order under gravity},}\ }\href@noop {} {\bibfield
  {journal} {\bibinfo  {journal} {Phys. Rev. Lett.}\ }\textbf {\bibinfo
  {volume} {107}},\ \bibinfo {pages} {058301:1--5} (\bibinfo {year}
  {2011})}\BibitemShut {NoStop}%
\bibitem [{\citenamefont {Ibrahim}\ and\ \citenamefont
  {Liverpool}(2015)}]{Liverpool2015}%
  \BibitemOpen
  \bibfield  {author} {\bibinfo {author} {\bibfnamefont {Y.}~\bibnamefont
  {Ibrahim}}\ and\ \bibinfo {author} {\bibfnamefont {T.B.}\ \bibnamefont
  {Liverpool}},\ }\bibfield  {title} {\enquote {\bibinfo {title} {The dynamics
  of a self-phoretic {Janus} swimmer near a wall},}\ }\href@noop {} {\bibfield
  {journal} {\bibinfo  {journal} {EPL}\ }\textbf {\bibinfo {volume} {111}},\
  \bibinfo {pages} {48008:1--6} (\bibinfo {year} {2015})}\BibitemShut {NoStop}%
\bibitem [{\citenamefont {Popescu}\ \emph
  {et~al.}(2017{\natexlab{b}})\citenamefont {Popescu}, \citenamefont {Uspal},
  \citenamefont {Tasinkevych},\ and\ \citenamefont {Dietrich}}]{Popescu2017b}%
  \BibitemOpen
  \bibfield  {author} {\bibinfo {author} {\bibfnamefont {M.N.}\ \bibnamefont
  {Popescu}}, \bibinfo {author} {\bibfnamefont {W.E.}\ \bibnamefont {Uspal}},
  \bibinfo {author} {\bibfnamefont {M.}~\bibnamefont {Tasinkevych}}, \ and\
  \bibinfo {author} {\bibfnamefont {S.}~\bibnamefont {Dietrich}},\ }\bibfield
  {title} {\enquote {\bibinfo {title} {Perils of ad hoc approximations for the
  activity function of chemically powered colloids},}\ }\href@noop {}
  {\bibfield  {journal} {\bibinfo  {journal} {Eur. Phys. J. E}\ }\textbf
  {\bibinfo {volume} {40}},\ \bibinfo {pages} {42:1--7} (\bibinfo {year}
  {2017}{\natexlab{b}})}\BibitemShut {NoStop}%
\bibitem [{\citenamefont {Sharifi-Mood}\ \emph {et~al.}(2016)\citenamefont
  {Sharifi-Mood}, \citenamefont {Mozaffari},\ and\ \citenamefont
  {C{\'o}rdova-Figueroa}}]{Koplik2016b}%
  \BibitemOpen
  \bibfield  {author} {\bibinfo {author} {\bibfnamefont {N.}~\bibnamefont
  {Sharifi-Mood}}, \bibinfo {author} {\bibfnamefont {A.}~\bibnamefont
  {Mozaffari}}, \ and\ \bibinfo {author} {\bibfnamefont {U.M.}\ \bibnamefont
  {C{\'o}rdova-Figueroa}},\ }\bibfield  {title} {\enquote {\bibinfo {title}
  {Pair interaction of catalytically active colloids: from assembly to
  escape},}\ }\href@noop {} {\bibfield  {journal} {\bibinfo  {journal} {J.
  Fluid Mech.}\ }\textbf {\bibinfo {volume} {798}},\ \bibinfo {pages}
  {910--954} (\bibinfo {year} {2016})}\BibitemShut {NoStop}%
\bibitem [{\citenamefont {Reigh}\ and\ \citenamefont {R.}(2015)}]{Reigh2015}%
  \BibitemOpen
  \bibfield  {author} {\bibinfo {author} {\bibfnamefont {S.Y.}\ \bibnamefont
  {Reigh}}\ and\ \bibinfo {author} {\bibfnamefont {Kapral}\ \bibnamefont
  {R.}},\ }\bibfield  {title} {\enquote {\bibinfo {title} {Catalytic dimer
  nanomotors: continuum theory and microscopic dynamics},}\ }\href@noop {}
  {\bibfield  {journal} {\bibinfo  {journal} {Soft Matter}\ }\textbf {\bibinfo
  {volume} {11}},\ \bibinfo {pages} {3149--3158} (\bibinfo {year}
  {2015})}\BibitemShut {NoStop}%
\bibitem [{\citenamefont {Popescu}\ \emph {et~al.}(2010)\citenamefont
  {Popescu}, \citenamefont {Dietrich}, \citenamefont {Tasinkevych},\ and\
  \citenamefont {Ralston}}]{Popescu2010}%
  \BibitemOpen
  \bibfield  {author} {\bibinfo {author} {\bibfnamefont {M.N.}\ \bibnamefont
  {Popescu}}, \bibinfo {author} {\bibfnamefont {S.}~\bibnamefont {Dietrich}},
  \bibinfo {author} {\bibfnamefont {M.}~\bibnamefont {Tasinkevych}}, \ and\
  \bibinfo {author} {\bibfnamefont {J.}~\bibnamefont {Ralston}},\ }\bibfield
  {title} {\enquote {\bibinfo {title} {Phoretic motion of spheroidal particles
  due to self-generated solute gradients},}\ }\href@noop {} {\bibfield
  {journal} {\bibinfo  {journal} {Eur. Phys. J. E}\ }\textbf {\bibinfo {volume}
  {31}},\ \bibinfo {pages} {351--367} (\bibinfo {year} {2010})}\BibitemShut
  {NoStop}%
\bibitem [{\citenamefont {Brady}(2011)}]{Brady2011}%
  \BibitemOpen
  \bibfield  {author} {\bibinfo {author} {\bibfnamefont {J.F.}\ \bibnamefont
  {Brady}},\ }\bibfield  {title} {\enquote {\bibinfo {title} {Particle motion
  driven by solute gradients with application to autonomous motion: Continuum
  and colloidal perspectives},}\ }\href@noop {} {\bibfield  {journal} {\bibinfo
   {journal} {J. Fluid Mech.}\ }\textbf {\bibinfo {volume} {667}},\ \bibinfo
  {pages} {216--259} (\bibinfo {year} {2011})}\BibitemShut {NoStop}%
\bibitem [{\citenamefont {de~Groot}\ and\ \citenamefont
  {Mazur}(1962)}]{Mazur_book}%
  \BibitemOpen
  \bibfield  {author} {\bibinfo {author} {\bibfnamefont {S.R.}\ \bibnamefont
  {de~Groot}}\ and\ \bibinfo {author} {\bibfnamefont {P.}~\bibnamefont
  {Mazur}},\ }\href@noop {} {\emph {\bibinfo {title} {Non-equilibrium
  Thermodynamics}}}\ (\bibinfo  {publisher} {North-Holland, Amsterdam},\
  \bibinfo {year} {1962})\BibitemShut {NoStop}%
\bibitem [{\citenamefont {Abramowitz}\ and\ \citenamefont
  {Stegun}(1972)}]{abram}%
  \BibitemOpen
  \bibfield  {author} {\bibinfo {author} {\bibfnamefont {M.}~\bibnamefont
  {Abramowitz}}\ and\ \bibinfo {author} {\bibfnamefont {I.R.}\ \bibnamefont
  {Stegun}},\ }\href@noop {} {\emph {\bibinfo {title} {{\rm Eds.} Handbook of
  mathematical functions}}}\ (\bibinfo  {publisher} {Dover, New York},\
  \bibinfo {year} {1972})\BibitemShut {NoStop}%
\bibitem [{\citenamefont {Ma}\ \emph {et~al.}(2015)\citenamefont {Ma},
  \citenamefont {Jannasch}, \citenamefont {Albrecht}, \citenamefont {Hahn},
  \citenamefont {Miguel-L{\'o}pez}, \citenamefont {Sch{\"a}ffer},\ and\
  \citenamefont {S{\'a}nchez}}]{Ma2015}%
  \BibitemOpen
  \bibfield  {author} {\bibinfo {author} {\bibfnamefont {X.}~\bibnamefont
  {Ma}}, \bibinfo {author} {\bibfnamefont {A.}~\bibnamefont {Jannasch}},
  \bibinfo {author} {\bibfnamefont {U.-R.}\ \bibnamefont {Albrecht}}, \bibinfo
  {author} {\bibfnamefont {K.}~\bibnamefont {Hahn}}, \bibinfo {author}
  {\bibfnamefont {A.}~\bibnamefont {Miguel-L{\'o}pez}}, \bibinfo {author}
  {\bibfnamefont {E.}~\bibnamefont {Sch{\"a}ffer}}, \ and\ \bibinfo {author}
  {\bibfnamefont {S.}~\bibnamefont {S{\'a}nchez}},\ }\bibfield  {title}
  {\enquote {\bibinfo {title} {Enzyme-powered hollow mesoporous {Janus}
  nanomotors},}\ }\href@noop {} {\bibfield  {journal} {\bibinfo  {journal}
  {Nano Lett.}\ }\textbf {\bibinfo {volume} {15}},\ \bibinfo {pages}
  {7043--7050} (\bibinfo {year} {2015})}\BibitemShut {NoStop}%
\bibitem [{\citenamefont {Eskandari}(2016)}]{Zahra2016}%
  \BibitemOpen
  \bibfield  {author} {\bibinfo {author} {\bibfnamefont {Z.}~\bibnamefont
  {Eskandari}},\ }\href@noop {} {\enquote {\bibinfo {title} {Head-on collisions
  of chemically active colloids},}\ } (\bibinfo {year} {2016}),\ \Eprint
  {http://arxiv.org/abs/unpublished} {unpublished} \BibitemShut {NoStop}%
\bibitem [{\citenamefont {Brown}\ and\ \citenamefont {Poon}(2014)}]{Brown2014}%
  \BibitemOpen
  \bibfield  {author} {\bibinfo {author} {\bibfnamefont {A.}~\bibnamefont
  {Brown}}\ and\ \bibinfo {author} {\bibfnamefont {W.}~\bibnamefont {Poon}},\
  }\bibfield  {title} {\enquote {\bibinfo {title} {Ionic effects in
  self-propelled {Pt-coated Janus} swimmers},}\ }\href@noop {} {\bibfield
  {journal} {\bibinfo  {journal} {Soft Matter}\ }\textbf {\bibinfo {volume}
  {10}},\ \bibinfo {pages} {4016--4027} (\bibinfo {year} {2014})}\BibitemShut
  {NoStop}%
\bibitem [{\citenamefont {C\'ordova-Figueroa}\ and\ \citenamefont
  {Brady}(2008)}]{Brady2008}%
  \BibitemOpen
  \bibfield  {author} {\bibinfo {author} {\bibfnamefont {Ubaldo~M.}\
  \bibnamefont {C\'ordova-Figueroa}}\ and\ \bibinfo {author} {\bibfnamefont
  {John~F.}\ \bibnamefont {Brady}},\ }\bibfield  {title} {\enquote {\bibinfo
  {title} {Osmotic propulsion: The osmotic motor},}\ }\href@noop {} {\bibfield
  {journal} {\bibinfo  {journal} {Phys. Rev. Lett.}\ }\textbf {\bibinfo
  {volume} {100}},\ \bibinfo {pages} {158303:1--4} (\bibinfo {year}
  {2008})}\BibitemShut {NoStop}%
\bibitem [{\citenamefont {Collins}(1961)}]{Collins1960}%
  \BibitemOpen
  \bibfield  {author} {\bibinfo {author} {\bibfnamefont {W.D.}\ \bibnamefont
  {Collins}},\ }\bibfield  {title} {\enquote {\bibinfo {title} {On some dual
  series equations and their application to electrostatic problems for
  spheroidal caps},}\ }\href@noop {} {\bibfield  {journal} {\bibinfo  {journal}
  {Math. Proc. Cambridge Phil. Soc.}\ }\textbf {\bibinfo {volume} {57}},\
  \bibinfo {pages} {367--384} (\bibinfo {year} {1961})}\BibitemShut {NoStop}%
\bibitem [{\citenamefont {Sneddon}(1966)}]{Sneddon_book}%
  \BibitemOpen
  \bibfield  {author} {\bibinfo {author} {\bibfnamefont {I.N.}\ \bibnamefont
  {Sneddon}},\ }\href@noop {} {\emph {\bibinfo {title} {Mixed boundary Value in
  Potential Theory}}}\ (\bibinfo  {publisher} {North-Holland},\ \bibinfo
  {address} {Amsterdam, The Netherlands},\ \bibinfo {year} {1966})\BibitemShut
  {NoStop}%
\bibitem [{\citenamefont {Samson}\ and\ \citenamefont
  {Deutch}(1978)}]{Deutch1978}%
  \BibitemOpen
  \bibfield  {author} {\bibinfo {author} {\bibfnamefont {R.}~\bibnamefont
  {Samson}}\ and\ \bibinfo {author} {\bibfnamefont {J.M.}\ \bibnamefont
  {Deutch}},\ }\bibfield  {title} {\enquote {\bibinfo {title}
  {Diffusion-controlled reaction rate to a buried active site},}\ }\href@noop
  {} {\bibfield  {journal} {\bibinfo  {journal} {J. Chem. Phys.}\ }\textbf
  {\bibinfo {volume} {68}},\ \bibinfo {pages} {285--290} (\bibinfo {year}
  {1978})}\BibitemShut {NoStop}%
\bibitem [{\citenamefont {Shoup}\ \emph {et~al.}(1981)\citenamefont {Shoup},
  \citenamefont {Lipari},\ and\ \citenamefont {Szabo}}]{Shoup1981}%
  \BibitemOpen
  \bibfield  {author} {\bibinfo {author} {\bibfnamefont {D.}~\bibnamefont
  {Shoup}}, \bibinfo {author} {\bibfnamefont {G.}~\bibnamefont {Lipari}}, \
  and\ \bibinfo {author} {\bibfnamefont {A.}~\bibnamefont {Szabo}},\ }\bibfield
   {title} {\enquote {\bibinfo {title} {Diffusion-controlled bimolecular
  reaction rates. {The} effect of rotational diffusion and orientation
  constraints},}\ }\href@noop {} {\bibfield  {journal} {\bibinfo  {journal}
  {Biophys. J.}\ }\textbf {\bibinfo {volume} {36}},\ \bibinfo {pages}
  {697--714} (\bibinfo {year} {1981})}\BibitemShut {NoStop}%
\bibitem [{\citenamefont {Shoup}\ and\ \citenamefont
  {Szabo}(1982)}]{Shoup1982}%
  \BibitemOpen
  \bibfield  {author} {\bibinfo {author} {\bibfnamefont {D.}~\bibnamefont
  {Shoup}}\ and\ \bibinfo {author} {\bibfnamefont {A.}~\bibnamefont {Szabo}},\
  }\bibfield  {title} {\enquote {\bibinfo {title} {Role of diffusion in ligand
  binding to macromolecules and cell-bound receptors},}\ }\href@noop {}
  {\bibfield  {journal} {\bibinfo  {journal} {Biophys. J.}\ }\textbf {\bibinfo
  {volume} {40}},\ \bibinfo {pages} {33--39} (\bibinfo {year}
  {1982})}\BibitemShut {NoStop}%
\bibitem [{\citenamefont {Traytak}(1994)}]{Traytak1994}%
  \BibitemOpen
  \bibfield  {author} {\bibinfo {author} {\bibfnamefont {S.D.}\ \bibnamefont
  {Traytak}},\ }\bibfield  {title} {\enquote {\bibinfo {title} {The steric
  factor in the time-dependent diffusion-controlled reactions},}\ }\href@noop
  {} {\bibfield  {journal} {\bibinfo  {journal} {J. Phys. Chem.}\ }\textbf
  {\bibinfo {volume} {98}},\ \bibinfo {pages} {7419--7421} (\bibinfo {year}
  {1994})}\BibitemShut {NoStop}%
\bibitem [{\citenamefont {Traytak}(1995)}]{Traytak1995a}%
  \BibitemOpen
  \bibfield  {author} {\bibinfo {author} {\bibfnamefont {S.D.}\ \bibnamefont
  {Traytak}},\ }\bibfield  {title} {\enquote {\bibinfo {title}
  {Diffusion-controlled reaction rate to an active site},}\ }\href@noop {}
  {\bibfield  {journal} {\bibinfo  {journal} {Chem. Phys.}\ }\textbf {\bibinfo
  {volume} {192}},\ \bibinfo {pages} {1--7} (\bibinfo {year}
  {1995})}\BibitemShut {NoStop}%
\bibitem [{\citenamefont {Traytak}\ and\ \citenamefont
  {Tachiya}(1995{\natexlab{a}})}]{Traytak1995b}%
  \BibitemOpen
  \bibfield  {author} {\bibinfo {author} {\bibfnamefont {S.D.}\ \bibnamefont
  {Traytak}}\ and\ \bibinfo {author} {\bibfnamefont {M.}~\bibnamefont
  {Tachiya}},\ }\bibfield  {title} {\enquote {\bibinfo {title}
  {Diffusion-controlled reaction rate to asymmetric reactants under {Coulomb}
  interaction},}\ }\href@noop {} {\bibfield  {journal} {\bibinfo  {journal} {J.
  Chem. Phys.}\ }\textbf {\bibinfo {volume} {102}},\ \bibinfo {pages}
  {9240--9247} (\bibinfo {year} {1995}{\natexlab{a}})}\BibitemShut {NoStop}%
\bibitem [{\citenamefont {Traytak}\ and\ \citenamefont
  {Tachiya}(1995{\natexlab{b}})}]{Traytak1995c}%
  \BibitemOpen
  \bibfield  {author} {\bibinfo {author} {\bibfnamefont {S.D.}\ \bibnamefont
  {Traytak}}\ and\ \bibinfo {author} {\bibfnamefont {M.}~\bibnamefont
  {Tachiya}},\ }\bibfield  {title} {\enquote {\bibinfo {title}
  {Diffusion-controlled reaction rate to an active site in an external electric
  field},}\ }\href@noop {} {\bibfield  {journal} {\bibinfo  {journal} {J. Chem.
  Phys.}\ }\textbf {\bibinfo {volume} {102}},\ \bibinfo {pages} {2760--2771}
  (\bibinfo {year} {1995}{\natexlab{b}})}\BibitemShut {NoStop}%
\bibitem [{\citenamefont {Traytak}\ and\ \citenamefont
  {Price}(2007)}]{Traytak2007}%
  \BibitemOpen
  \bibfield  {author} {\bibinfo {author} {\bibfnamefont {S.D.}\ \bibnamefont
  {Traytak}}\ and\ \bibinfo {author} {\bibfnamefont {W.S.}\ \bibnamefont
  {Price}},\ }\bibfield  {title} {\enquote {\bibinfo {title} {Exact solution
  for anisotropic diffusion-controlled reactions with partially reflecting
  conditions},}\ }\href@noop {} {\bibfield  {journal} {\bibinfo  {journal} {J.
  Chem. Phys.}\ }\textbf {\bibinfo {volume} {127}},\ \bibinfo {pages}
  {184508:1--8} (\bibinfo {year} {2007})}\BibitemShut {NoStop}%
\bibitem [{\citenamefont {Malgaretti}\ \emph {et~al.}(2018)\citenamefont
  {Malgaretti}, \citenamefont {Popescu},\ and\ \citenamefont
  {Dietrich}}]{Paolo2018}%
  \BibitemOpen
  \bibfield  {author} {\bibinfo {author} {\bibfnamefont {P.}~\bibnamefont
  {Malgaretti}}, \bibinfo {author} {\bibfnamefont {M.N.}\ \bibnamefont
  {Popescu}}, \ and\ \bibinfo {author} {\bibfnamefont {S.}~\bibnamefont
  {Dietrich}},\ }\bibfield  {title} {\enquote {\bibinfo {title}
  {Self-diffusiophoresis induced by fluid interfaces},}\ }\href@noop {}
  {\bibfield  {journal} {\bibinfo  {journal} {Soft Matter}\ }\textbf {\bibinfo
  {volume} {14}},\ \bibinfo {pages} {1375--1388} (\bibinfo {year}
  {2018})}\BibitemShut {NoStop}%
\bibitem [{\citenamefont {Wagner}\ and\ \citenamefont
  {Ripoll}(2017)}]{Ripoll2017}%
  \BibitemOpen
  \bibfield  {author} {\bibinfo {author} {\bibfnamefont {M.}~\bibnamefont
  {Wagner}}\ and\ \bibinfo {author} {\bibfnamefont {M.}~\bibnamefont
  {Ripoll}},\ }\bibfield  {title} {\enquote {\bibinfo {title} {Hydrodynamic
  front-like swarming of phoretically active dimeric colloids},}\ }\href@noop
  {} {\bibfield  {journal} {\bibinfo  {journal} {EPL}\ }\textbf {\bibinfo
  {volume} {119}},\ \bibinfo {pages} {66007:1--7} (\bibinfo {year}
  {2017})}\BibitemShut {NoStop}%
\bibitem [{\citenamefont {Leshansky}\ \emph {et~al.}(2007)\citenamefont
  {Leshansky}, \citenamefont {Kenneth}, \citenamefont {Gat},\ and\
  \citenamefont {Avron}}]{Leshansky2007}%
  \BibitemOpen
  \bibfield  {author} {\bibinfo {author} {\bibfnamefont {A.M.}\ \bibnamefont
  {Leshansky}}, \bibinfo {author} {\bibfnamefont {O.}~\bibnamefont {Kenneth}},
  \bibinfo {author} {\bibfnamefont {O.}~\bibnamefont {Gat}}, \ and\ \bibinfo
  {author} {\bibfnamefont {J.E.}\ \bibnamefont {Avron}},\ }\bibfield  {title}
  {\enquote {\bibinfo {title} {A frictionless microswimmer},}\ }\href@noop {}
  {\bibfield  {journal} {\bibinfo  {journal} {New J. Phys.}\ }\textbf {\bibinfo
  {volume} {9}},\ \bibinfo {pages} {145:1--17} (\bibinfo {year}
  {2007})}\BibitemShut {NoStop}%
\bibitem [{\citenamefont {Theers}\ \emph {et~al.}(2016)\citenamefont {Theers},
  \citenamefont {Westphal}, \citenamefont {Gompper},\ and\ \citenamefont
  {Winkler}}]{Winkler2016}%
  \BibitemOpen
  \bibfield  {author} {\bibinfo {author} {\bibfnamefont {M.}~\bibnamefont
  {Theers}}, \bibinfo {author} {\bibfnamefont {E.}~\bibnamefont {Westphal}},
  \bibinfo {author} {\bibfnamefont {G.}~\bibnamefont {Gompper}}, \ and\
  \bibinfo {author} {\bibfnamefont {R.~G.}\ \bibnamefont {Winkler}},\
  }\bibfield  {title} {\enquote {\bibinfo {title} {Modeling a spheroidal
  microswimmer and cooperative swimming in a narrow slit},}\ }\href@noop {}
  {\bibfield  {journal} {\bibinfo  {journal} {Soft Matter}\ }\textbf {\bibinfo
  {volume} {12}},\ \bibinfo {pages} {7372--7385} (\bibinfo {year}
  {2016})}\BibitemShut {NoStop}%
\end{thebibliography}%

\end{document}